\documentclass[useAMS,usenatbib,letterpaper]{mn2e}

\usepackage{natbib}
\usepackage{nicefrac}
\usepackage{lscape}
\usepackage{epstopdf}
\usepackage{graphicx}
\usepackage{amssymb,amsmath}
\usepackage{url}
\usepackage{multirow}
\usepackage{longtable}
\usepackage{multicol,supertabular}
\usepackage{caption}
\usepackage{array}
\usepackage{pdfpages}

\DeclareMathOperator{\sech}{sech}
\title[Distribution of Galactic Wolf-Rayet stars]
{Spatial distribution of Galactic Wolf-Rayet stars and implications for the global population}

\author[C. K. Rosslowe and P. A. Crowther]{C. K. 
Rosslowe\thanks{Email: chris.rosslowe@sheffield.ac.uk}
 and P. A. Crowther\\
Dept of Physics and Astronomy, University of Sheffield, Hicks Building, Hounsfield Road, S3 7RH, United Kingdom}
\begin{document}

\date{}

\pagerange{\pageref{firstpage}--\pageref{lastpage}} \pubyear{2014}

\maketitle

\label{firstpage}

\begin{abstract} 
We construct revised near-infrared absolute magnitude calibrations for 126
Galactic Wolf-Rayet (WR) stars at known distances, based in part upon recent large
scale spectroscopic surveys. Application to 246 WR stars located in the field,
permits us to map their Galactic distribution. As anticipated, WR stars
generally lie in the thin disk ($\sim$40 pc half width at half maximum) between
Galactocentric radii 3.5--10 kpc, in accordance with other star formation
tracers. We highlight 12 WR stars located at vertical distances of $\geq$300\,pc
from the midplane. Analysis of the radial variation in WR subtypes exposes a
ubiquitously higher $N_{WC}/N_{WN}$ ratio than predicted by stellar evolutionary
models accounting for stellar rotation. Models for non-rotating stars or
accounting for close binary evolution are more consistent with observations. We
consolidate information acquired about the known WR content of the Milky Way
to build a simple model of the complete population. We derive
observable quantities over a range of wavelengths, allowing us to estimate a
total number of $1200^{+300}_{-100}$ Galactic WR stars, implying an average 
duration of ${\sim}\,0.25\,$Myr for the WR phase at the current Milky Way star formation
rate. Of relevance to future spectroscopic surveys, we use this model WR
population to predict follow-up spectroscopy to $K_S\,{\simeq}\,13$\,mag will
be necessary to identify 95\% of Galactic WR stars. We anticipate that ESA's
\textit{Gaia} mission will make few additional WR star discoveries via
low-resolution spectroscopy, though will significantly refine existing distance
determinations. Appendix A provides a complete inventory of 322 Galactic WR
stars discovered since the VIIth catalogue (313 including Annex), including a
revised nomenclature scheme.
\end{abstract}

\begin{keywords}
stars: Wolf-Rayet - stars: evolution - stars: massive - stars: 
distances - infrared: stars - Galaxy: disc
\end{keywords}

\renewcommand{\thefootnote}{\roman{footnote}}

\section{Introduction}

Massive stars exert a major influence on their immediate surroundings, and play
a dominant role in the evolution of their host galaxies. Wolf-Rayet (WR) stars
represent the ultimate, short-lived (${<}\,1$Myr) evolutionary phase of only the
most massive ($M_i\,{>}\,25\,\textrm{M}_\odot$) O-stars (see
\citealt{crowther07}). They possess dense and fast stellar winds, giving them
characteristic strong and broad emission line spectra. Their distinctive
spectral appearance befits them as effective tracers of high-mass star formation
both in the Galaxy (e.g., \citealt{kurtev07}; \citealt{davies12b}) and at
extra-galactic distances \citep{schaerer98}. Through their powerful winds and
likely fate as Type Ib/c supernovae, they are important sources of nuclear
processed material to the interstellar medium (ISM; \citealt{esteban95};
\citealt{dray03}), and are capable of influencing further episodes of star
formation on local \citep{shetty08,kendrew12} and galactic \citep*{hopkins11}
scales. However, the postulated link between WR stars and H-free Type Ib/c SN
remains unestablished \citep{eldridge13}, hence it is crucial we better our 
knowledge of the uniquely resolvable population in the Milky Way.

Strong mass-loss in WR stars proceeds to unveil successive layers of nuclear
processed material, such that examples are seen with spectra dominated by
nitrogen (WN), carbon (WC), and oxygen (WO). WC and WO stars are universally
H-free and He-rich, whereas the surface H mass fraction of WN stars varies from zero
to $X_H{\simeq}\,50\%$. A subset of H-rich WN stars display weak hydrogen 
emission and \emph{intrinsic} hydrogen absorption lines, referred to here as 
WNha stars (see \citealt{crowther11}).
These stars are almost uniquely found in young clusters,
suggesting current masses ${>}65\,M_\odot$ from cluster turn-offs and higher 
luminosities than their core He-burning counterparts \citep{crowther95a}, and hence are
very rare. They differ from `classical' He core-burning WR stars in that they
represent an extension of the upper main-sequence, and are thought to be H-burning. 
We treat these objects separately from other WN
stars when calibrating absolute magnitudes.

Our knowledge of the Galaxy's Wolf-Rayet content rests on the successive
achievements of tailored imaging surveys. The use of narrow-band selection
techniques was pioneered by \citet{massey83} and \citet{moffat83} to identify
extra-galactic WR stars, taking advantage of strong WR star emission lines at
optical wavelengths. \citet{shara91,shara99} applied similar methods to push the
extent of the known Galactic population beyond 5\,kpc from the Sun, and
extension to near-IR wavelengths has facilitated yet deeper investigation of the
Galactic disk \citep{shara09, shara12}. Another distinctive feature of WR stars
- the near-IR excess caused by free-free emission in their winds - has been
exploited to yield further discoveries (\citealt{homeier03};
\citealt{hadfield07}; \citealt*{mauerhan11}). These efforts, alongside several
serendipitous discoveries (e.g., \citealt{clark05}, \citealt{mauerhan10a}), have
brought the recognised Galactic WR star population to ${\sim}635$ as of March
2014\footnote{\url{http://pacrowther.staff.shef.ac.uk/WRcat/}}. 

Several attempts have been made to determine the total number of WR stars in the
Galaxy. \citet{maeder82} used the then-known 157 WR stars to arrive at a total
of ${\sim}$1,200 by assuming the surface density of WR stars to vary with
Galactocentric radius ($R_G$) in the same way as giant HII-regions, including a
dearth at $R_G\,{\lesssim}\,3$kpc. To emphasise the need for IR investigation,
\citet{shara99} created a model WR star population featuring a stellar disk of
exponentially increasing density towards $R_G\,{=}\,0$. From this they inferred
a total of 2,500 Galactic WR stars, or 1,500 if few WR stars inhabit the region
$R_G\,{\lesssim}\,3$kpc, as the decline in gas density suggests (barring the
inner 500\,pc). \citet{vdH01} arrived at a much higher 6,500 WR stars by
extrapolating the surface density of local WR stars (7kpc${<}\,R_G{<}\,12$kpc)
across the entire disk, neglecting the decrease in star formation rate interior
to $R_G\,{\sim}\,3$kpc. Most recently, in the light of numerous WR star
discoveries in IR surveys, \citet{shara09} presented an updated population model
- still featuring an exponential disk of stars - yielding a total of 6,400. The
same work also suggested that observations of WR stars as faint as
$K\,{\simeq}\,15.5$ mag are necessary to achieve a completeness of 95\%.

The Galaxy provides a range of environments over which to test various
predictions of massive star evolution, which has long been expected to depend on
metallicity (Z). As the winds of hot stars are driven by the transfer of photon
momentum to metal-lines (see \citet*{puls08} for a recent review), and mass-loss
dominates the evolutionary fate of the most massive stars, we expect to observe
differences between the population of evolved massive stars in the metal-rich
Galactic Centre (GC) regions, and that of the metal-poor outer Galaxy.
\citet{smith68} first observationally demonstrated evidence for differences in
WR populations between the Milky Way and Magellanic Clouds. \citet{crowther02}
showed that WC subtype variations are primarily a consequence of denser stellar
winds at higher metallicity, while WN stars have long been known to be a more
heterogeneous group. Increased mass-loss is predicted to have two main effects
of WR surface properties; more efficient removal of outer (hydrogen-rich) layers
will lead to quicker progression through post main-sequence evolutionary
phases i.e., from WN to WC stages, and the accelerated spin-down of a star due
to loss of angular momentum will inhibit various internal mixing processes, with
implications for the lifetimes of evolutionary phases \citep{maeder00}.
Systematic testing of such predictions requires statistically significant,
unbiased samples of evolved massive stars, currently only available through IR
investigation of the Galactic disk. 

An improved set of IR tools are necessary to reveal and characterise 
the full Galactic WR population, allowing accurate distances and 
classifications to be obtained. In Section~\ref{sec:absmag_cal} 
we introduce improved near-IR absolute magnitude-spectral type calibrations for WR stars. 
In Section~\ref{sec:3d_dist} these calibrations are applied to estimate 
distances to the majority of the known WR population, from which the radial 
variation of WR subtypes is obtained, allowing a comparison with evolutionary 
model predictions. In Section~\ref{sec:model_wrpop} we develop a toy model to estimate 
the global WR population of the Milky Way. 
Finally, we make predictions about the detectability of WR 
stars, which may be of interest to those planning future surveys. Our findings 
are summarised in Section~\ref{sec:conc}. Appendix A lists all 322 
Galactic Wolf-Rayet stars discovered since the VIIth catalogue of \citet{vdH01} and
its Annex \citep{vdH06}.


\section{Calibration of IR Absolute Magnitudes for WR stars}
\label{sec:absmag_cal}

\citet{vdH01} reviewed and updated the $v$-band absolute magnitude
for Galactic Wolf-Rayet stars. However,
the accuracy and usefulness of this relation is limited by the relatively small
number of WR stars observable at optical wavelengths. Recent discoveries of
visibly obscured WR stars provide a much larger sample, from which 
broad-band calibrations in the near-IR may be obtained. 
In this section we present a calibration of absolute magnitudes over the
wavelength range $1\mbox{--}8\mu$m for each WR spectral type, extending earlier
results by \citet{crowther06b} via additional WR stars located within star
clusters that have been identified within the past decade.

\subsection{Calibration sample}

Adopted distances and spectral types for the WR stars used for our IR absolute
magnitude calibration are shown in Tables \ref{tab:wc_cal}, \ref{tab:wn_cal} \&
\ref{tab:bin_cal} for WC/WO stars, WN stars, and WR stars in binary systems with
OB companions respectively. This sample is drawn from an updated online
catalogue of Galactic WR stars$^1$ and totals $126$, with $91$ inhabiting
clusters, $26$ in OB associations and $9$ appearing `isolated'. By subtype, $85$
of these are Nitrogen (WN) type, $40$ Carbon (WC) type, and $1$ Oxygen (WO)
type. For OB associations that have been historically well studied at optical
wavelengths, membership is taken from \citet{lundstrom84}. For WR stars in
visually obscured clusters we generally accept the membership conclusions of the
discovering author(s), except where noted.

For most star clusters and associations considered, there is typically more than
one distance measurement to be found in the literature. Where these measurements
are in general agreement we favour methods of OB-star spectrophotometry over
main-sequence fitting. A small number of WR stars, in relative isolation, have
kinematic distances derived from velocity measurements of an associated nebula;
we accept these distances but remain wary of kinematic distance estimates in
general because of their sensitivity to the assumed Galactic rotation curve. 

Where multiple consistent distance estimates are found
in the literature we take an average of the reported distances - weighted by the
square of the inverse uncertainty reported on each (e.g., Westerlund 1 and
Car OB1) - indicated by multiple references in Tables \ref{tab:wc_cal} -- \ref{tab:bin_cal}. 
Cases in which inconsistent distances
have been reported in the literature are discussed further in Appendix~\ref{app:distance}.

\newpage
\begin{landscape}
\begin{table}
\caption{Apparently single dust-free WO and WC stars, and dust-producing WC stars of known distance used to calibrate near-IR absolute magnitudes by spectral type. New nomenclature is explained in Appendix \ref{app:new_wr}. }
\centering
\begin{tabular}{ l c c c c c c c c c c c c c}
\hline\\[-2.0ex]
Sp. Type & WR\# &  Cluster & Association & Distance (kpc) & Ref & $J$ & $H$ & $K_S$ & Ref & $A^J_{K_S}$ & $A^H_{K_S}$ & $\bar{A}_{K_S}$ & $M_{K_S}$ \\[1.0ex]
\hline\\[-2.0ex]
\textbf{WO2}   & 142  & Berkeley 87 &           & $1.23\pm0.04$ & 1 & 9.54 & 8.89 & 8.60 & a,a,a & 0.40 & 0.41 & $0.41\pm0.01$ & $-2.26\pm0.07$ \\
\textbf{WC4}   & 144  &  & Cyg OB2              & $1.40\pm0.08$ & 2 & 9.41 & 8.59 & 7.71 & a,a,a & 0.51 & 0.43 & $0.48\pm0.02$ & $-3.50\pm0.13$ \\
\textbf{WC5}   & 111  &  & Sgr OB1              & $1.9\pm0.2$  & 3, 4 & 7.28 & 7.14 & 6.51 & a,a,a & 0.07 & 0.07 & $0.07\pm0.01$ & $-4.95\pm0.23$ \\
               & 114  &  & Ser OB1              & $2.05\pm0.09$ & 5, 6 & 8.98 & 8.43 & 7.61 & a,a,a & 0.36 & 0.34 & $0.35\pm0.02$ & $-4.30\pm0.10$ \\
\textbf{WC6}   & 23   &            & Car OB1   & $2.6\pm0.2$  & 7, 8 & 7.89 & 7.60 & 7.06 & a,a,a & 0.10 & ... & $0.10\pm0.03$ & $-5.12\pm0.17$ \\
               & *48-4 & Danks 1 &               & $4.16\pm0.60$ & 9 & 13.16 & 11.82 & 10.78 & a,a,a & 0.84 & 0.65 & $0.75\pm0.03$ & $-3.06\pm0.32$ \\
               & 154  & & Cep OB1               & $3.5\pm1.0$ & 10 & 9.30 & 9.01 & 8.29 & a,a,a & 0.19 & 0.20 & $0.19\pm0.01$ & $-4.62\pm0.62$ \\
\textbf{WC7}   & 14   & & Anon Vel a            & $2.0\pm0.1$ & 11 & 7.49 & 7.25 & 6.61 & a,a,a & 0.12 & 0.07 & $0.11\pm0.01$ & $-5.00\pm0.11$ \\
               & 68   &           & Cir OB1     & $3.6\pm0.3$ & 12 & 9.90 & 9.39 & 8.75 & a,a,a & 0.25 & 0.08 & $0.20\pm0.02$ & $-4.23\pm0.18$ \\
\textbf{WC8}   & 48-3 & (G305.4+0.1)/Danks 1  & & $4.16\pm0.60$ & 9 & 10.75 & 9.57 & 8.77 & a,a,a & 0.74 & 0.59 & $0.67\pm0.03$ & $-4.99\pm0.32$ \\
               & 48-2  & Danks 2 &               & $4.16\pm0.60$ & 9 & 10.83 & 9.83 & 8.98 & a,a,a & 0.67 & 0.66 & $0.67\pm0.03$ & $-4.78\pm0.32$ \\
               & 77g  & Westerlund 1 &          & $4.0\pm0.2$  & 13, 14 & 11.81 & 10.40 & 9.53 & b,b,b & 0.88 & 0.69 & $0.81\pm0.04$ & $-4.29\pm0.13$ \\
               & 102k & Quintuplet &            & $8.00\pm0.25$ & 15, 16 & 16.71 & 13.45 & 11.19 & d,d,a & 2.49 & 2.55 & $2.50\pm0.08$ & $-5.83\pm0.12$ \\
               & *124-1 & Glimpse 20 &           & $4.45\pm0.65$ & 17 & ... & 10.38 & 9.19 & a,a & ... & 1.14 & $1.14\pm0.10$ & $-5.20\pm0.33$ \\
               & 135  &     & Cyg OB3           & $1.9\pm0.2$ & 18 & 7.23 & 7.11 & 6.66 & a,a,a & 0.07 & 0.10 & $0.08\pm0.01$ & $-4.81\pm0.23$ \\
\textbf{WC9}   & 77p & Westerlund 1 &           & $4.0\pm0.2$  & 13, 14 & 10.12 & 9.09 & 8.29 & b,b,b & 0.76 & 0.76 & $0.76\pm0.05$ & $-5.48\pm0.16$ \\
               & 101f &     GC &                & $8.00\pm0.25$ & 15, 16 & 18.78 & 15.43 & 13.11 & e,e,e & 2.65 & 2.78 & $2.72\pm0.04$ & $-4.12\pm0.08$ \\
               & 101oa &     GC &               & $8.00\pm0.25$ & 15, 16 & 18.48 & 15.43 & 13.01 & e,e,e & 2.56 & 2.92 & $2.72\pm0.04$ & $-4.23\pm0.08 $ \\
               & 102h & Quintuplet &            & $8.00\pm0.25$ & 15, 16 & 16.62 & 13.51 & 11.34 & d,d,a & 2.47 & 2.58 & $2.50\pm0.08$ & $-5.68\pm0.12$ \\[1.0ex]
\textbf{WC8d}  & 53 &             &             & $4.00\pm1.00$ & 19     & 8.75  & 7.92  & 6.84  & a,a,a & ...  & ... & $0.29\pm0.09^\ddagger$ & $-6.43\pm0.55$ \\
               & 102e & Quintuplet &            & $8.00\pm0.25$ & 15, 16  & 17.5 & 13.3 & 10.4 & d,d,d & ... & ... & $3.1\pm0.5^\dagger$ &  $-7.22\pm0.51$ \\ 
               & 102f & Quintuplet &            & $8.00\pm0.25$ & 15, 16 & ...   & ...   & 10.4 & c & ...   & ... & $3.1\pm0.5^\dagger$ &  $-7.22\pm0.51$ \\
               & 113 &            &             & $2.0\pm0.2$ & 20     & 7.02 & 6.28 & 5.49 & a,a,a & ... & ... & $0.38\pm0.01^\ddagger$ & $-6.37\pm0.22$ \\ 
\textbf{WC9d}  & 65 &              & Cir OB1    & $3.6\pm0.3$ & 12 & 8.46 & 7.28 & 6.17 & a,a,a & ... & ... & $0.91\pm0.04^\ddagger$ & $-7.45\pm0.19$ \\
               & 77aa & Westerlund 1 &          & $4.00\pm0.25$ & 13,14 & 10.04 & 8.21 & 6.72 & b,b,b &... & ... & $1.01\pm0.14^\dagger$ & $-7.30\pm0.20$ \\ 
               & 77b  & Westerlund 1 &          & $4.00\pm0.25$ & 13,14 & 9.69 & 7.84 & 6.41 & b,b,b & ... & ... & $1.01\pm0.14^\dagger$ & $-7.61\pm0.20$ \\
               & 77i & Westerlund 1  &          & $4.00\pm0.25$ & 13,14 & 10.13 & 7.64 & 6.90 & b,b,b & ... & ... & $1.01\pm0.14^\dagger$ & $-7.12\pm0.22$ \\
               & 77l & Westerlund 1  &          & $4.00\pm0.25$ & 13,14 & 10.31 & 8.56 & 7.38 & b,b,b & ... & ... & $1.01\pm0.14^\dagger$ & $-6.64\pm0.20$ \\
               & 77m &  Westerlund 1 &        & $4.00\pm0.25$ & 13,14 & 11.26 & 9.51 & 8.23 & b,b,b & ... & ... & $1.01\pm0.14^\dagger$ & $-5.79\pm0.20$ \\
               & 77n &  Westerlund 1 &        & $4.00\pm0.25$ & 13,14 & 9.85 & 7.97 & 7.28 & b,b,b & ... & ... & $1.01\pm0.14^\dagger$ & $-6.74\pm0.20$ \\
               & 95 & Trumpler 27 &             & $2.5\pm0.5$ & 21 & 8.29 & 6.67 & 5.27 & a,a,a & ... & ... & $0.66\pm0.03^{\dagger b}$ & $-7.38\pm0.44$ \\
               & 101ea &  GC    &               & $8.00\pm0.25$ & 15,16 & 17.79 & 13.46 & 10.50 & f,f,f & ... & ... & $3.2\pm0.2$ & $-7.22\pm0.22$ \\
               & 102-2 &  Quintuplet &          & $8.00\pm0.25$ & 15,16 & ... & ... & 10.30 & c & ... & ... & $3.1\pm0.5^\dagger$ & $-7.32\pm0.51$ \\
               & 102-3 &  Quintuplet &          & $8.00\pm0.25$ & 15,16 & 15.49 & 11.71 & 9.32 & d,d,d & ... & ... & $3.1\pm0.5^\dagger$ & $-8.30\pm0.51$ \\
               & 104 &              &           & $2.6\pm0.7$ & 4 & 6.67 & 4.34 & 2.42 & a,a,a & ... & ... & $0.86\pm0.02^\ddagger$ & $-10.44\pm0.64$ \\[0.5ex]
\hline\\[-2.3ex]
\multicolumn{14}{c}{Continued on next page...} \\[0.5ex]
\hline
\end{tabular}
\label{tab:wc_cal}
\end{table}
\end{landscape}

\newpage
\begin{landscape}
\begin{center}
\begin{table}
\centering
\begin{tabular}{ l c c c c c c c c c c c c c}
\hline\\[-2.3ex]
\multicolumn{14}{c}{Continued from previous page} \\[0.5ex] 
\hline\\[-2.0ex]
Sp. Type & WR\# &  Cluster & Association & Distance (kpc) & Ref & $J$ & $H$ & $K_S$ & Ref & $A^J_{K_S}$ & $A^H_{K_S}$ & $\bar{A}_{K_S}$ & $M_{K_S}$ \\[1.0ex] 
\hline\\[-2.0ex]
               & 111a & SGR 1806-20 &           & $8.70\pm1.65$ & 22 & ... & 13.76 & 11.60 & g,g & ... & ... & $3.0\pm0.3^{\dagger}$ & $-6.10\pm0.51$ \\
               & 118-1 & Quartet &               & $6.3\pm0.2$ & 17 & 13.22 & 10.14 & 8.09 & a,a,a & ... & ... & $1.6\pm0.4^{\dagger}$ & $-7.51\pm0.41$ \\[0.5ex]
\hline
\end{tabular}
\end{table}
\end{center}
*Indicates updated spectral types based on an improved near-IR classification scheme. \\
Distance references: (1)\citet{turner06}, (2)\citet{rygl12}, (3)\citet{melnik09}, (4)\citet{tuthill08}, (5)\citet{hillenbrand93}, (6)\citet{djurasevic01}, (7)\citet{smith06}, (8)\citet*{hur12}, (9)\citet{davies12a}, (10)\citet{cappa10}, (11)\citet{lundstrom84}, (12)\citet{vazquez95}, (13)\citet{kothes07}, (14)\citet{koumpia12}, (15)\citet{reid09}, (16)\citet{gillessen13}, (17)\citet{messineo09}, (18)\citet{reid11}, (19)\citet*{martin07}, (20)\citet{esteban95b}, (21)\citet{crowther06b}, (22)\citet{bibby08}. \\
Photometry references: (a)2MASS, (b)\citet{crowther06b}, (c)\citet*{liermann09}, (d)\citet*{figer99}, (e)\citet*{dong12}, (f)\citet{eikenberry04}, (g)\citet{bibby08}. \\
Extinction: $^\dagger$Average of parent cluster, $^\ddagger \mathrm{A}_v$ taken from \citet{vdH01} and converted using $\mathrm{A}_K\,{=}\,0.12\mathrm{A}_v$.  
\end{landscape}

\newpage
\begin{landscape}
\begin{table}
\caption{Apparently single WN stars (and those with an insignificant companion contribution at IR wavelengths) of known distance used to calibrate near-IR absolute magnitudes by spectral type. New nomenclature is explained in Appendix \ref{app:new_wr}. }
\centering
\begin{tabular}{ l c c c c c c c c c c c c c}
\hline\\[-2.0ex]
Spect Type & WR\# &  Cluster & Association & Distance(kpc) & Ref & $J$ & $H$ & $K_S$ & Ref & $A^J_{K_S}$ & $A^H_{K_S}$ & $\bar{A}_{K_S}$ & $M_{K_S}$ \\[1.0ex] 
\hline\\[-2.0ex]
\textbf{WN2b}  & 2    &            & Cas OB1    & $2.4\pm0.8$  & 23 & 10.04 & 9.78 & 9.45 & a,a,a & 0.33 & 0.52 & $0.40\pm0.02$ & $-2.86\pm0.72$ \\
\textbf{WN3b}  & 46   &            & Cru OB4    & $4.00\pm0.85$ & 24, 25 & 10.20 & 10.08  & 9.83 & a,a,a & 0.23 & 0.39 & $0.27\pm0.01$ & $-3.45\pm0.46$ \\
\textbf{WN4b}  & 1    &            & Cas OB7    & $2.3\pm0.5$  & 3 & 8.21 & 7.86 & 7.48 & a,a,a & 0.17 & 0.15 & $0.17\pm0.01$ & $-4.49\pm0.47$ \\
               & 6    &            &            & $1.80\pm0.27$ & 26 & 6.35 &  6.23 & 5.89 & a,a,a & 0.04 & 0.09 & $0.05\pm0.01$ & $-5.44\pm0.33$ \\
               & 7    &            &            & $5.5\pm0.5$  & 27 & 9.97 & 9.67 & 9.27 & a,a,a & 0.16 & 0.19 & $0.17\pm0.01$ & $-4.60\pm0.20$ \\
               &  18  &            & Car OB1   & $2.6\pm0.2$  & 7, 8 & 8.57 & 8.21 & 7.68 & a,a,a & 0.25 & 0.36 & $0.27\pm0.02$ & $-4.67\pm0.17$ \\
               & 35b  & Sher 1     &            & $10.0\pm1.4$   & 28 & 10.95 & 10.35 & 9.76 & a,a,a & 0.39 & 0.46 & $0.41\pm0.03$ & $-5.65\pm0.31$ \\
\textbf{WN6b}  & *102c & Quintuplet &            & $8.00\pm0.25$ & 15, 16 & ... & 13.12 & 11.53 & d,d & ... & 1.78 & $1.78\pm0.38$ & $-4.77\pm0.44$ \\
               & 111c & SGR 1806-20 &           & $8.70\pm1.65$ & 22 & ... &  14.03 & 12.16 & f,f & ... & 2.25 & $2.25\pm0.15$ & $-4.79\pm0.44$ \\
               & 134  &            & Cyg OB3    & $1.9\pm0.2$  & 18 & 6.72 & 6.52 & 6.13 & a,a,a & 0.10 & 0.17 & $0.12\pm0.01$ & $-5.39\pm0.23$ \\
               & 136  &            &  Cyg OB1   & $1.3\pm0.2$  & 29 & 6.13 & 5.90 & 5.56 & a,a,a & 0.09 & 0.10 & $0.09\pm0.01$ & $-5.10\pm0.33$ \\
\textbf{WN7b}  & 77sc & Westerlund 1 &          & $4.0\pm0.2$  & 13, 14 & 10.34 & 9.11 & 8.37 & b,b,b & 0.76 & 0.66 & $0.74\pm0.05$ & $-5.38\pm0.16$ \\
               & 111-2 & Cl 1813-178 &           & $3.6\pm0.7$  & 30 & 9.62 & 8.60 & 7.94 & a,a,a & 0.62 & 0.55 & $0.59\pm0.03$ & $-5.44\pm0.42$ \\
\textbf{WN3}  & 152  &            & Cep OB1    & $3.5\pm1.0$  & 10 & 10.49 & 10.32 & 10.04 & a,a,a & 0.26 & 0.43 & $0.31\pm0.02$ & $-2.99\pm0.62$ \\
\textbf{WN5}  & 77e  & Westerlund 1 &          & $4.0\pm0.2$  & 13, 14 & 11.70 & 10.30 & 9.70 & b,b,b & 0.87 & 0.62 & $0.79\pm0.06$ & $-4.10\pm0.16$ \\
               & 77q  & Westerlund 1 &          & $4.0\pm0.2$  & 13, 14 & 11.92 & 10.84 & 10.26 & b,b,b & 0.70 & 0.59 & $0.67\pm0.04$ & $-3.42\pm0.13$ \\
               & 77sd & Westerlund 1 &          & $4.0\pm0.2$  & 13, 14 & 12.36 & 11.08 & 10.25 & b,b,b & 0.92 & 0.94 & $0.93\pm0.04$ & $-3.69\pm0.13$ \\
\textbf{WN6}  & 67   & Pismis 20    & Cir OB1  & $3.6\pm0.3$  & 12 & 9.28 & 8.86 & 8.45 & a,a,a & 0.31 & 0.35 & $0.32\pm0.02$ & $-4.65\pm0.18$ \\
               & 77sb & Westerlund 1 &          & $4.0\pm0.2$  & 13, 14 & 11.00 & 9.98 & 9.45 & b,b,b & 0.65 & 0.52 & $0.61\pm0.04$ & $-4.17\pm0.12$ \\
               & 77a  & Westerlund 1 &          & $4.0\pm0.2$  & 13, 14 & 11.72 & 10.67 & 10.00 & b,b,b & 0.73 & 0.72 & $0.73\pm0.04$ & $-3.74\pm0.13$ \\
               & 77s  & Westerlund 1 &          & $4.0\pm0.2$  & 13, 14 & 10.77 &  9.72 & 9.20 & b,b,b & 0.66 & 0.51 & $0.61\pm0.04$ & $-4.42\pm0.12$ \\
               & 85  &           &              & $2.8\pm1.1$  & 31 & ... & 7.94 & 7.48 & a,a & ... & 0.43 & $0.43\pm0.16$ & $-5.19\pm0.87$ \\
               & *101o & GC          &          & $8.00\pm0.25$ & 15, 16 & 17.94 & 14.13 & 11.60 & e,e,e & 3.00 & 3.20 & $3.11\pm0.04$ & $-6.02\pm0.08$ \\
               & 115  &            & Ser OB1    & $2.05\pm0.09$ & 5, 6 & 7.99 & 7.42 & 6.95 & a,a,a & 0.41 & 0.44 & $0.42\pm0.02$ & $-5.03\pm0.10$ \\
\textbf{WN7}  & 75ba &         &               & $4.1\pm0.4$  & 32 & 10.22 & 9.29 & 8.91 & a,a,a & 0.56 & 0.39 & $0.51\pm0.04$ & $-4.67\pm0.23$ \\
               & 77r & Westerlund 1 &           & $4.0\pm0.2$  & 13, 14 & 11.63 & 10.31 & 9.61 & b,b,b & 0.90 & 0.83 & $0.87\pm0.04$ & $-4.27\pm0.13$ \\
               & 77j & Westerlund 1 &           & $4.0\pm0.2$  & 13, 14 & 11.36 & 9.97 & 9.28 & b,b,b & 0.93 & 0.82 & $0.89\pm0.04$ & $-4.62\pm0.13$ \\
               & 77d & Westerlund 1 &           & $4.0\pm0.2$  & 13, 14 & 11.06 & 9.83 & 9.26 & b,b,b & 0.80 & 0.65 & $0.74\pm0.04$ & $-4.49\pm0.13$ \\
               & *77sa & Westerlund 1 &          & $4.0\pm0.2$  & 13, 14 & 12.11 & 10.75 & 10.04 & b,b,b & 0.92 & 0.85 & $0.89\pm0.04$ & $-3.86\pm0.13$ \\
               & 78 & NGC 6231 &   Sco OB1      & $1.64\pm0.03$ & 33 & 5.44 & 5.27 & 4.98 & a,a,a & 0.16 & 0.25 & $0.18\pm0.01$ & $-6.27\pm0.05$ \\
               & 87 & Halven-Moffat 1 &         & $3.3\pm0.3$  & 34 & 8.00 & 7.45 & 7.09 & a,a,a & 0.37 & 0.36 & $0.37\pm0.02$ & $-5.88\pm0.20$ \\
               & *101ai & GC      &             & $8.00\pm0.25$ & 15, 16 & ... &  14.33 & 12.12 & e,e & ... & 2.84 & $2.84\pm0.07$ & $-5.23\pm0.10$ \\
               & *111-4 & Cl 1813-178 &          & $3.6\pm0.7$  & 30 & 10.31 & 9.27 & 8.66 & a,a,a & 0.72 & 0.70 & $0.71\pm0.03$ & $-4.84\pm0.42$ \\
\textbf{WN8}  & 12\textbullet & Bochum 7 &                & $4.2\pm2.1$ & 35 & 8.62 & 8.26 & 7.87 & a,a,a & 0.29 & 0.40 & $0.32\pm0.02$ & $-5.57\pm1.09$ \\
               & *48-7 & Danks 1  &              & $4.16\pm0.60$ & 9 & 9.81 & 8.48 & 7.65 & a,a,a & 0.97 & 1.01 & $0.99\pm0.04$ & $-6.43\pm0.32$ \\
               & 66   &       & Cir OB1         & $3.6\pm0.3$ & 12 & 8.93 & 8.48 & 8.15 & a,a,a & 0.31 & 0.31 & $0.31\pm0.02$ & $-4.94\pm0.18$ \\[0.5ex]
\hline\\[-2.3ex]
\multicolumn{14}{c}{Continued on next page...} \\[0.5ex]
\hline
\end{tabular}
\label{tab:wn_cal}
\end{table}
\end{landscape}

\newpage
\begin{landscape}
\begin{table}
\centering
\begin{tabular}{ l c c c c c c c c c c c c c}
\hline\\[-2.3ex]
\multicolumn{14}{c}{Continued from previous page} \\[0.5ex] 
\hline\\[-2.0ex]
Spect Type & WR\# &  Cluster & Association & Distance(kpc) & Ref & $J$ & $H$ & $K_S$ & Ref & $A^J_{K_S}$ & $A^H_{K_S}$ & $\bar{A}_{K_S}$ & $M_{K_S}$ \\ \hline
               & 77c & Westerlund 1 &           & $4.0\pm0.2$  & 13, 14 & 10.89 & 9.57 & 8.86 & b,b,b & 0.90 & 0.85 & $0.88\pm0.04$ & $-5.03\pm0.13$ \\
               & 77h & Westerlund 1 &           & $4.0\pm0.2$  & 13, 14 & 10.75 & 9.42 & 8.76 & b,b,b & 0.89 & 0.77 & $0.84\pm0.04$ & $-5.09\pm0.13$ \\
               & 89 & Halven-Moffat 2 &         & $3.3\pm0.3$  & 34 & 7.39 & 6.96 & 6.58 & a,a,a & 0.32 & 0.39 & $0.34\pm0.02$ & $-6.36\pm0.20$ \\
               & 101b & GC       &              & $8.00\pm0.25$ & 15, 16 & ...   & 13.53 & 11.43 & e,e & ... & 2.69 & $2.69\pm0.06$ & $-5.77\pm0.09$ \\
               & 101nc & GC       &             & $8.00\pm0.25$ & 15, 16 & 17.38 & 14.23 & 11.91 & e,e,e & 2.60 & 2.99 & $2.79\pm0.04$ & $-5.40\pm0.08$ \\
               & *101oc & GC       &             & $8.00\pm0.25$ & 15, 16 & 18.66 & 14.93 & 12.61 & e,e,e & 2.89 & 2.99 & $2.94\pm0.04$ & $-4.85\pm0.08$ \\
               & *101dd & GC       &             & $8.00\pm0.25$ & 15, 16 & 18.96 & 15.43 & 13.01 & e,e,e & 2.84 & 3.12 & $2.98\pm0.04$ & $-4.49\pm0.08$ \\
               & 102a   & Arches   &             & $8.00\pm0.25$ & 15, 16 & 17.17 & 13.26 & 11.02 & h,a,a & 2.94 & 2.88 & $2.91\pm0.05$ & $-6.40\pm0.10$ \\
               & *102ae & Arches   &             & $8.00\pm0.25$ & 15, 16 & 15.43 & 12.40 & 10.62 & i,i,i & 2.28 & 2.25 & $2.27\pm0.02$ & $-6.16\pm0.07$ \\
               & *102af & Arches   &             & $8.00\pm0.25$ & 15, 16 & 15.97 & 12.81 & 10.88 & i,i,i & 2.42 & 2.46 & $2.44\pm0.03$ & $-6.08\pm0.07$ \\
               & *102ah & Arches   &             & $8.00\pm0.25$ & 15, 16 & 15.23 & 12.03 & 10.07 & i,i,i & 2.45 & 2.50 & $2.48\pm0.03$ & $-6.92\pm0.07$ \\
               & *102al & Arches   &             & $8.00\pm0.25$ & 15, 16 & 15.11 & 12.09 & 10.24 & i,i,i & 2.31 & 2.34 & $2.33\pm0.02$ & $-6.60\pm0.07$ \\
               & 124 &            &             & $3.35\pm0.67$ & 36 & 8.58 & 8.18 & 7.73 & a,a,a & 0.34 & 0.47 & $0.39\pm0.02$ & $-5.28\pm0.44$ \\
\textbf{WN9}  & *48-6 & (G305.4+0.1)/Danks 1 &  & $4.16\pm0.60$ & 9 & 10.21 & 8.57 & 7.58 & a,a,a & 1.19 & 1.24 & $1.21\pm0.05$ & $-6.73\pm0.32$ \\
               & *48-10 & Danks 1 &               & $4.16\pm0.60$ & 9 & 9.42 & 8.15 & 7.48 & a,a,a & 0.86 & 0.79 & $0.83\pm0.04$ & $-6.45\pm0.32$ \\
               & 48-9 & Danks 1 &               & $4.16\pm0.60$ & 9 & 8.26 & 7.27 & 6.61 & a,a,a & 0.72 & 0.77 & $0.74\pm0.03$ & $-7.22\pm0.32$ \\
               & 77k & Westerlund 1 &           & $4.0\pm0.2$  & 13, 14 & 9.08 & 7.72 & 7.19 & b,b,b & 0.84 & 0.59 & $0.75\pm0.04$ & $-6.57\pm0.13$ \\
               & *101m & GC        &             & $8.00\pm0.25 $ & 15, 16 & 16.58 & 13.53 & 11.32 & e,e,e & 2.50 & 2.84 & $2.67\pm0.03$ & $-5.86\pm0.08$ \\
               & *101e & GC        &             & $8.00\pm0.25 $ & 15, 16 & 15.87 & 12.73 & 10.41 & e,e,e & 2.60 & 2.99 & $2.79\pm0.04$ & $-6.90\pm0.08$ \\
               & *102aa & Arches   &             & $8.00\pm0.25$ & 15, 16 & ... & ... & 11.18  & j & ... & ... & $2.48\pm0.37^\dagger$ & $-5.82\pm0.39$ \\
               & *102ad & Arches   &             & $8.00\pm0.25$ & 15, 16 & 15.86 & 12.44 & 10.35 & i,i,i & 2.63 & 2.69 & $2.66\pm0.03$ & $-6.83\pm0.08$ \\
               & *102ag & Arches   &             & $8.00\pm0.25$ & 15, 16 & 15.67 & 12.45 & 10.46 & i,i,i & 2.48 & 2.55 & $2.52\pm0.03$ & $-6.58\pm0.07$ \\
               & *102ai & Arches   &             & $8.00\pm0.25$ & 15, 16 & ... & 12.24 & 10.34 & a,a & ... & 2.41 & $2.41\pm0.19$ & $-6.59\pm0.22$ \\
               & *102aj & Arches   &             & $8.00\pm0.25$ & 15, 16 & 16.55 & 13.56 & 11.79 & i,i,i & 2.26 & 2.24 & $2.25\pm0.02$ & $-4.98\pm0.07$ \\
               & *102bb & Arches   &             & $8.00\pm0.25$ & 15, 16 & 15.58 & 12.36 & 10.36 & i,i,i & 2.48 & 2.54 & $2.52\pm0.03$ & $-6.67\pm0.07$ \\
               & *102bc & Arches   &             & $8.00\pm0.25$ & 15, 16 & ... & 13.14 & 11.20 & i,i &... & 2.47 & $2.47\pm0.13$ & $-5.79\pm0.16$ \\
               & 102d & Quintuplet  &           & $8.00\pm0.25$ & 15, 16 & 15.58 & 12.40 & 10.50 & d,d,c & 2.41 & 2.42 & $2.42\pm0.09$ & $-6.43\pm0.15$ \\
               & 102hb & Quintuplet  &          & $8.00\pm0.25$ & 15, 16 & 14.19 & 10.90 & 9.60 & d,d,c & 2.18 & 1.61 & $2.01\pm0.09$ & $-6.93\pm0.15$ \\
               & 102i & Quintuplet  &           & $8.00\pm0.25$ & 15, 16 & 14.77 & 11.67 & 10.22 & d,a,a & 2.16 & 1.81 & $1.98\pm0.05$ & $-6.27\pm0.10$ \\
               & 102j & Quintuplet  &           & $8.00\pm0.25$ & 15, 16 & 14.66 & 11.77 & 10.23 & a,a,a & 2.10 & 1.94 & $2.03\pm0.03$ & $-6.32\pm0.08$ \\
               & 105  &         & Sgr OB1       & $1.9\pm0.2$ & 3, 4 & 7.04 & 6.25 & 5.73 & a,a,a & 0.56 & 0.58 & $0.57\pm0.02$ & $-6.24\pm0.23$ \\
\textbf{WN6ha} & 20a1 &         &               & $8.0\pm1.0$ & 37 & 9.61 & 8.84 & 8.34 & a,a,a & 0.61 & 0.65 & $0.63\pm0.03$ & $-6.80\pm0.38$ \\
               & 20a2 &         &               & $8.0\pm1.0$ & 37 & 9.61 & 8.84 & 8.34 & a,a,a & 0.61 & 0.65 & $0.63\pm0.03$ & $-6.80\pm0.38$ \\
               & 24 & Coll 228          & Car  OB1     & $2.6\pm0.2$ & 7, 8 & 6.10 & 6.01 & 5.82 & a,a,a & 0.14 & 0.23 & $0.16\pm0.01$ & $-6.42\pm0.17$ \\
               & 25\textbullet & Trumpler 16       & Car  OB1     & $2.6\pm0.2$ & 7, 8 & 6.26 & 5.97 & 5.72 & a,a,a & 0.26 & 0.31 & $0.28\pm0.02$ & $-6.63\pm0.17$ \\    
               & 43A1 & NGC 3603 &                & $7.6\pm0.4$ & 38 & 8.57 & 8.36 & 7.78 & k,k,k & 0.38 & 0.77 & $0.49\pm0.05$ & $-7.11\pm0.16$ \\
               & 43A2 & NGC 3603 &                & $7.6\pm0.4$ & 38 & 8.98 & 8.77 & 8.19 & k,k,k & 0.38 & 0.77 & $0.49\pm0.04$ & $-6.70\pm0.14$ \\     
\hline\\[-2.3ex]
\multicolumn{14}{c}{Continued on next page...} \\[0.5ex]
\hline
\end{tabular}
\end{table}
\end{landscape}

\newpage
\begin{landscape}
\begin{center}
\begin{table}
\centering
\begin{tabular}{ l c c c c c c c c c c c c c}
\hline\\[-2.3ex]
\multicolumn{14}{c}{Continued from previous page} \\[0.5ex] 
\hline\\[-2.0ex]
Spect Type & WR\# &  Cluster & Association & Distance(kpc) & Ref & $J$ & $H$ & $K_S$ & Ref & $A^J_{K_S}$ & $A^H_{K_S}$ & $\bar{A}_{K_S}$ & $M_{K_S}$ \\[1.0ex] 
\hline\\[-2.0ex]
               & 43B & NGC 3603 &                & $7.6\pm0.4$ & 38 & 7.78 & 7.70 & 7.08 & k,k,k & 0.34 & 0.83 & $0.47\pm0.03$ & $-7.80\pm0.13$ \\
              & 43C\textbullet & NGC 3603 &                & $7.6\pm0.4$ & 38 & 8.49 & 8.13 & 7.81 & k,k,k & 0.33 & 0.41 & $0.35\pm0.03$ & $-6.95\pm0.13$ \\
\textbf{WN7ha} & 22\textbullet &           & Car        & $2.6\pm0.2$ & 7, 8 & 5.71 & 5.58 & 5.39 & a,a,a & 0.17 & 0.26 & $0.20\pm0.03$ & $-6.81\pm0.17$ \\
               & *125-3 & Mercer 23 &            & $6.5\pm0.3$ & 39 & 8.65 & 7.84 & 7.33 & l,l,l & 0.64 & 0.70 & $0.67\pm0.02$ & $-7.40\pm0.10$ \\
\textbf{WN9ha} & 79a & NGC 6231 &   Sco OB1     & $1.64\pm0.03$ & 33 & 5.15 & 5.09 & 4.90 & a,a,a & 0.14 & 0.25 & $0.17\pm0.02$ & $-6.34\pm0.05$ \\
               & 79b &    &    KQ Sco           & $3.5\pm0.5$ & 40 & 6.76 & 6.62 & 6.48 & a,a,a & 0.16 & 0.19 & $0.17\pm0.01$ & $-6.41\pm0.31$ \\[0.5ex]
\hline
\end{tabular}
\end{table}
\end{center}
*Indicates updated spectral types based on an improved near-IR classification scheme. \\
\textbullet Spectroscopic binary systems with a dominant WR component at IR wavelengths ($F^{WR}/F^{sys}>\nicefrac{2}{3}$). Binary detections: (WR\,12)\citet{fahed12}, (WR\,25)\citet{gamen06}, (WR\,43C)\citet{schnurr08}, (WR\,22)\citet{schweickhardt99}. \\ 
Distance references (1-22 as in Table \ref{tab:wc_cal}): (23)\citet{arnal99}, (24)\citet*{crowther95b}, (25)\citet*{tovmassian96}, (26)\citet{howarth95}, (27)\citet{cappa99}, (28)\citet*{moffat91}, (29)\citet{garmany92}, (30)\citet{messineo11}, (31)\citet{vazquez05}, (32)\citet*{cohen05}, (33)\citet{sana06}, (34)\citet{vazquez01}, (35)\citet*{corti07}, (36)\citet*{marchenko10}, (37)\citet{rauw07}, (38)\citet{melena09}, (39)\citet{hanson10}, (40)\citet{bohannan99}.\\ 
Photometry references (a-g as in Table \ref{tab:wc_cal}): (h)\citet{cotera99}, (i)\citet*{espinoza09}, (j)\citet{martins08}, (k)\citet*{harayama08}, (l)\citet{hanson10}. \\
Extinction: $^\dagger$Average of parent cluster.
\end{landscape}

\newpage
\begin{landscape}
\begin{table}
\caption{WR stars in confirmed binary systems (WR+non-WR) used for absolute magnitude-spectral type calibrations.}
\centering
\begin{tabular}{l c c c c l c c l l}
\hline\\[-2.0ex]
WR\# & Spect Type  & Cluster/Association & Distance(kpc) & Ref & JHK$_S^{sys}$ & Ref & Flux ratio         & Extinction & $M^{WR}$ \\ 
      &            &                     &               &     &               &     &  $F^{WR}/F^{sys}$  &            &          \\[1.0ex] 
\hline\\[-2.0ex]
11  & WC8+O7.5III  &              & $0.342\pm0.035$ & 41  & $J=2.12$   & m & $0.45\pm0.32$     & 0.00                   & $M_J=-4.68\pm0.81$ \\
    &              &              &                 &     & $H=2.17$   & m & $0.48\pm0.30$     & 0.00                   & $M_H=-4.70\pm0.72$ \\
    &              &              &                 &     & $K_S=1.98$ & m & $0.60\pm0.23$ & 0.00                   & $M_{K_S}=-5.14\pm0.48$ \\[2ex]
77o & WN7o+?       & Westerlund 1 & $4.0\pm0.2$     & 13, 14 & $J=10.34$ & b & $0.59\pm0.10$     & $2.98\pm0.20$     & $M_J=-5.08\pm0.26$ \\
    &              &              &                 &     & $K_S=8.37$ & b & $0.80\pm0.10$ & $0.96\pm0.05$ & $M_{K_S}=-5.36\pm0.23$ \\[2ex]
79  & WC7+O5-8V    &  Sco OB1     & $1.64\pm0.03$   & 33  & $J=5.96$   & a & $0.41\pm0.05$     & $0.48\pm0.03$     & $M_J=-4.62\pm0.14$ \\
    &              &              &                 &     & $K_S=5.39$ & a & $0.45\pm0.05$ & $0.16\pm0.01$ & $M_{K_S}=-4.97\pm0.13$ \\[2ex]
93  & WC7+O7-9     &  Pismis 24   & $2.0\pm0.2$     & tw. & $K_S=5.87$ & a & $0.73\pm0.72$ & $0.58\pm0.03^\ddagger$ & $M_{K_S}=-5.88\pm1.10$ \\[2ex]
127 & WN5+O8.5V   &  Vul OB2     & $4.41\pm0.12$   & 42  & $J=9.18$   & a & $0.58\pm0.17$     & $0.56\pm0.09$   & $M_{J}=-4.00\pm0.65$ \\
    &              &              &                 &     & $H=9.02$   & a & $0.59\pm0.13$     & $0.31\pm0.05$   & $M_{H}=-3.93\pm0.26$ \\
    &              &              &                 &     & $K_S=8.76$ & a & $0.67\pm0.09$ & $0.18\pm0.03$ & $M_{K_S}=-4.21\pm0.16$ \\[2ex]
133 & WN5+O9I       &  NGC 6871     & $2.14\pm0.07$   & 43  & $J=6.32$   & a & $0.22\pm0.05$     & $0.55\pm0.05$   & $M_{J}=-3.86\pm0.19$ \\
    &              &              &                 &     & $K_S=6.25$ & a & $0.23\pm0.05$ & $0.18\pm0.02$ & $M_{K_S}=-4.04\pm0.19$ \\[2ex]
137 & WC7+O9       &  Cyg OB1     & $1.3\pm0.2$     & 29  & $J=7.00$   & n & $0.41\pm0.12$ & $0.59\pm0.07$ & $M_J=-3.19\pm0.47$ \\
    &              &              &                 &     & $K_S=6.43$ & n & $0.46\pm0.13$   & $0.19\pm0.02$ & $M_{K_S}=-3.49\pm0.46$ \\[2ex]
139 & WN5+O6III-V &  Cyg OB1     & $1.3\pm0.2$     & 29  & $J=6.70$   & a & $0.44\pm0.06$     & $0.59\pm0.06$   & $M_{J}=-3.57\pm0.36$ \\
    &              &              &                 &     & $K_S=6.33$ & a & $0.50\pm0.07$ & $0.19\pm0.02$ & $M_{K_S}=-3.69\pm0.36$ \\[2ex]
141 & WN5+O5III-V &  Cyg OB1     & $1.3\pm0.2$     & 29  & $J=7.34$   & a & $0.65\pm0.07$     & $0.45\pm0.15^\dagger$   & $M_{J}=-3.21\pm0.34$ \\[2ex]
157 & WN5+?       &  Markarian 50 & $3.46\pm0.35$  &  44 & $J=8.22$   & a & $0.47\pm0.10$  & $0.90\pm0.15$   & $M_J=-4.53\pm0.30$ \\
    &              &              &                 &     & $K_S=7.73$ & a & $0.68\pm0.10$  & $0.29\pm0.04$ & $M_{K_S}=-4.88\pm0.36$ \\[0.5ex]
\hline\\[-2.0ex]
\end{tabular}
\label{tab:bin_cal}
\end{table}
Distance references (1-40 as in Tables \ref{tab:wc_cal} \& \ref{tab:wn_cal}): (41)\citet{leeuwen07}, (42)\citet{turner80}, (43)\citet{malchenko09}, (44)\citet*{baume04}. \\ 
Photometry references (a-l as in Tables \ref{tab:wc_cal} \& \ref{tab:wn_cal}): (m)\citet{williams90b}, (n)\citet{williams01}. \\ 
Spectral types: (WR$\,11$) \citet{demarco99}, (WR$\,77$o) \citet{crowther06b}, (WR$\,79$) \citet*{smith90}, (WR$\,93$) \citet*{lortet84}, (WR$\,127$) \citet*{chevrotiere11}, (WR$\,133$) \citet{underhill94}, (WR$\,137$) \citet{williams01}, (WR$\,139$) \citet*{marchenko94}, (WR$\,141$) \citet*{marchenko98}, (WR$\,157$) \citet{smith96}. \\
Extinction: $^\ddagger\mathrm{A}_v$ taken from \citet{vdH01} and converted using $\mathrm{A}_K\,{=}\,0.12\mathrm{A}_v$. $^\dagger$Average extinction taken from WR$\,136$ and WR$\,139$, also members of Cyg OB1.
\end{landscape}


\subsection{Classification of WR stars from near-infrared spectra}

Spectral types of approximately $50\%$ of our calibration sample have been
obtained from optical spectroscopy following \citet*{smith96} for WN subtypes and
\citet*{crowther98} for WC and WO subtypes. For the remaining objects, we
reassess published spectral types based upon their near-IR ($1\mbox{--}5\mu$m)
spectra, using an updated version of the scheme from \citet{crowther06b}. 
Updated spectral types are shown in Tables 1--3. 
Diagnostics involve emission line equivalent width ratios drawn from adjacent
ionisation stages of the same atomic species. Full details will be presented
elsewhere, but we shall briefly discuss the methodology here.

\subsubsection{WN diagnostics}

For WN stars, ratios of He\,{\sc ii}/He\,{\sc i} lines provide the primary
classification diagnostics; particularly He\,{\sc ii} $\,1.012\mu$m/He\,{\sc
i} $1.083\mu$m in the J-band and He\,{\sc ii}+Br$\gamma 2.165\mu$m/He\,{\sc ii}
$2.189\mu$m in the K-band. Degeneracies in these primary line
ratios between spectral types are lifted by considering various weaker lines. For
example, we find WN$7\mbox{--}9$ types can be distinguished by
considering the strength of Si\,{\sc iv} $1.190\mu$m relative to He\,{\sc
ii} $1.163\mu$m, while an inspection of spectral morphology in the K-band
permits WN4--6 stars to be distinguished 
using N\,{\sc v} $2.100\mu$m and N\,{\sc iii} $2.116\mu$m.

\subsubsection{WC diagnostics}

For WC stars, ratios of C\,{\sc ii-iv} provide the primary
classification diagnostics, with C\,{\sc iv} $1.191\mu$m/C\,{\sc iii} $0.972\mu$m 
in the J-band permitting a consistent classification to
optical lines. He\,{\sc ii} $1.012\mu$m/He\,{\sc i} $1.083\mu$m also prove useful 
for classification, although the H-band contains little diagnostic information. 
However, we find the very strong C\,{\sc
iv} $1.736\mu$m line to be useful for recognising the dilution effects of hot
circumstellar dust (see Section \ref{sec:dustywc}). In the K-band, the ratio of
C\,{\sc iv} $2.070\mbox{--}2.084\mu$m to C\,{\sc iii}+He\,{\sc i} $2.112\mbox{--}2.114\mu$m 
serves as a good ionisation diagnostic, but is incapable of discriminating
between WC$4\mbox{--}6$ stars.

\subsubsection{Accuracy of near IR spectral types}

To gain an insight into the reliability of our revised near-IR classification 
scheme, we have carried out blind-tests using WR stars with optically 
derived spectral types, and find J and K-band diagnostics provide 
the highest level of consistency. 

For WN stars, an exact 3D spectral type \citep{smith96} can 
be achieved from low-resolution J through K spectra, with solely the 
J-band proving criteria for identifying the presence of Hydrogen. We find 
spectra in J or K alone yield spectral types with an accuracy of ${\pm}\,1$, and 
H-band diagnostics are accurate to ${\pm}\,2$. 

For WC stars, we find an exact spectral type can be assigned solely from a
J-band spectrum. Our K-band diagnostics are capable of ${\pm}\,1$ spectral type
amongst early (WC$4\mbox{--}6$) types, and exact classification for late
(WC$7\mbox{--}9$). We only find it possible to distinguish between WCE/WCL using
spectra from H, L or M bands. For both WN and WC stars, ionisation types at
either extreme (WN9,3,2; WC9) are conspicuous in spectral appearance, and can be
identified with a higher degree of certainty, usually by inspection of the
spectral morphology alone. Stars are only included in our calibration sample if
we are confident within ${\pm}\,1$ of their spectral types.  

\subsection{Photometry and extinction}
\label{sec:photom-ext}

In general we take JHK$_S$ photometry for each WR star from the Two Micron
All-Sky Survey (2MASS) point source catalogue \citep{skrutskie06}, plus IRAC
$[3.6]\mbox{--}[8.0](\mu$m) photometry from the GLIMPSE survey
\citep{benjamin03} for sufficiently isolated sources in the surveyed field. 

We require a minimum quality flag of C where 2MASS photometry is used. Many cluster
and association members are located in fields too crowded for 2MASS to be
useful. In such cases we turn to dedicated observations with higher spatial
resolution of the stellar groupings in question (Tables
\ref{tab:wc_cal}--\ref{tab:bin_cal}). 

We have attempted to ensure consistency in the near-IR photometry used. 
For example, observations of WR stars in the Galactic Centre region are assembled by
\citet{dong12}, consisting of Hubble Space Telescope snapshot imaging plus multiple
ground-based observations. In this case, to maintain consistency with other
assembled photometry, we construct and apply a simple algorithm to convert the
\citeauthor{dong12} JHK$_S$ values into the 2MASS photometric system (following
their equations $7\mbox{--}9$). However, in general we regard the slight
differences between ground-based filter systems as insignificant, as they have a
much smaller effect on calculated absolute magnitudes than that of distance
uncertainties.

\begin{table}
\caption{Intrinsic colours adopted for each type of WR star, primarily from
\citet{crowther06b}, supplemented with unpublished stellar atmospheric model results for 
additional types considered here (e.g. WO).}
\centering
\begin{tabular}{ l r r }
\hline\\[-2.0ex]
Sp. type & (J--K$_S$)$_0$ & (H--K$_S$)$_0$ \\ [1.0ex] 
\hline\\[-2.0ex]
WO2  & 0.11 & 0.00 \\
WC4--7 & 0.62 & 0.58 \\
WC8   & 0.43 & 0.38 \\
WC9   & 0.23 & 0.26 \\
WN4--7b & 0.37 & 0.27 \\
WN2--4 & --0.11 & --0.03 \\
WN5--6 & 0.18 & 0.16 \\
WN7--9 & 0.13 & 0.11 \\
WN5--6ha & --0.015 & 0.03 \\
WN7--9ha & --0.04 & 0.005 \\
\hline
\end{tabular}
\label{tab:int_col} 
\end{table}

We calculate an extinction towards each calibration star by evaluating the colour 
excesses E$_{J-K_S}$ and E$_{H-K_S}$, utilising intrinsic JHK$_S$ Wolf-Rayet colours given by 
\citet{crowther06b}, updated in Table \ref{tab:int_col}. 
Two values of K$_S$-band extinction follow;
\begin{equation}
A^J_{K_S} =  E_{J-K_S}\times(\frac{A_J}{A_{K_S}}-1)^{-1},
\label{eq:ak1}
\end{equation}  
and
\begin{equation}
A^H_{K_S} =  E_{H-K_S}\times(\frac{A_H}{A_{K_S}}-1)^{-1}.
\label{eq:ak2}
\end{equation}  
The second terms in Equations \ref{eq:ak1} and \ref{eq:ak2} require knowledge of
the near-IR extinction law. 

\setlength\extrarowheight{1pt}
\begin{table*}
\centering
\begin{minipage}{15cm}
\caption{Calibrated near-IR absolute magnitudes for Galactic WR stars. 
The number of objects used to arrive at each value is indicated in adjacent parentheses. 
Two uncertainties are shown with each value; formal errors (parenthesised, 
Equation \ref{eq:cal_err1}) do not account for the intrinsic spread in magnitude within a 
WR spectral type - estimated to be ${\sim}\,0.3\,$mag - which is incorporated into the final 
(nonparenthesised) uncertainty. }
\centering
\begin{tabular}{lcccccc}
\hline\\[-2.0ex]
Sp. type & $\bar{M}_J$ & (N) & $\bar{M}_H$ & (N) & $\bar{M}_{K_S}$ & (N) \\[1.0ex] 
\hline\\[-2.0ex]
WO2 & $-2.15\pm(0.08)\,0.31$ & (1) & $-2.26\pm(0.08)\,0.31$ &(1) & $-2.26\pm(0.07)\,0.31$& (1)  \\
WC4 & $-2.88\pm(0.13)\,0.33$ &(1) & $-2.92\pm(0.13)\,0.33$ &(1) & $-3.50\pm(0.13)\,0.33$ &(1)  \\ 
WC5 & $-3.80\pm(0.27)\,0.40$ &(2) & $-3.84\pm(0.27)\,0.40$ &(2) & $-4.40\pm(0.25)\,0.39$ &(2)  \\
WC6 & $-4.03\pm(0.61)\,0.68$ &(3) & $-4.06\pm(0.62)\,0.69$ &(3) & $-4.66\pm(0.61)\,0.68$ &(3)  \\ 
WC7 & $-4.21\pm(0.20)\,0.36$ &(4) & $-4.25\pm(0.20)\,0.36$ &(2) & $-4.84\pm(0.21)\,0.36$ &(5)  \\
WC8	& $-4.26\pm(0.24)\,0.38$ &(6) & $-4.35\pm(0.22)\,0.37$ &(7) & $-5.04\pm(0.28)\,0.41$ &(7)  \\
WC9	& $-4.42\pm(0.39)\,0.49$ &(4) & $-4.17\pm(0.32)\,0.44$ &(4) & $-4.57\pm(0.38)\,0.48$ &(4)  \\ 
WC8d & $-5.53\pm(0.25)\,0.39$ &(3) & $-5.83\pm(0.23)\,0.38$ &(3) & $-6.57\pm(0.27)\,0.41$ &(4)  \\
WC9d & $-6.34\pm0(.25)\,0.39$ &(12) & $-6.63\pm(0.21)\,0.37$ &(13) & $-7.06\pm(0.20)\,0.36$& (14) \\[1.0ex]
WN2b & $-2.97\pm(0.73)\,0.79$ &(1) & $-2.89\pm(0.73)\,0.78$ &(1) & $-2.86\pm(0.72)\,0.78$ &(1) \\ 
WN3b & $-3.56\pm(0.46)\,0.55$ &(1) & $-3.48\pm(0.46)\,0.55$ &(1) & $-3.45\pm(0.46)\,0.55$ &(1)  \\
WN4b & $-4.48\pm(0.23)\,0.38$ &(5) & $-4.58\pm(0.23)\,0.38$ &(5) & $-4.85\pm(0.23)\,0.38$ &(5) \\
WN5b & $-4.70\pm(0.16)\,0.34^a$&(0) & $-4.74\pm(0.16)\,0.34^a$ &(0)& $-5.00\pm(0.16)\,0.34^a$ & (0) \\ 
WN6b & $-4.93\pm(0.23)\,0.38$ &(2) & $-4.90\pm(0.22)\,0.37$ &(4) & $-5.16\pm(0.22)\,0.37$ &(4)  \\
WN7b & $-5.02\pm(0.16)\,0.34$ &(2) & $-5.12\pm(0.19)\,0.36$ &(2) & $-5.38\pm(0.15)\,0.34$ &(2) \\ 
WN3 & $-3.10\pm(0.62)\,0.69$ &(1) & $-3.02\pm(0.62)\,0.69$ &(1) & $-2.99\pm(0.62)\,0.69$ &(1) \\ 
WN4 & $-3.36\pm(0.32)\,0.44^a$& (0) & $-3.33\pm(0.32)\,0.44^a$ &(0)& $-3.39\pm(0.32)\,0.44^a$  & (0)\\
WN5 & $-3.63\pm(0.16)\,0.34$ &(8) & $-3.66\pm(0.15)\,0.34$ &(4) & $-3.86\pm(0.15)\,0.34$ &(7)  \\
WN6 & $-4.47\pm(0.30)\,0.43$ &(6) & $-4.74\pm(0.34)\,0.45$ &(7) & $-4.94\pm(0.34)\,0.46$ &(7)  \\ 
WN7 & $-5.32\pm(0.34)\,0.45$ &(9) & $-5.01\pm(0.28)\,0.41$ &(9) & $-5.49\pm(0.30)\,0.42$ &(10)  \\
WN8 & $-5.94\pm(0.19)\,0.35$ &(15) & $-5.78\pm(0.19)\,0.36$ &(16) & $-5.82\pm(0.20)\,0.36$ &(16)  \\
WN9 & $-6.18\pm(0.18)\,0.35$ &(15) & $-6.19\pm(0.16)\,0.34$ &(17) & $-6.32\pm(0.15)\,0.33$ &(18) \\
WN6ha & $-6.98\pm(0.17)\,0.34$ &(8) & $-6.94\pm(0.19)\,0.36$ &(8) & $-7.00\pm(0.18)\,0.35$ &(8) \\
WN7ha & $-7.33\pm(0.25)\,0.39$ &(2) & $-7.26\pm(0.27)\,0.40$ &(2) & $-7.24\pm(0.28)\,0.41$ &(2)  \\
WN9ha & $-6.38\pm(0.07)\,0.31$ &(2) & $-6.33\pm(0.07)\,0.31$ &(2) & $-6.34\pm(0.05)\,0.30$ &(2)  \\[1.0ex]
\hline\\[-2.0ex]
\end{tabular}
\\$^a$Average of adjacent types since no stars of this type are available for calibration. 
\label{tab:cal_jhk}
\end{minipage}
\end{table*}

Due to the growing body of evidence suggesting a
difference in dust properties toward the GC, we employ two different Galactic
near-IR extinction laws. For stars in the GC region
($358^{\circ}{<}\,l\,{<}\,2^{\circ}, | b |\,{<}\,1^{\circ}$) we use the line-derived
extinction law of \citet{fritz11} ($\nicefrac{A_J}{A_{K_S}}{=}\,3.05{\pm}\,0.07,
\nicefrac{A_H}{A_{K_S}}{=}\,1.74{\pm}\,0.03$). For all other Galactic sight lines we
use the law of \citet{stead09} ($\nicefrac{A_J}{A_{K_S}}{=}\,3.1{\pm}\,0.2,
\nicefrac{A_H}{A_{K_S}}{=}\,1.71{\pm}\,0.09$) -- an updated form of that provided by
\citet{indebetouw05}. For the purpose of calculating the
absolute magnitude of each calibration star, we take an average,
$\bar{A}_{K_S}$, from Equations \ref{eq:ak1} and \ref{eq:ak2}. Since extinction in 
the IRAC bands is lower, we opt for a more straightforward approach and use the 
relations given by \citet{indebetouw05}, independent of sight line.  

A minority of stars in our calibration sample only have single-band
photometry available, preventing an extinction calculation by colour excess.
For these objects we resort to one of two alternatives; we adopt the average
$A_{K_S}$ calculated for other O or WR stars in the cluster/association if
sufficient numbers are available, or we take $A_v$ as listed in \citet{vdH01}
and convert this using $A_{K_S}{\simeq}\,0.11A_V{\simeq}\,0.12A_v$ \citep{rieke85}. If neither is possible we exclude the star from our sample.

\subsubsection{Correction for binary companions}
\label{sec:binary}

For cluster/OB association WR stars in spectroscopically confirmed binary
systems, we attempt to correct for the contribution of companion(s) to systemic
magnitudes, allowing an absolute magnitude calculation for the WR component.
Depending on the information available about the companion star(s), we follow
one of two methods to apply these corrections.

\begin{table*}
\centering
\begin{minipage}{17cm}
\caption{Calibrated mid-IR (Spitzer IRAC filters) absolute magnitudes for Galactic WR stars. 
The number of objects used to arrive at each value is indicated in adjacent parentheses. 
Two uncertainties are shown with each value; formal errors (parenthesised, 
Equation \ref{eq:cal_err1}) do not account for the intrinsic spread in magnitude within a 
WR spectral type - estimated to be ${\sim}\,0.3\,$mag - which is incorporated into the final 
(nonparenthesised) uncertainty. }
\centering
\begin{tabular}{lcccccccc}
\hline\\[-2.0ex]
WR & $\bar{M}_{[3.6]}$&(N) & $\bar{M}_{[4.5]}$&(N) & $\bar{M}_{[5.8]}$&(N) & $\bar{M}_{[8.0]}$&(N) \\ 
\mbox{Sp. type} &           &                      &                     &                      \\[1.0ex]
\hline\\[-2.2ex]
WC$5\mbox{--}6$  & $-4.34\pm(0.35)\,0.46$& (2) & $-4.75\pm(0.35)\,0.46$ &(2) & $-5.02\pm(0.28)\,0.41$ &(3) & $-5.32\pm(0.29)\,0.41$ &(3) \\
WC$7\mbox{--}9$  & $-5.96\pm(0.29)\,0.42$& (4) & $-6.27\pm(0.33)\,0.45$ &(4) & $-6.06\pm(0.33)\,0.45$ &(4) & $-6.27\pm(0.34)\,0.46$ &(4) \\
WC8d             & ...& & ...& & $-8.18\pm(0.55)\,0.63$ &(1) & $-8.47\pm(0.55)\,0.63$ &(1) \\
WC9d             & $-6.88\pm(0.48)\,0.57$ &(1) & $-7.25\pm(0.50)\,0.58$ &(1) & $-9.29\pm(0.16)\,0.34$ &(4) & $-9.36\pm(0.16)\,0.34$ &(4) \\[1.0ex]
WN3b             & $-3.59\pm(0.46)\,0.55$ &(1) & $-3.84\pm(0.46)\,0.55$ &(1) & $-4.11\pm(0.46)\,0.55$ &(1) & $-4.50\pm(0.46)\,0.55$ &(1) \\
WN$6\mbox{--}7$b & $-5.51\pm(0.41)\,0.51$ &(2) & $-5.88\pm(0.40)\,0.50$ &(2) & $-6.14\pm(0.38)\,0.49$ &(2) & $-6.41\pm(0.44)\,0.53$ &(2) \\
WN$4\mbox{--}6$  & $-4.18\pm(0.13)\,0.33$ &(1) & $-4.42\pm(0.14)\,0.33$ &(1) & $-4.71\pm(0.15)\,0.34$ &(1) & $-5.05\pm(0.15)\,0.33$ &(1) \\
WN$7\mbox{--}9$  & $-5.96\pm(0.39)\,0.49$ &(5) & $-6.23\pm(0.36)\,0.47$ &(6) & $-6.53\pm(0.28)\,0.41$ &(12) & $-6.79\pm(0.30)\,0.43$& (11) \\
WN7ha            & ...& & $-7.74\pm(0.12)\,0.32$ &(1) & $-7.87\pm(0.12)\,0.32$ &(1) & $-8.22\pm(0.12)\,0.32$ &(1) \\
WN9ha            & ...& & ...& & $-6.79\pm(0.31)\,0.43$ &(1) & $-6.90\pm(0.31)\,0.43$ &(1)\\[1.0ex]
\hline
\end{tabular}
\label{tab:cal_irac}
\end{minipage}
\end{table*}

If the companion is an OB star of known spectral type, we use the synthetic
photometry of \citet{martins06}, or \textit{Hipparcos}-based absolute magnitudes
\citep{wegner06}, to correct for its contribution. Otherwise we determine a
WR/companion continuum flux ratio by considering dilution of the WR emission
lines in the bands that line measurements are available. The \emph{single star}
emission line strengths used to determine WR/companion continuum flux ratios are
presented in Appendix~\ref{app:em_lines}. If the companion is not an OB-star or
is insufficiently bright to notably dilute WR emission lines, it will not make a
significant contribution to the combined light. Ten of the WR stars in our
calibration sample, detailed in Table \ref{tab:bin_cal}, have been corrected by
one of these methods.  

Two systems in our calibration sample are WNha+WNha binaries. There are 
no known `classical' WR+WR binaries, highlighting the sensitivity of post-MS evolution 
to initial mass. The fact that WNha+WNha binaries \emph{are} observed emphasises 
their similarity to massive O-stars. We separate the light
contributions of individual stars to each system by considering mass ratios
derived by spectroscopic monitoring of their orbits. The stars making up
WR$\,20$a are of identical spectral type and have very similar masses
\citep{rauw05}, hence we assume an equal light contribution from each star in
the J, H and K$_S$-bands, and alter the systemic photometry accordingly. Similarly,
WR$\,43$A in the NGC$\,3603$ cluster is comprised of two stars with very high
measured masses, $116{\pm}\,31\,\textrm{M}_\odot$ and $89{\pm}\,16\,\textrm{M}_\odot$
($q{=}\,0.8{\pm}\,0.2$; \citealt{schnurr08}). Using the mass-luminosity
relationship for very massive stars ($M{>}\,80\,\textrm{M}_\odot$) provided by
\citet{yusof13}, we arrive at a light ratio of $1.46{\pm}\,0.47$ for this system
in all bands, assuming identical spectral energy distributions (SEDs). We
include the stars of WR$\,20$a and WR$\,43$A under the WN6ha spectral type 
in Table \ref{tab:wn_cal}.

\subsubsection{Treatment of dust-forming WC stars}
\label{sec:dustywc}

The majority of WC9, and a diminishing fraction of earlier WC stars, show evidence of periodic
or persistent circumstellar dust production (e.g. WR$\,140$,
\citealt{williams90a}). Episodes of dust formation occur at perihelion passage in
eccentric WC+OB systems, whereas circular orbits allow persistent dust
production, enhancing the near-IR flux of the system dramatically. For
completeness, we perform near to mid-IR absolute magnitude calibrations for WC8d and WC9d
spectral types based on the $18$ stars at known distances. However, we do not make
any attempt to remove the light contribution of potential companion stars;
firstly because their $K_S$-band flux is usually insignificant compared to that
of the hot circumstellar dust, and secondly because dust production seems to be
inextricably linked to the presence of these companions \citep{crowther03}. 

Thermal emission from hot circumstellar dust dominates the IR
colours of dusty WC systems, prohibiting extinction determination via near-IR
colour excess. For the dusty systems in our calibration sample we either adopt
an average $A_{K_S}$ for the relevant cluster/association, or its $A_v$ from
\citet{vdH01} and convert this to the K$_S$-band according to $A_{K_S}\,{\simeq}\,0.12A_v$.

We make an exception of WR$\,137$ - a member of the Cyg OB1 association comprising
WC7 and O9(${\pm}\,0.5$) type stars - which displays periodic dust formation episodes
concurrent with its $13\,$yr orbit. \citet{williams01} present JHK$_S$
photometry for this system during a quiescent phase ($1992\mbox{--}4$),
allowing us to derive a K$_S$-band flux ratio ($F_{WR}/F_{sys}\,{=}\,0.46\,{\pm}\,0.13$) using
line strengths measured from spectra taken during quiescence 
(W.D.~Vacca, priv. comm.), and remove the O-star light. The WC7 component is 
included in Table \ref{tab:bin_cal}.

\subsection{Callibration method and uncertainties}
\label{sec:cal_method}

\begin{figure*}
\includegraphics[width=0.8\textwidth]{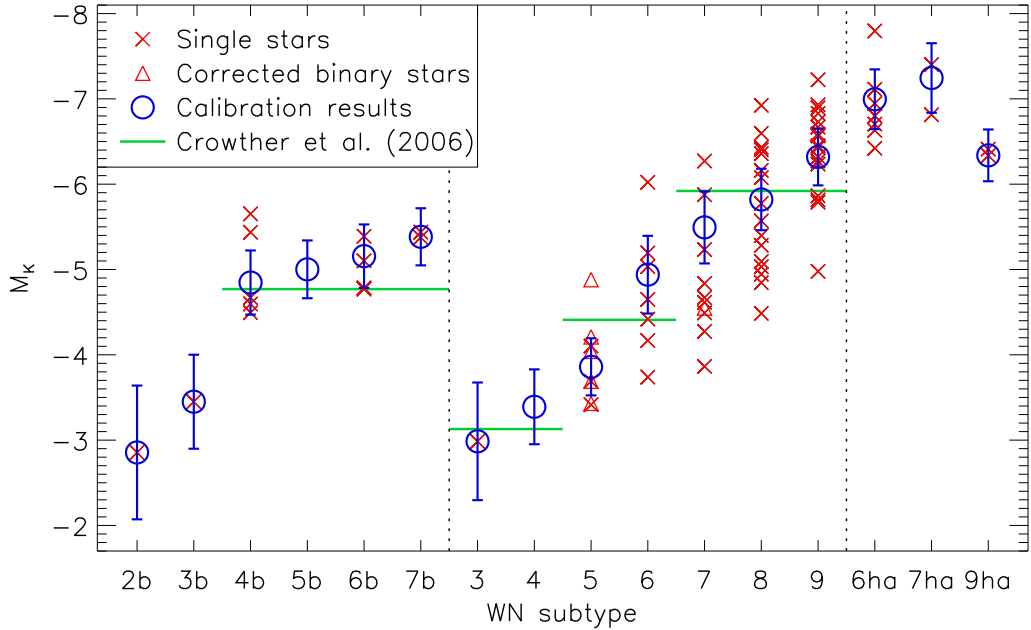}
\caption{Calibration of $M_{K_S}$ for WN spectral types. Broad-line stars are on the left, weak-line (including `WN\#o' and `WN\#h' stars) in the centre, and `WN\#ha' stars to the right. Individual single stars are represented by small (red) crosses, and stars that have been corrected for a companion(s) by (red) triangles.  Larger (blue) symbols show the weighted average for each type with a combination of statistical error (Equations \ref{eq:cal_err1}, \ref{eq:cal_err2} \& \ref{eq:cal_err3}) and the estimated intrinsic spread ($0.3\,$mag) in $M_{K_S}$ within a WR spectral type. Horizontal lines represent the previous calibrations of \citet{crowther06b}.}
\label{fig:wn_cal}
\end{figure*}

\begin{figure}
\includegraphics[width=0.48\textwidth]{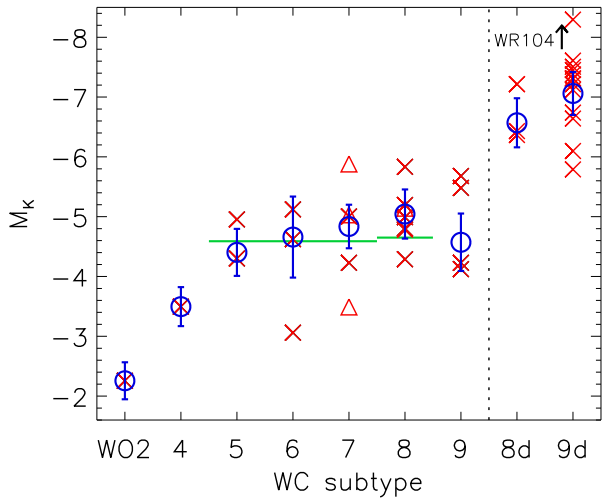}
\caption{Calibration of $M_{K_S}$ for WO and WC subtypes. To the right of the vertical dotted line are subtypes associated with circumstellar dust. WR$\,104$ (WC9d+B) has $M_{K_S}\,{=}\,-10.4$, outside the range displayed here. Symbols same as Figure \ref{fig:wn_cal}.}
\label{fig:wc_cal}
\end{figure}

The results of our near-IR absolute magnitude calibrations are presented in Table 
\ref{tab:cal_jhk}, with mid-IR calibrations shown in Table~\ref{tab:cal_irac}.
Figures~\ref{fig:wn_cal} and \ref{fig:wc_cal} present the K$_S$-band absolute magnitudes for
WN and WC stars, respectively. We use a weighted mean 
method to arrive at an average absolute magnitude for each WR spectral type, computed by
\begin{equation}
\bar{M} = \sum_{i=1}^{n} \frac{p_iM_i}{p},
\label{eq:mag_wm}
\end{equation} 
using weights
\begin{equation}
p_i = \frac{1}{s_i^2},\qquad~p = \sum_{i=1}^{n}p_i, 
\label{eq:mag_w}
\end{equation} 
where $s_i$ is the error in absolute magnitude ($M_i$) - invariably dominated by
distance uncertainty - calculated for each of the $n$ WR stars of that type.
We calculate a formal error ($\sigma$) on each calibrated absolute magnitude
value by combining two uncertainty estimates for weighted data:
\begin{equation}
\sigma = \sqrt{\sigma_1^2 + \sigma_2^2},
\label{eq:cal_err1}
\end{equation}
where
\begin{equation}
\sigma_1 = \frac{1}{\sqrt{p}},
\label{eq:cal_err2}
\end{equation} 
and
\begin{equation}
\sigma_2 = \sqrt{\frac{ \sum_{i=1}^{n}p_i(M_i-\bar{M})^2 }{p(n-1)}}.
\label{eq:cal_err3}
\end{equation} 
This combination is chosen because $\sigma_1$ depends only on 
$s_i$ and does not consider the spread in $M_i$, which is taken into 
consideration by $\sigma_2$. 

This weighted average approach favours objects with the most accurately
determined distances, but the formal uncertainty given by Equation
\ref{eq:cal_err1} does not account for any \emph{intrinsic} scatter in WR star
luminosity within a spectral type. Such a scatter is expected as a WR spectral
type does not represent a perfectly uniform class of objects, but the division
of smoothly varying Wolf-Rayet properties at spectroscopically identifiable boundaries.
Therefore one expects each subclass to encompass a range in mass, temperature
and luminosity. 

Westerlund 1, the Arches and the GC cluster individually contain enough stars of
a single spectral type to evaluate some basic measures of spread, effectively
eliminating the scatter introduced by distance uncertainties when considering
one type across multiple clusters. In Table \ref{tab:scat} we show the range and
standard deviation in $M_{K_S}$ within a WR spectral type. Ranges are typically
${<}\,1$, although WN9 stars in the Arches cluster show a larger range due to
the anomalously faint WR$\,102$aj; we classify this star based on a K-band
spectrum published by \citet{martins08} and thus can only claim a $\pm1$
accuracy on the WN9 spectral type. Typical standard deviations of
$0.3\mbox{--}0.6\,$mag in $M_{K_S}$ suggest that WR absolute magnitudes
intrinsically vary by at least ${\pm}\,0.3$ within a spectral type. Following
this result, we add an uncertainty of $0.3\,$mag in quadrature to the result of
Equation \ref{eq:cal_err1} (see Tables \ref{tab:cal_jhk} - \ref{tab:cal_irac}),
and recommend the adoption of this combined uncertainty when applying these
calibrated absolute magnitudes to WR stars in the field.  

\setlength\extrarowheight{0pt}
\begin{table}
\caption{Intrinsic scatter in absolute magnitude within a WR spectral type}
\begin{tabular}{l p{2.8cm} c c}
\hline\\[-2.0ex]
Cluster & Stars (WR\#) & $M_{K_S}$ range & $\sigma_{M_{K}}$ \\[1.0ex] 
\hline\\[-2.0ex]
\multicolumn{4}{c}{\textbf{WN7}}  \\[1ex]
Westerlund 1 & 77d, 77j, 77r, 77sa & $0.76$ & $0.33$ \\
\multicolumn{4}{c}{\textbf{WN8}}  \\[1ex]
Arches & 102a, 102ae, 102af, 102ah, 102al & $0.84$ & $0.34$ \\
GC Cluster & 101b, 101dd, 101nc, 101oc & $1.28$ & $0.57$ \\
\multicolumn{4}{c}{\textbf{WN9}}  \\[1ex]
Arches & 102ad, 102ag, 102ai, 102aj, 102bb, 102bc & $1.85$ & $0.72$ \\[1.0ex]
\hline
\end{tabular}
\label{tab:scat} 
\end{table}
    
Throughout the rest of this paper we favour discussion and application of the 
$M_{K_S}$ calibration as these results are affected by lower (and more accurately 
determined) extinctions than those in J and H-bands, and are derived using the 
largest sample. 
For completeness, in the cases of WN4 
and WN5b stars (unrepresented in our calibration sample), we take average values 
in each band from the adjacent ionisation types. 

\subsection{Results of near to mid-IR absolute magnitude calibrations}
\label{sec:cal_results}

Both strong and weak-line WN stars show a monotonic increase in intrinsic near-IR 
brightness with ionisation type. This is largely due to cooler, late-type WN stars 
having smaller bolometric corrections. WN stars displaying intrinsic absorption 
features (the `ha' stars) are the most luminous at 
these wavelengths, as a consequence of their high masses. Our results show good agreement with the calibrations of 
\citet{crowther06b} for weak-line WN stars, but suggest slightly higher 
IR luminosities for strong-line WN4\mbox{--}7b. We note that for 
ionisation types 6\mbox{--}7, strong and weak line stars have similar $M_{K_S}$. 
One would expect an enhanced contribution from free-free excess in the denser winds 
of `b' stars to make them brighter at IR wavelengths than weak-line stars. However, the strong-line stars of these ionisation types have 
higher effective temperatures \citep{hamann06}, so the enhanced IR emission from free-free 
excess is counteracted by larger bolometric corrections at these subtypes.       

The lower number of WC stars available reveal a less obvious variation in
$M_{K_S}$ with ionisation type, yet a monotonic increase in near-IR luminosity
does appear to be present for WC4\mbox{--}8. As expected, dusty WC stars
display a large range in $M_{K_S}$ due to varying quantities of dust and the
range of orbital phases sampled amongst periodic dust forming systems. 

The limited area and resolution of the GLIMPSE survey results in far fewer stars
available for absolute magnitude calibration at $3.6\mbox{--}8.0\mu$m. Hence, in
some cases spectral types showing only small differences in $M_{K_S}$ are grouped
together to provide more robust estimates. For all WR types with GLIMPSE
coverage we observe a brightening across $[3.6]\mbox{--}[8.0]$, gradual in most
cases except dusty WC stars which exhibit a dramatic $\Delta
M\,{\sim}\,2$ between [4.5] and [5.8], owing to hot circumstellar dust
emission.   

\subsubsection{The effects of a different Galactic Center extinction law}
\label{sec:gc_law}

The results presented in this paper are produced by applying the line-derived
\citet{fritz11} $1\mbox{--}19\mu$m law - with $A\,{\propto}\,\lambda^{-\alpha},
\alpha\,{=}\,2.13\pm0.08$ over the JHK$_S$ range - to WR stars residing in
the GC ($358^{\circ}{<}\,l\,{<}\,2^{\circ}, | b |\,{<}\,1^{\circ}$).
Alternatively, \citet{nishiyama09} propose a shallower law ($\alpha\,{=}\,2.0$). It
can be seen in Equations \ref{eq:ak1} and \ref{eq:ak2} that a shallower law
would lead to lower derived extinctions by the colour excess method. We perform 
a second set of absolute magnitude calibrations using the \citeauthor{nishiyama09} law 
to quantify its effect on our results. The biggest change is seen in our 
calibrated absolute magnitudes for late-type WN and WC stars, as these dominate 
in the inner Galaxy. Compared to values presented in Table~\ref{tab:cal_jhk},
adopting the \citeauthor{nishiyama09} law changes $\bar{M}_{J}$, $\bar{M}_H$, and $\bar{M}_{K_S}$ of WN9 stars by
$-0.30$, $-0.31$ and $-0.30$ mag, respectively; WN8 stars by $-0.26$,
$-0.31$, $-0.32$ mag, and WC9 stars by
$-0.27$, $-0.32$, $-0.32$ mag. All differences are within our adopted
uncertainties (Table~\ref{tab:cal_jhk}), and hence are not significant. However, 
as we proceed to obtain further results based on these values
we monitor their effects and comment where discrepancies arise.


\section{The Observed Galactic WR star Distribution}
\label{sec:3d_dist}

The sample of WR stars involved in our absolute magnitude calibrations
represents approximately 20\% of the current known Galactic population. The
remainder either have poorly defined spectral types, uncertain binary status,
or in a majority of cases do not reside in an identified cluster or association
for which a distance measurement is available. For convenience, we shall refer
to any WR star not in our calibration sample (i.e., Tables \ref{tab:wc_cal}--\ref{tab:bin_cal}) as a `field' star. 

Up until recently it was widely accepted that most stars formed in clusters
\citep{lada03}, so the low frequency of WR stars presently in clusters arose via
dynamical ejection or rapid cluster dissolution. It is now recognised that a
high fraction of star formation occurs in relatively low density regions
\citep{bressert10}, so the low fraction of WR stars in clusters does not require
an unusually high rate of ejection. \cite{smith14} compare the association of WR
stars (and Luminous Blue Variables) in the Milky Way and Magellanic Clouds with
O stars. They find WR stars to be less clustered than O-type stars, with WC 
stars in particular showing weak spatial coincidence with O-stars and H-rich WN 
stars. 

Our WN and WC calibration samples echo this finding, with approximately
half as many WC stars residing in clusters or associations as WN stars. 
The typical velocity dispersion of cluster stars is not high
enough to account for the isolation of WC stars considering their age. The
relative isolation of WC stars challenges the commonly accepted evolutionary
paradigm that this phase follows the WN phase in the lives of the most massive
stars. Two alternative scenarios may explain the locations of WC stars; either
they descend from a lower initial mass regime than other WR types
\citep{sander12}, or the runaway fraction of WC and H-free WR stars in general
is higher. Further detailed modelling of cluster collapse and the ejection of
massive stars is needed to explain these emerging statistics.  

In this section we present an analysis of the spatial distribution of WR stars,
where distances to $246$ field WR stars are obtained by application of our
absolute magnitude calibrations. Runaway WR stars are discussed further in Section \ref{sec:highz}.

\begin{figure*}
\includegraphics[width=0.8\textwidth]{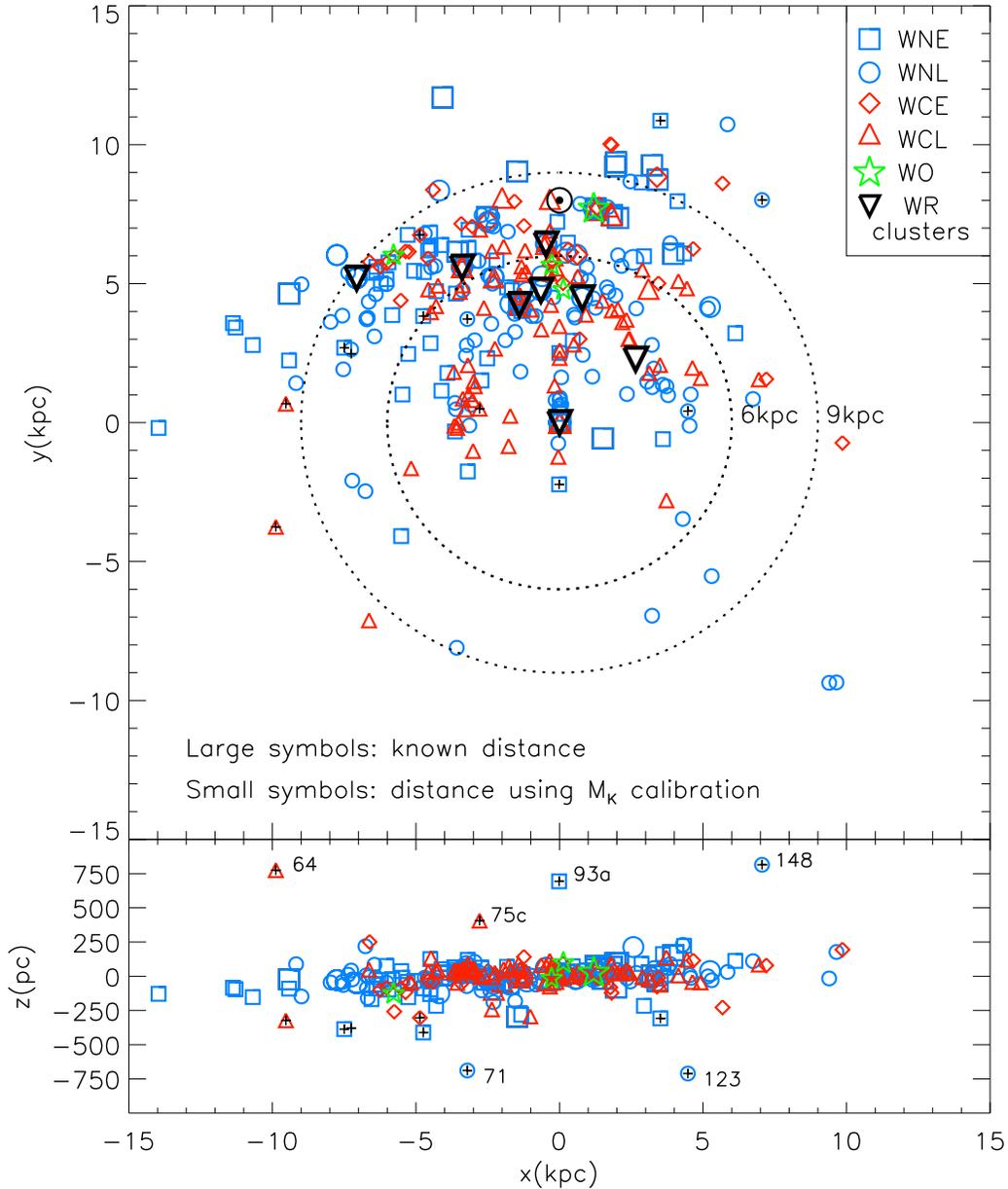}
\caption{Positions of 354 WR stars projected on the Galactic plane (top) and 
viewed edge-on (bottom) in Cartesian coordinates, with the GC at (0,0,0). Galactic longitude increases anti-clockwise about the Sun, which is represented by the standard symbol. Stars with known distances (calibration sample) are represented by larger symbols, whereas those with photometric distances (field sample) are represented by smaller symbols. Black crosses indicate stars located at $|z|\,{>}\,300\,$kpc and the most extreme cases are labelled with WR number. Dotted lines at R$_G{=}\,6\,$kpc and $9\,$kpc delineate the three chosen metallicity zones. From left to right, the displayed clusters from which ${>}\,1$ WR stars 
are taken are: NGC 3603 [--7.07, 5.20], Danks 1\&2 [--3.39, 5.59], Westerlund 1 [--1.40, 4.25], 
Havlen-Moffat 1\&2 [--0.65, 4.76], NGC 6231 [--0.47, 6.43], GC [0.0, 0.0], 
Arches \& Quintuplet [0.02, 0.0], Cl 1813-178 [0.79, 4.49] and Quartet [2.65, 2.29].}
\label{fig:xyzdist}
\end{figure*}

\subsection{Applying $\mathbf{M_{K_S}}$ Calibrations}

Photometry and the derived spatial information for $246$ field WR stars is
given in Tables \ref{tab:field-wr-sin}--\ref{tab:field-wr-bin}. For any non-dusty field WR star with a
well-defined spectral type and no evidence of a significant binary companion -
either spectroscopically or through dilution of near-IR emission lines - we
simply apply our $M_{K_S}$ calibrations to obtain a distance. For these
straightforward cases, we once again use 2MASS photometry and calculate
extinctions by the method described in Section \ref{sec:photom-ext}. 

\begin{table*}
\centering
\begin{minipage}{15cm}
\caption{Calculated spatial locations of the $228$ `field' WR stars showing no conclusive evidence for an IR-bright companion, to which our calibrated absolute magnitudes have been assigned. Shown for each object are the adopted spectral type, 2MASS photometry (unless stated), derived $K_S$-band extinction, heliocentric distance ($d$), Galactocentric radius ($R_G$) and vertical distance from the Galactic midplane ($z$). A full version of this table is available online, and further details 
of stars discovered following the VIIth catalogue (WRXXX-\#) are provided in Appendix A.}
\centering
\begin{tabular}{ l c c c c c c c c c }
\hline\\[-2.0ex]
WR\# & Sp. Type & ref & J & H & K$_S$ & $\bar{\mathrm{A}}_{K_S}$ & $d$(kpc) & R$_G$(kpc) & z(pc) \\[1.0ex]
\hline\\[-2.0ex]
3 & WN3ha & 1 & 10.24 & 10.13 & 10.01  & $0.18\pm0.02$ & $4.53\pm1.15$ & $11.41\pm1.18$ & $-308\pm83$ \\
4 & WC5 & 2 & 8.75 & 8.57 & 7.88 & $0.13\pm0.01$ & $2.69\pm0.49$ & $10.15\pm0.55$ & $-121\pm25$ \\
5 & WC6 & 2 & 8.63 & 8.34 & 7.65 & $0.16\pm0.02$ & $2.69\pm0.84$ & $10.18\pm0.88$ & $-82\pm32$ \\
13 & WC6 & 2 & 10.14 & 9.64 & 8.86 & $0.29\pm0.02$ & $4.42\pm1.38$ & $9.46\pm1.40$ & $-40\pm19$ \\
15 & WC6 & 2 & 7.85 & 7.34 & 6.60 & $0.28\pm0.02$ & $1.57\pm0.49$ & $8.11\pm0.55$ & $-10\pm9$ \\
16 & WN8h & 2 & 6.97 & 6.71 & 6.38 & $0.24\pm0.02$ & $2.77\pm0.46$ & $7.95\pm0.52$ & $-104\pm20$ \\
17 & WC5 & 2 & 9.93 & 9.74 & 9.17 & $0.07\pm0.02$ & $5.02\pm0.91$ & $8.31\pm0.95$ & $-303\pm59$ \\
17-1 & WN5b & 3 & 11.73 & 10.38 & 9.53 & $0.85\pm0.05$ & $5.43\pm0.86$ & $8.58\pm0.89$ & $-41\pm10$ \\
19 & WC5+O9 & 4,5 & 9.75 & 9.13 & 8.53 & $0.20\pm0.02$ & $3.52\pm0.64$ & $7.93\pm0.69$ & $-54\pm13$ \\
19a & WN7(h) & 2 & 9.07 & 8.13 & 7.50 & $0.71\pm0.04$ & $2.41\pm0.47$ & $7.78\pm0.53$ & $-23\pm8$ \\
... & ...  & ... & ...  & ...  & ...  & ...           & ...          & ...          & ... \\
\hline\\[-2.0ex]
\end{tabular}\par
\label{tab:field-wr-sin}
Spectral types: 
(1)\citet{marchenko04}, 
(2)\citet{vdH01}, 
(3) this work, 
(4)\citet{crowther98},
(5)\citet{williams09b}, ...
\end{minipage}
\end{table*}


\setlength\extrarowheight{4pt}
\begin{table*}
\begin{minipage}{17cm}
\caption{Calculated spatial locations of $18$ binary `field' WR stars to which our calibrated absolute magnitudes have been assigned. Shown for each object are the adopted spectral type(s), systemic 2MASS photometry (unless stated), calculated K$_S$-band (unless stated otherwise in parentheses) WR/system flux ratio, derived $K_S$-band extinction, heliocentric distance ($d$), Galactocentric radius ($R_G$) and vertical distance from the Galactic midplane ($z$).}
\centering
\begin{tabular}{ l c c c c c c c c c c }
\hline\\[-3.5ex]
WR\# & Sp. Type & ref & J$^{sys}$ & H$^{sys}$ & K$_S^{sys}$ & $F^{WR}/F^{sys}$ & $\bar{\mathrm{A}}_{K_S}$ & $d$(kpc) & R$_G$(kpc) & z(pc) \\[1.0ex]
\hline\\[-3.5ex]
29 & WN7h+O5I & 1 & 9.91 & 9.46 & 9.12 & $0.49\pm0.16$ & $0.51\pm0.19$ & $9.48\pm2.56$ & $10.27\pm2.57$ & $-148\pm45$ \\
30 & WC6+O6-8 & 2 & 10.06 & 9.76 & 9.21 & $0.75_{-0.66}^{+0.25}$ & $0.25^{+0.37}_{-0.25}$ & $6.11\pm3.44$ & $8.29\pm3.45$ & $-258\pm157$ \\
30a & WO4+O5-5.5 & 3 & 10.25 & 9.83 & 9.56 & $0.13\pm0.04$ & $0.40\pm0.04$ & $5.67\pm1.18$ & $8.17\pm1.21$ & $-117\pm28$ \\
31 & WN4+O8V & 2 & 9.17 & 8.96 & 8.69 & $0.46\pm0.23$ & $0.30\pm0.22$ & $3.34\pm1.12$ & $7.63\pm1.15$ & $21\pm1$ \\
35a & WN6h+O8.5V & 4 & 10.47 & 9.98 & 9.65 & $0.80_{-0.33}^{+0.20}$ & $0.37\pm0.14$ & $7.81\pm1.99$ & $9.13\pm2.00$ & $12\pm2$ \\
41 & WC5+OB & 2 & 11.53 & 10.98 & 10.12 & $0.93_{-0.53}^{+0.07}$ & $0.47\pm0.27$ & $6.69\pm2.06$ & $8.40\pm2.07$ & $-101\pm37$ \\
42 & WC7+O7V & 5 & 7.59 & 7.52 & 7.08 & $0.53\pm0.10$(J) & $0.14\pm0.03$ & $2.96\pm0.53$ & $7.47\pm0.58$ & $-6\pm5$ \\
47 & WN6+O5.5 & 6 & 8.32 & 7.92 & 7.55 & $0.93_{-0.12}^{+0.07}$(J) & $0.37\pm0.03$ & $2.12\pm0.44$ & $7.10\pm0.50$ & $11\pm2$ \\
50 & WC7+OB & 2 & 9.75 & 9.38 & 8.81 & $0.86^{+0.14}_{-0.59}$ & $0.21^{+0.37}_{-0.21}$ & $5.23\pm2.18$ & $6.49\pm2.19$ & $45\pm11$ \\
51 & WN4+OB? & 2 & 10.90 & 10.33 & 9.89 & $0.77^{+0.23}_{-0.30}$ & $0.72^{+0.86}_{-0.72}$ & $3.70\pm1.80$ & $6.55\pm1.82$ & $35\pm7$ \\
63 & WN7+OB & 2 & 8.60 & 8.07 & 7.64 & $0.89^{+0.11}_{-0.44}$ & $0.43\pm0.27$ & $3.68\pm1.23$ & $5.85\pm1.26$ & $-6\pm8$ \\
86 & WC7+B0III & 7 & 7.44 & 7.14 & 6.67 & $0.77\pm0.26$ & $0.31\pm0.02$ & $1.97\pm0.47$ & $6.05\pm0.53$ & $83\pm15$ \\
125 & WC7ed+O9III & 2 & 9.26$^8$ & 8.74$^8$ & 8.25$^8$ & $0.46\pm0.07$ & $0.25\pm0.12$ & $5.45\pm1.05$ & $6.56\pm1.08$ & $120\pm19$ \\
138 & WN5+OB & 2 & 6.97 & 6.80 & 6.58 & $0.61\pm0.11$ & $0.27\pm0.10$ & $1.38\pm0.26$ & $7.76\pm0.36$ & $46\pm5$ \\
143 & WC4+Be & 9 & 8.58 & 8.10 & 7.46 & $0.53\pm0.17$ & $0.60\pm0.06$ & $1.33\pm0.30$ & $7.82\pm0.39$ & $18\pm1$ \\
151 & WN4+O5V & 2 & 9.76 & 9.36 & 9.01 & $0.73\pm0.10$ & $0.40\pm0.11$ & $2.93\pm0.65$ & $9.10\pm0.70$ & $91\pm16$ \\
155 & WN6+O9II-Ib & 2 & 7.48 & 7.34 & 7.16 & $0.70\pm0.07$ & $0.14\pm0.08$ & $2.56\pm0.56$ & $9.02\pm0.61$ & $-38\pm13$ \\
158 & WN+OB? & 2 & 8.64 & 8.20 & 7.81 & $0.84_{-0.28}^{+0.16}$ & $0.40\pm0.03$ & $6.46\pm1.38$ & $12.23\pm1.40$ & $31\pm3$ \\[1.0ex]
\hline\\[-3.0ex]
\end{tabular}
\label{tab:field-wr-bin}
(1)\citet{gamen09}, 
(2)\citet{vdH01},
(3)\citet{gosset01}, 
(4)\citet{gamen14},
(5)\citet*{davis81},
(6)\citet{fahed12},
(7)\citet{lepine01},
(8)\citet{williams92}: average of quiescent photometry in 1989,
(9)\citet{varricatt06}.
\end{minipage}
\end{table*}


Regarding rare WO stars, although only 
one star (WR$\,142$, WO2) is available for calibration, we apply the absolute 
magnitude of this star to the other three field WO stars in the Galaxy, spanning 
WO$\mbox{1--4}$ spectral types. 

We find the spectral type and binary status of many field WR stars to be
uncertain. The majority of the field sample are heavily reddened stars
discovered by near-IR surveys, with typically only a $K$-band (and occasionally
$H$-band) spectrum available in the literature; inclusion of these stars in our
analysis is subject to a spectral type being attributable by our near-IR
classification scheme to the the required ${\pm}\,1$. Of the field WR stars
included in this distribution analysis, we modify the previously claimed
spectral types of ${\sim}25\%$ of those with \emph{only} IR data available,
indicated in Tables \ref{tab:field-wr-sin}--\ref{tab:field-wr-bin}. For the remaining
75\%, we either agree with previous spectral types based on near-IR spectra, or
adopt optically assigned spectral types (always assumed to be reliable).

Further difficulty is encountered when trying to determine the binary status of 
each WR star, particularly late-WC stars, as IR line-dilution could be a result of 
thermal emission from circumstellar dust \emph{or} the continuum of a bright 
companion. For WR stars of types other than WC7-9 showing evidence of binarity, if a flux 
ratio of the WR component to its companion is determinable in either J, H 
or K$_S$, it is straightforward to adjust the corresponding 2MASS photometry 
to that of the WR component alone. We follow one of two procedures to obtain 
near-IR flux ratios for suspected binary systems, depending on whether or not 
the companion responsible for line-dilution has been classified. 

In the case of SB2 systems 
where both stars have been explicitly classified, we determine flux ratios in the 
JHK$_S$ bands using the absolute magnitudes of \citet{martins06} (O-stars) or 
\citet{wegner06} (B-stars) along with our calibrated WR star absolute magnitudes. 
The advantage of this method is it enables us to correct photometry over all 
bands, allowing an extinction calculation by our favoured colour excess 
method. This method is also applicable to dusty WC stars, 
providing photometry is available from a quiescent period.

If a (non WC7-9) WR star shows diluted near-IR emission lines - but the nature
of the diluting source is unclear - we estimate the flux ratio using single star
emission line strengths (Appendix \ref{app:em_lines}). The dilution of optical
lines can also be used, as the uniformity of OB-star
intrinsic (V-K$_S$) colours \citep{martins06} makes it straightforward to translate a
V-band flux ratio to the near-IR. However, the paucity of published line
strengths typically results in a flux ratio only being determinable in one
near-IR band. When this is the case, we either adopt 
$A_v$ from the literature (indicated in Tables
\ref{tab:field-wr-sin}--\ref{tab:field-wr-bin}) or exclude the star from our analysis. 

To determine the nature of late-WC stars displaying diluted emission lines, we
incorporate photometry from the WISE all-sky survey \citep{wright10} allowing us
to construct a simplistic $1\mbox{--}22\mu$m SED. We interpret a peak energy
output at ${\gtrsim}\,5\mu$m as evidence for circumstellar dust emission, and
stars displaying this are excluded from our analysis as we cannot determine
their distances accurately. We identify only one line-diluted late WC, WR$\,42$
(WC7+O7V), to be conclusively dust-free, and include this star in the 
binary field sample (Table~\ref{tab:field-wr-bin}) with $J$-band photometry 
corrected for the companion.

Our field WR star sample consists of $246$ objects; $18$ of these are corrected
for a companion by the line dilution technique, and $3$ are corrected for a
spectroscopically classified companion. We note that with the currently known
population standing at ${\sim}\,635$, approximately $260$ WR stars are
unaccounted for in our calibration and field samples. Of these, the majority
have uncertain spectral types, and lack spectra of sufficient quality (or
spectral range) to obtain the required precision. Also excluded are stars with
inconclusive evidence for a companion, and subtypes for which we cannot
assign reliable near-IR absolute magnitudes (dusty WC stars, WN/C stars,
WN10--11 stars).

A complete list of WR stars discovered between the Annex to the VIIth WR
catalogue \citep{vdH01, vdH06} and March 2014 is provided in
Appendix~\ref{app:new_wr}, which also highlights those for which distances have
been obtained from the present study, together with an explanation of the
revised nomenclature.

\subsection{Spatial properties}
\begin{figure}
\includegraphics[width=0.48\textwidth]{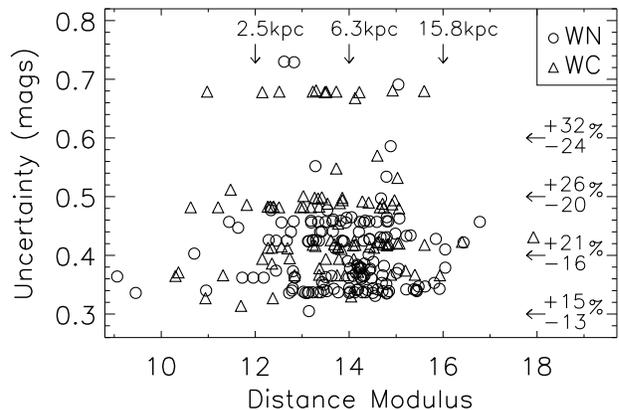}
\caption{Distance modulus uncertainty for field WR stars, with equivalent distances indicated by vertical arrows and percentage errors by horizontal arrows.}
\label{fig:dist_err}
\end{figure}

The Galactic locations of 354 WN and non-dusty WC stars comprising our
calibration and field samples are shown in Figure \ref{fig:xyzdist}.
Uncertainties on distance moduli of field WR stars are displayed in Figure
\ref{fig:dist_err}, where it can be seen that $\Delta DM\,{\sim}\,0.4$ mag
typically applies, and minimum distance uncertainties are approximately
${\pm}14\%$.

\subsubsection{Radial distribution}

\begin{figure}
\includegraphics[width=0.48\textwidth]{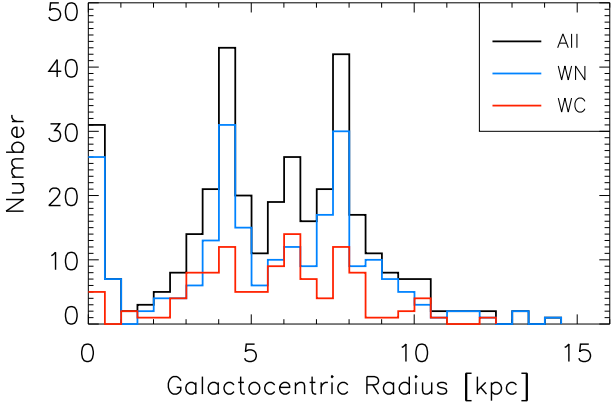}
\caption{The distribution of Galactocentric radii for 354 WR stars, in $0.5\,$kpc bins} 
\label{fig:rad_wr}
\end{figure}

Figure \ref{fig:rad_wr} shows the radial distribution of $354$ WR stars in the 
Milky Way. As expected, the majority of WR stars are located at Galactocentric
distances of 3.5--10 kpc, with an additional peak at $R_G\,{<}\,500\,$pc owing to significant 
star formation within the central molecular zone (CMZ), reminiscent of OB-star forming
regions \citep{bronfman00}. 

Two conspicuous sub-peaks, consisting mostly of WN stars, occur at
$R_G\,{\sim}\,4.5\,$kpc and ${\sim}\,7.5\,$kpc. Both may be viewed as 
superpositions on the underlying WR population, the innermost and outermost are largely
attributable to the WR content of Westerlund 1 (${\sim}\,20\,$WR) and the nearby 
Cygnus star-forming region (${\sim}\,15\,$WR, $l\,{\sim}\,75^\circ$,
$d\,{=}\,1.3\mbox{--}1.9\,$kpc) respectively.

\subsubsection{z - distribution}
\label{sec:zdist}

\begin{figure}
\includegraphics[width=0.48\textwidth]{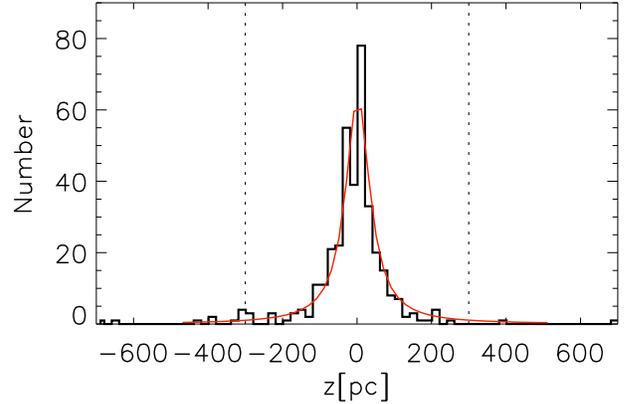}
\caption{z distribution of 354 Galactic WR stars shown in 20\,pc bins (thick black line) with a fitted 
three-parameter Lorentzian function (thin red line, Equation \ref{eq:lorentz}). Stars at 
$|z|{>}\,300$pc (outside dotted lines) are listed in Table \ref{tab:highz}.}
\label{fig:zfit}
\end{figure}

Figure~\ref{fig:zfit} shows that, as expected, WR stars are largely
confined to the thin disk. This strict confinement to $z\,{=}\,0$ resembles a Cauchy 
distribution. Indeed, a non-linear least squares fit of
a Lorentzian function (Equation \ref{eq:lorentz}) matches well the distribution of
vertical heights (z) of WR stars from the Galactic plane,
\begin{equation}
N(z) = A\Big[\frac{\gamma^2}{(z-z_o)^2+\gamma^2}\Big],
\label{eq:lorentz}
\end{equation}
where $\gamma$ is the half width at half maximum (HWHM), $z_o$ is the location 
of the peak and $A$ is an intensity. Assuming the Sun lies $20\,$pc above the Galactic 
plane \citep{humphreys95}, our fit yields $\gamma\,{=}\,39.2\,$pc and $z_o\,{=}\,1.9\,$pc (Figure~\ref{fig:zfit}).   

Unlike other young stellar population tracers, we find no evidence for
flaring of the WR star disk with increasing $R_G$, although this
is likely due to the small number of WR stars identified beyond the solar
circle. \citet*{paladini04} perform a Gaussian fit to the $z$-distribution of $456$
Galactic H\,{\sc ii} regions interior to the solar circle, finding $\sigma\,{\simeq}\,52\,$pc
(full width half maximum, FWHM$\,{\sim}\,125\,$pc). Although not identical to the form
of our fit, this distribution is broader than what we observe for WR stars. The
thickness of the OB star forming disk interior to the solar circle is measured
by \citet{bronfman00}, who find a FWHM of $30\mbox{--}50\,$pc - slightly narrower than the
WR star disk - flaring to ${>}\,200\,$pc beyond $R_G\,{=}\,12\,$kpc.

\subsubsection{WR stars at large distances from the Galactic disk}
\label{sec:highz}

A small fraction of WR stars are found at high vertical distances from the
Galactic disk. Additional details of the 12 WR stars at $|z|\,{>}\,300\,$pc
(${\gtrsim}\,7{\gamma}$) are shown in Table \ref{tab:highz}, all of which are
presumably runaways from star formation sites in the thin disk. We
include WR\,124 in Table~\ref{tab:highz} since it has previously been identified
as an extreme runaway by its high peculiar radial velocity ($156\,$kms$^{-1}$;
\citealt*{moffat82}). Here we briefly discuss the possible events leading to
their runaway status, and summarise the evidence for each.

First, we address the possibility that some of these objects are much
fainter (thus less distant) WR-like central stars of planetary nebulae (CSPN).
Both WC-like ([WC], e.g., \citealt{depew11}) and WN-like ([WN], e.g.,
\citealt{miszalski12} and \citealt{todt13}) CSPN have been observed in the field,
although [WC]-type are far more common. These objects are almost identical in
spectral appearance to their high-mass analogues \citep{crowther06a}, yet are
intrinsically fainter by several magnitudes.

We conduct a search for nebulosity around each $|z|\,{>}\,300\,$pc WR star by
inspection of SuperCOSMOS H$\alpha$ images \citep{parker05} and any other
published H$\alpha$ imaging. Identification of a surrounding nebula cannot alone
prove any of these objects to be CSPN, as some WR stars are seen to posess
ejecta nebulae \citep{stock10}, yet it would provide a strong indication.
Nebulosity is only observed around WR\,71, which are known to be a
genuinely massive, potential WR+compact object binary system \citep{isserstedt83}. We therefore conclude that none of
these 12 high-$z$ WR stars are  incorrectly classified CSPN.

\setlength\extrarowheight{1pt}
\begin{table*}
\centering
\begin{minipage}{15cm}
\caption{Properties of 12 WR stars observed at $|z|{>}\,300\,$pc, plus WR\,124 which is known to be an extreme runaway.}
\centering
\begin{tabular}{l l c c c c c}
\hline\\[-2.0ex]
WR\# & Sp. type & z(pc) & Binary status & L$_X$ (ergs s$^{-1}$) & H$\alpha$ Nebula? & Natal Cluster \\[1.0ex] \hline\\[-2.0ex]
17 & WC5 & $-303\pm59$  &  single   &   not detected$^d$  &   ...   &  ...   \\
3  & WN3ha & $-308\pm83$   &  single   &  $2.5\times10^{32}$ $^c$ &  ...  &   ...    \\ 
56 & WC7 & $-323\pm58$ &  single  &   not detected$^d$   &  No  &   ...   \\
54 & WN5 & $-378\pm61$ &  single  &  not detected$^d$   & No &    ...   \\
49  & WN5(h) & $-386\pm63$ &  single   &  not detected$^d$  &  No  &   ...   \\
75c & WC9 & $406\pm78$ &  single  &   ...   &  No  &      ...    \\
61 & WN5 & $-411\pm66$ &  single  &  $<5.0\times10^{30}$ $^c$   &  No  &  G305? \\
71 & WN6 & $-689\pm139$ &  Binary? (SB1)$^a$ &  not detected$^d$   &  Yes$^e$  &      ...       \\
93a & WN3 & $694\pm214$ & single   &  ...  &  No  &     GC?    \\
123 & WN8 & $-711\pm120$ &  single  &  not detected$^d$   &  No &      ...       \\
64 & WC7 & $775\pm127$ &  single  &  not detected$^d$   &   No  &   ...    \\
148 & WN8h & $814\pm131$ & Binary (SB1)$^b$ &  $<1.6\times10^{32}$ $^c$ &  ...  &      ...       \\[1.0ex]
124 & WN8h & $213\pm39$ & single & $<2.0\times10^{32}$ $^c$ & Yes$^e$ & ... \\[0.5ex]
\hline\\[-2.5ex]
\end{tabular}\par
$^a$\citet{isserstedt83}, $^b$\citet{drissen86}, $^c$\citet{oskinova05}, \mbox{$^d$(ROSAT, \citealt*{pollock95})}, \mbox{$^e$\citet{stock10}}.
\label{tab:highz}
\end{minipage}
\end{table*}

There are two leading mechanisms by which massive stars can be ejected from
their birthplaces; the binary supernova scenario where a massive binary system
becomes unbound after a SN explosion \citep{blaauw61}, and the
dynamical ejection scenario where close encounters in a dense cluster can eject
massive single or binary stars \citep*{poveda67}.

A WR star at $|z|\,{=}\,700\,$pc (similar to the highest observed), assuming
$z\,{=}\,0$ at birth and a time since ejection of 5\,Myr (typical WR star age),
would require an average velocity in the $z$ direction of $140\,$kms$^{-1}$. In the
case of dynamical interaction between massive single and binary stars, a typical ejection
velocity is given by $v_{ej}^2\,{=}\,GM_b/a$ ($M_b\,{=}$ total mass of binary
with semi-major axis $a$) according to \citet{fujii11}, and the ejected star
usually has the lowest mass of the three. By this reasoning, assuming a
$M\,{\gtrsim}\,25\,\textrm{M}_\odot$ WR progenitor limit, a $M_b{=}50\,\textrm{M}_\odot$ ($160\,\textrm{M}_\odot$) 
binary with a period up to 170d (550d) would be capable of ejecting a WR progenitor star
with at least $v_{ej}\,{=}\,140\,$kms$^{-1}$.
Ejection of the binary system is also possible in such an interaction, which one
might expect thereafter to be associated with considerable hard X-ray flux from the collision of
stellar winds. In Table \ref{tab:highz} we include available X-ray observations for these 12
stars, showing that only WR$\,3$ is conspicuous, lying on the
$L_X/L_{bol}$ relation for spectroscopic O-star binaries \citep{oskinova05}.
However, \citet{marchenko04} find no evidence for short period (${<}\,2$yr)
radial velocity changes, concluding that WR$\,3$ is likely a single star.

Alternatively, the locations of these stars may be explained by momentum gained
from the supernova explosion of a companion. \citet{dray05} estimate that
\nicefrac{2}{3} of massive runaways are produced this way. \citet*{isserstedt83}
show that kick velocities of ${\sim}\,150\,$kms$^{-1}$ may be imparted on a
surviving star, and that this star and the resulting supernova remnant may
remain bound if less than half the total system mass is lost during the
supernova. Therefore, one would expect a fraction of massive runaway stars to
have compact companions. Indeed, WR$\,148$ is an SB1 \citep{drissen86}
and the strongest Galactic candidate after Cyg-X3 for a WR$+$compact object
binary. Low amplitude photometric and spectroscopic variations have been
observed in WR\,71 and WR\,124 \citep{isserstedt83,moffat82}, suggesting they
may also be SB1 systems with small mass functions. However, the absence of X-ray
emission from accretion onto a compact object remains unexplained in all 3
cases. We note that the WR$+$OB binary fraction amongst this sample is very low,
and quite possibly zero.

It has been suggested that WN8\mbox{--}9 subtypes are more frequently 
observed as WR runaways \citep*{moffat89}. If we consider only the most extreme 
examples, i.e., $|z|{>}\,500\,$pc plus WR\,124, it 
can be seen from table \ref{tab:highz} that 3/6 are of the WN8 subtype. 
Although numbers are small, a preference seems to exist for WN8 runaways.
\citet{moffat89} argue that a WN8 spectral appearance may arise from mass 
accretion from a binary companion. Combined with the unusually low WR$+$OB binary 
fraction and the low-mass companion of WR\,148, this evidence favours a binary 
supernova origin for the extreme runaway WR population.

Two of the runaway stars listed in Table \ref{tab:highz}, WR$\,61$ and
WR$\,93$a, are observed at similar Galactic longitudes to the G$\,305$ complex
and GC clusters, respectively. Considering our typical distance uncertainties
(Figure \ref{fig:dist_err}), ejection from these massive star forming regions is
a possible explanation for their large distance from the Galactic plane.

\subsection{Subtype distributions across the Galactic metallicity gradient}

Here we assess how WR subtypes vary across the Milky Way disk and compare this
to the predictions of metallicity (Z)-dependent evolutionary models. By
including WR stars in the Large and Small Magellanic Clouds (LMC and SMC), we
may probe massive star evolution over a metallicity range
Z$\,{=}\,0.002\mbox{--}0.04$. A constant star formation rate is implicitly
assumed in all regions considered, allowing us to relate the relative numbers of
WR subtypes observed with the relative duration of associated phases.

We proceed by dividing the Galaxy into three broad zones of super-solar
($R_G\,{<}\,6\,$kpc), solar ($6{<}\,R_G{<}\,9\,$kpc), and sub-solar
($R_G\,{>}\,9\,$kpc) metallicity. Based on the HII-region metallicity analysis
of \citet{balser11}, we assign approximate Oxygen abundances
$(\log[\nicefrac{O}{H}]\,{+}\,12)$ of 8.85, 8.7 and 8.55 (${\pm}\,0.1$dex) to
each zone respectively; each value is arrived at by inspection of their figure 8
and a derived (azimuthally averaged) $\log[\nicefrac{O}{H}]$ gradient of
${-}\,0.05\,{\pm}\,0.02\,$dex kpc$^{-1}$. According to our distribution analysis
of $354$ WR stars, we find $187$, $132$ and $35$ to inhabit the super-solar,
solar, and sub-solar metallicity zones respectively. Additionally, there are 148
known WR stars in the LMC (\citealt{breysacher99}; \citealt{neugent12} 
and references therein; \citealt{massey14}) for which the Oxygen
abundance is $(\log[\nicefrac{O}{H}]\,{+}\,12)\,{=}\,8.38$ \citep*{rolleston02},
and 12 WR stars in the SMC \citep{massey01b,massey03} for which
$(\log[\nicefrac{O}{H}]\,{+}\,12)\,{=}\,8.13$ \citep{rolleston03}.

As recently highlighted by \citet{groh14}, there is not a straightforward
correspondence between spectroscopic and evolutionary phases in massive stars,
particularly post-main sequence. Spectroscopically, any WN showing surface
Hydrogen (WN\#h or (h)) or with ionisation type ${\geq}\,7$ is identified as
late-type (WNL), while H-free WN of ionisation type ${\leq}\,6$ or those
displaying broad emission lines (WN\#b) are early-type (WNE). We follow these
definitions here, noting that the lack of near-IR hydrogen diagnostics is
unlikely to significantly affect our measured $N_{WNE}/N_{WNL}$, as Galactic
$\mathrm{WN}\,{\leq}\,7$ stars are generally H-free \citep*{hamann06}, so
division by ionisation type alone is sufficient. This assumption is less
applicable in the lower metallicity regions of the outer Galaxy, however, low
extinction in these directions means optical (hydrogen) diagnostics are commonly
available. We include the WNha stars as WNL when evaluating subtype number 
ratios; their definition as such has minimal effect as only 18/235 WN considered 
belong to this class. The division in WC stars is more straightforward, with
WC$4\mbox{--}6$ defined as early (WCE) and WC$7\mbox{--}9$ as late-type (WCL).
In Table \ref{tab:ratios} we show the subtype breakdown of WR stars observed in
each Galactic metallicity region.

In stellar models, stars have historically been matched with the aforementioned
spectroscopic WR types using basic surface abundance and effective temperature
($T_{eff}$) criteria. For example, \citet{meynet05} employ
$T_{eff}\,{>}\,10^4\,$K and $X_H{<}\,0.4$ as the definition of a WR star in
their models, while \citet*{eldridge08} add a further constraint of
log$(L/\textrm{L}_\odot)\,{>}\,4.9$. This $T_{eff}$ boundary is too low, since
even the coolest WR stars (WN$8\mbox{--}9$) are found to have
log$(T_{eff})\,{\simeq}\,4.6$ \citep{hamann06}. A surface temperature of
$10^4\,$K is more typical of late-B/early-A supergiants \citep{przybilla06}.
Also, recent spectroscopic analysis of WC stars by \citet*{sander12} indicate
that some WC9 stars are very close to this lower luminosity limit. The
transition between eWNE and eWNL phases (where `e' denotes the definition in
evolutionary models) is regarded to occur when $X_H{<}\,10^{-5}$, and the eWC
phase begins when carbon dominates nitrogen by mass \citep{meynet05}. By
computing model spectra from evolutionary models, \citet{groh14} have shown that
spectroscopic WNE and WNL lifetimes can differ radically from eWNE and eWNL
lifetimes, as a star may have a WNE spectrum while retaining some surface
hydrogen, hence this is a poor indicator. The problem is not so severe regarding
the transition from WN to WC stars, as the change in surface carbon abundance is
a rapid process, meaning the eWC phase corresponds well to the spectroscopic WC
phase.

\setlength\extrarowheight{0pt}
\begin{table*}
\centering
\caption{Observed Wolf-Rayet number ratios in the Galaxy, LMC and SMC. Galactic WC stars counted here show no evidence of circumstellar dust, results incorporating an estimated 28\% of neglected (dusty) WC stars are parenthesised. The 4 WO stars are counted as WCE. An indication of uncertainty is given assuming $\sqrt{N}$ errors on each count.}
\begin{minipage}{15cm}
\centering
\begin{tabular}{ p{1.8cm} c c c c c c }
\hline\\[-2.0ex]
Region                              & $N_{WR}$ & $N_{WC}$      & $N_{WN}$ & $N_{WC}/N_{WN}$ & $N_{eWNE}/N_{eWNL}$ & $N_{(WCE{+}WO)}/N_{WCL}$ \\
\mbox{($\log[\nicefrac{O}{H}]+12$)} &          & ($N_{WCd}$)   &          & ($N_{(WC{+}WCd)}/N_{WN}$)   &        &                        \\[1.0ex] 
\hline\\[-2.0ex]
Inner Galaxy                        & 187      & $63$          & $124$    & $0.51\pm0.08$   & $0.23\pm0.05$       & $0.05\pm0.03$          \\
\mbox{($8.85\pm0.1$)}               &          & (${\sim}\,22$) &       &   $(0.69)$         &                       &          \\ [1ex]
Mid Galaxy                          & 132      & $46$          & $86$  & $0.53\pm0.10$ & $0.79\pm0.17$  & $1.00\pm0.28$ \\
\mbox{($8.7\pm0.1$)}                &          & (${\sim}\,16$) &       &   $(0.73)$           &                  &       \\ [1ex]
Outer Galaxy                        & 35       & $10$          & $25$  & $0.40\pm0.16$ & $1.27\pm0.58$ & $1.5\pm0.97$ \\
\mbox{($8.55\pm0.1$)}               &          &  (${\sim}\,4$) &       &   $(0.57)$   &               &     \\ [1ex]
LMC                                 &  148     & 26 & 122 & $0.21\pm0.05$ & $1.93\pm0.37$  & 26/0 \\
\mbox{($8.38\pm0.05$)}              &            &    &       &               &                &        \\ [1ex]
SMC \mbox{($8.13\pm0.05$)} &         12 & $1$   & $11$  & $0.10\pm0.09$ & $11/0$  & $1/0$ \\[0.5ex]
\hline
\end{tabular}\par
\end{minipage}
\label{tab:ratios}
\end{table*}

\subsubsection{Incompleteness of the sample}
Before comparing observed numbers with evolutionary model predictions,
it is necessary to comment on two selection effects - one in our distribution
analysis and one inherent to WR star surveys - that affect the star counts
we present. 

\begin{figure*}
\includegraphics[width=0.8\textwidth]{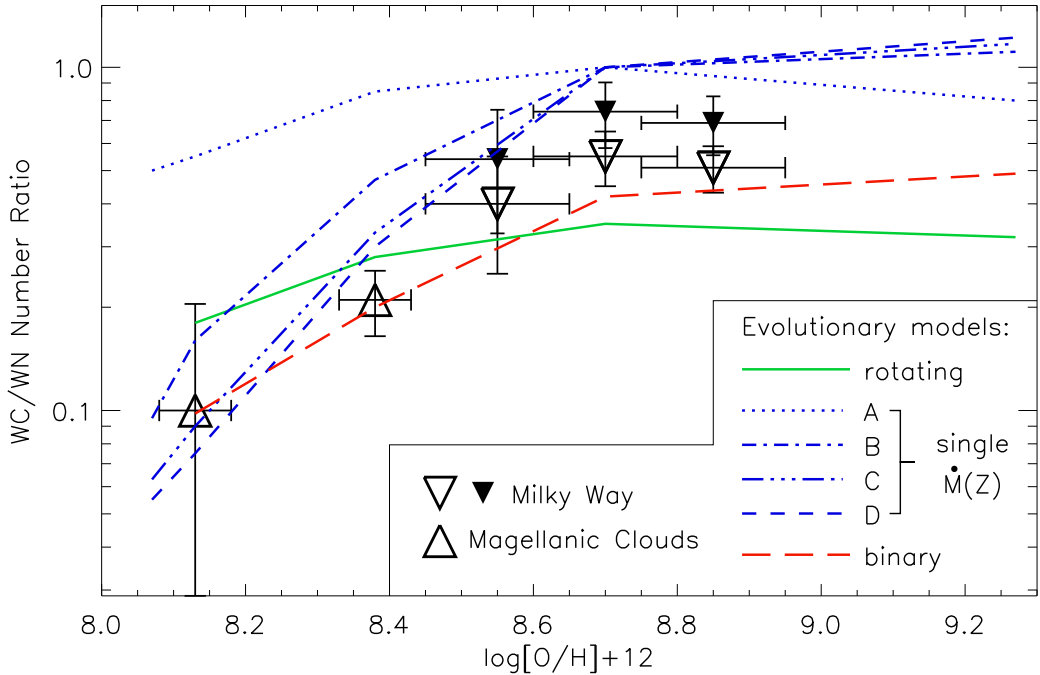}
\caption{Number ratio of WC/WN stars in the LMC and SMC (triangles), and 
across three Milky Way regions (upsidedown triangles) where results omitting dusty WC stars 
(as in Table \ref{tab:ratios}) are plotted as larger open symbols and smaller 
filled symbols represent the case where 28\% of all WC stars posses hot circumstellar 
dust (Section \ref{sec:local_wr}). Solid (green) line shows the predictions of 
\citet{meynet05} for rotating single stars. Long-dashed (red) line shows the 
predictions of \citet{eldridge08} for a population of massive binaries. All other lines (blue) 
represent non-rotating single-star evolutionary predictions of \citet{eldridge06} for four 
different $\dot{\mathrm{M}}{-}\mathrm{Z}$ dependencies. Errors on the number ratios shown are estimated assuming $\sqrt{N}$ 
counting errors in $N_{WC}$ and $N_{WN}$. An uncertainty of 0.1 dex is assigned to each Galactic O/H value.}
\label{fig:wcwn_z}
\end{figure*}

Firstly, as we cannot assume reliable near-IR absolute magnitudes for dusty WC
stars they have been excluded from our distribution analysis. Thus, in Table
\ref{tab:ratios} we count only those WC stars showing no evidence for
circumstellar dust. However, by inspection of a local volume-limited
(${<}\,3\,$kpc) sample of WR stars (see Section \ref{sec:local_wr} for full
details), we estimate that 27(${\pm}\,9$)\% of WC stars shown evidence of
circumstellar dust. To account for the effect of these neglected stars, we plot
in Figure \ref{fig:wcwn_z} a set of enhanced WC/WN number ratios along with the
values shown in Table \ref{tab:ratios}. As this information is only
available in the solar neighbourhood, we are forced to assume an
unvarying fraction of dust-forming WC stars across the whole Galaxy. Late-type 
WC stars are known to dominate at higher metallicity, and it is
predominantly these that are seen forming dust, hence we expect this
fraction in reality to be higher towards the Galactic Centre. The
slight downturn in $N_{WC}/N_{WN}$ at $R_G{<}\,6\,$kpc could be due to a higher
number of late WC stars omitted from our analysis in this region.    

Secondly, the two most widely employed WR star survey techniques are both least
effective at identifying weak-lined WNE stars. Narrow-band IR imaging
surveys are biased against WNEs due to low
photometric excesses from their weak emission lines. The IR excess emission from
free-free scattering - exploited by broad-band selection techniques - is also 
weaker in WNE stars as their wind densities are
lower than other WR subtypes. For these reasons, and
considering their modest IR luminosities, we expect WNE stars to be slightly
under-represented in our total sample, especially beyond the solar
neighbourhood. Therefore, future observations will likely refine the numbers
presented here by marginally decreasing $N_{WC}/N_{WN}$ and increasing
$N_{WNE}/N_{WNL}$.  

\subsubsection{Comparison to evolutionary predictions}
In Figure \ref{fig:wcwn_z} we plot observed $N_{WC}/N_{WN}$ in different regions
of the Galaxy, and the Magellanic Clouds, alongside the predictions of various
evolutionary models for massive stars. Rotating single-star models are taken
from \citet{meynet05}, non-rotating single-star models with various $\dot{M}(Z)$
dependencies are taken from \citet{eldridge06}, and finally predictions for a
population of non-rotating massive binaries exhibiting metallicity dependent
mass-loss are taken from \citet{eldridge08}
(BPASS\footnote{\url{http://www.bpass.org.uk/}}). 

All models predict an increasing number of WC compared to WN stars
with metallicity, due to increasingly rapid exposure of nuclear burning products
caused by stronger stellar winds in more metal-rich environments. Our analysis
shows only a modest variation of $N_{WC}/N_{WN}$ ($0.4\mbox{--}0.55$) across the Galactic disk,
whereas the ratio drops considerably to $0.2\mbox{--}0.1$ at LMC and SMC
metallicities. At all metallicities the observed $N_{WC}/N_{WN}$ lies between the predictions from a population of binary stars \citep{eldridge08} 
and single non-rotating stars. The addition of an 
estimated 28\% of neglected (dusty) WC stars at Galactic metallicities does not significantly alter this. However, 
the predictions of evolutionary models including rotation lie ubiquitously lower 
than our observations at Galactic metallicities. Fast
rotation has the effect of lengthening WR lifetimes, manifest predominantly in
the eWNL phase, thus reducing $N_{WC}/N_{WN}$ \citep{meynet05}. However, it is 
not expected that all massive stars are formed rotating as quickly 
($v^{rot}_i{=}\,300$\,km\,s$^{-1}$) as those generated in these models \citep{penny09}. 

\begin{figure}
\includegraphics[width=0.48\textwidth]{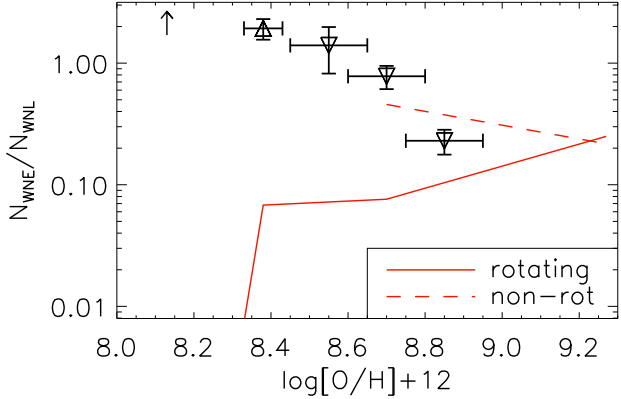}
\caption{Number ratio of WNE/WNL stars across three metallicity zones in the 
Milky Way, LMC and SMC. Red lines show eWNE/eWNL 
predictions from rotating (solid) and non-rotating (dashed) evolutionary models \citep{meynet05}.}
\label{fig:early-late_wn}
\end{figure}

Figure \ref{fig:early-late_wn} shows the number ratio of early to late WN
stars in each Galactic metallicity zone, as well as in the LMC and SMC. An
increase in the proportion of WNE can be seen with decreasing metallicity, and
no WNL stars are known in the SMC. Contrary to this, the rotating models of
\citet{meynet05} produce a \emph{shorter} relative eWNE phase at lower
metallicity, due to less efficient removal of the H-rich stellar envelope during
prior evolutionary phases. Furthermore, rotationally induced mixing
allows stars to become WR earlier in their evolution and experience an extended
eWNL phase. The extreme sensitivity of the eWNE/eWNL transition to the chosen
hydrogen surface abundance criterion clearly has a major
influence on predictions \citep{groh14}. Therefore we interpret the disparity shown in figure
\ref{fig:early-late_wn} largely as a symptom of these definitions rather than a
serious conflict with evolutionary theory. 

Our results show that the WN phase of
WR stars at sub-solar metallicities is almost entirely spent with a WNE spectral
appearance, whereas the WNL spectral type endures longer on average at
super-solar metallicities.

\section{Modelling the total WR star population of the Milky Way}
\label{sec:model_wrpop}

With knowledge of how WR subtypes vary with Galactocentric radius, and the
intrinsic near-IR brightness of each subtype, we are in a position to model the
observational properties of the whole Galactic WR population. To this end we 
develop a 3D, azimuthally symmetric ``toy'' model of the WR population - described in the following section - that is scalable to
different total numbers of WR stars. For varying numbers of WR stars, we apply a
simple Galactic dust distribution to redden the population, and derive
magnitude distributions in various bands for comparison with the observed
population, allowing us to estimate the total number of Galactic WR stars.   

\subsection{Populating the model}
\label{sec:model_stars}

We do not attempt to incorporate complex structural features such as spiral arms
or the Galactic bar into this model WR population, as our aim is to derive basic
observational characteristics of the whole population, smoothing over any local
enhancements. We therefore distribute model WR stars in an azimuthally
symmetric disk, with the same thickness the observed population. We generate the
$z$-coordinate of each star so they are Cauchy distributed, by computing 
\begin{equation}
z_i = \gamma \tan(\mathrm{\pi} r_{01}-\frac{1}{2}),
\label{eq:z_cauchy}
\end{equation}
where $r_{01}$ is a randomly generated number between 0 and 1 and $\gamma$ is 
the observed HWHM ($39.2\,$pc, Section \ref{sec:zdist}). The $z$-distribution is truncated at 
$z\,{=}\,{\pm}\,1\,$kpc in accordance with the most extreme runaway stars observed. 

We construct a model WR disk composed of 24 annuli of $0.5\,$kpc in width spanning radii 
$R_G{=}\,3\mbox{--}15\,$kpc. Rather than using our observed radial distribution of
WR stars to dictate the relative number in each annulus, we utilise
a normalised version of the radial HII-region distribution presented by
\citet{paladini04}, since we consider HII-regions to be a more complete tracer of
hot young stars over a larger Galactic extent. Within each annulus, the model
stars are randomly placed between the lower and upper radii at a random angle
($\theta$) between 0 and $2\,\pi$ around the model disk. In these coordinates the
Sun is located at $z\,{=}\,20\,$pc, $\theta\,{=}\,0^\circ$, and $R_G{=}\,8000\,$pc. 

\begin{figure}
\includegraphics[width=0.48\textwidth]{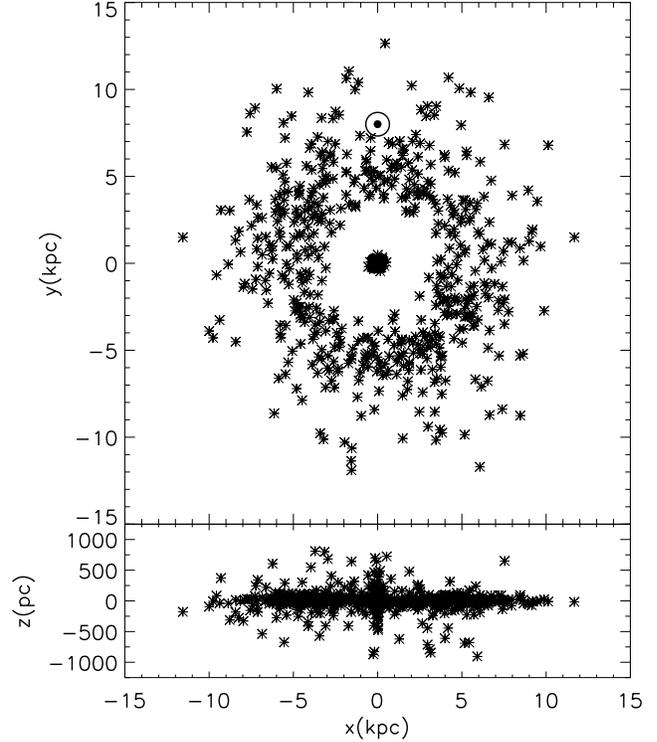}
\caption{An example model WR population mimicking that of the Galaxy, containing 550 stars at $R_G{=}\,3\mbox{--}15\,$kpc and 250 at $R_G{<}\,3\,$kpc, shown in Cartesian coordinates. The location of the Sun is indicated by the standard symbol at (0,$8\,$kpc) in the top panel.}
\label{fig:model_xyzdist}
\end{figure}

Interior to $R_G\,{=}\,3\,$kpc, where star formation is suppressed 
(aside from the CMZ) we include a fixed number of WR stars in
every model. \citet{dong12} report on a Pa-$\alpha$ survey covering the central
${\sim}\,0.6^{\circ}$ (${\sim}\,80\,$pc) of the Galaxy, including the three
massive clusters (Arches, Quintuplet and GC). Within this region they
identify as many emission line sources (evolved massive star candidates) outside
of these clusters as within them; given that ${\sim}\,80$ WR stars are known to reside
in these clusters, we estimate $160$ WR stars present in the \citeauthor{dong12} survey
area. The CMZ is approximately $3^{\circ}$ ($400\,$pc) across, and the density of
gas in this extreme environment is strongly centrally peaked, so that
approximately 40\% of the CMZ gas lies within the \citeauthor{dong12} survey area
\citep{ferriere08}. Assuming the non-cluster population of WR stars roughly
follows the amount of molecular gas, this would imply a further ${\sim}\,100$
WR stars in the CMZ. The inner Galaxy ($R_G\,{<}\,3\,$kpc) contains little star 
formation outside the CMZ, we therefore populate this area in our models with 250 WR stars,
following a Gaussian distribution centred on $R_G{=}\,0$ with $\sigma\,{=}\,200\,$pc. An
example model is displayed in Figure \ref{fig:model_xyzdist}.

The number of WR stars in each radial bin of our model population is divided into four
components representing WNE, WNL, WCE and WCL. The relative numbers of each WR 
type match those observed (Table \ref{tab:ratios}), varying from the model GC
where late-types dominate to $R_G{=}\,15\,$kpc where early-types are in the
majority. We assign absolute magnitudes of $M_{K_S}\,{=}\,{-}4.31$, ${-}6.01$, ${-}4.45$ and
${-}4.89$ to WNE, WNL, WCE, and WCL types respectively, based on averaging our calibration results
(Section \ref{sec:cal_results}). We do not include WO or WN/C stars as they constitute
a negligible fraction of the population.

\subsection{A dust model for the Milky Way}
We include two dust components in our model WR star disk; one associated with
molecular (H$_2$) gas, the other with atomic H gas. We apply the same
dust-to-gas mass ratio for each component. This assumption follows
\citet{bohlin78} who derive a \emph{total} neutral hydrogen to colour excess
ratio, implying the each atom of H is responsible for a set amount of
extinction, whether in molecular or atomic form. Both dust components are
included as two dimensional functions in $R_G$ and $z$, motivated by the spatial
distribution measured for their respective gas species. These gas measurements
are taken from \citet{nakanishi06} and \citet{nakanishi03} for molecular (traced
using CO) and atomic gas respectively.

Functions describing each dust component have the form 
$D_m(R_G)\,{\times}\,D_h(R_G,z)$, where $D_m(R_G)$ describes the dependence of midplane 
density on Galactocentric radius, and $D_h(R_G,z)$ describes how the density 
drops with vertical distance from the midplane. 

Both dust components are included 
with a vertical dependence of the form $D_h(R_G,z)\,{=}\,\sech^2(\zeta)$, where
\begin{equation}
\zeta(R_G,z) = \log(1+\sqrt{2})\frac{z}{z_{\nicefrac{1}{2}}(R_G)}.
\label{eq:dust_zeta}
\end{equation}
and $z_{\nicefrac{1}{2}}(R_G)$ is the height at which the density falls to
half of the midplane value, which increases linearly with galactocentric radius
for both gas species. For molecular gas, $z_{\nicefrac{1}{2}}(R_G{=}0)\,{=}\,25\,$pc
increasing to $90\,$pc at $R_G{=}\,10\,$kpc \citep{nakanishi06}. For atomic gas,
$z_{\nicefrac{1}{2}}(R_G{=}0)\,{=}\,100\,$pc increasing to $500\,$pc at
$R_G{=}\,15\,$kpc \citep{nakanishi03}. 

\begin{figure}
\includegraphics[width=0.48\textwidth]{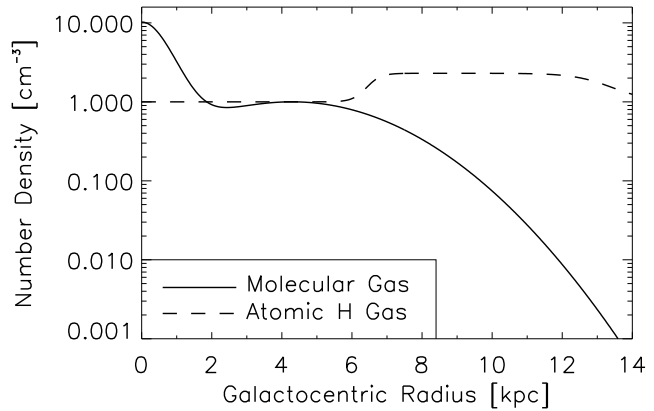}
\caption{Midplane number density of molecular $\mathrm{H}_2$ (solid) and 
atomic H (dashed) gas, used to govern our model dust distribution, as a function of Galactocentric radius (Equations \ref{eq:mol_mid} and \ref{eq:atom_mid}, respectively).} 
\label{fig:dust_dist}
\end{figure}

To represent the midplane density of molecular gas, we construct the following function:
\begin{equation}
\begin{split}
D_m^{mol}(R_G)=N_0^{mol}\sech^2\Big(\frac{R_G}{800pc}\Big)... \\
 ...+\exp\Big[\frac{-(R_G-4300pc)^2}{2(2500pc)^2}\Big] [\mathrm{cm}^{-3}],
\label{eq:mol_mid}
\end{split}
\end{equation}
which is shown in Figure \ref{fig:dust_dist}. The scaling values in Equation
\ref{eq:mol_mid} and $N_0^{mol}\,{=}\,10\,\mathrm{cm}^{-3}$ are chosen to
reproduce the maps of \citet{nakanishi06}. 

To represent the midplane density of atomic gas, we employ a summation of two 
step functions:
\begin{equation}
\begin{split}
D_m^{atom}(R_G)=N_0^{atom}\Big[ 1 + \frac{1.3}{1+\exp(\frac{-(R_G-6500pc)}{200pc})}...  \\
 ...-\frac{1.3}{1+\exp(\frac{-(R_G-13200pc)}{550pc})}\Big] [\mathrm{cm}^{-3}]
\label{eq:atom_mid}
\end{split}
\end{equation}
as shown in Figure \ref{fig:dust_dist}. Once again, we choose the
scaling values and $N_0^{atom}{=}\,0.08\,\mathrm{cm}^{-3}$ to reproduce the
maps of \citet{nakanishi03}.

Finally, we integrate the total dust function,
\begin{equation}
D(R_G,z)= (D_m^{mol}\times D_h^{mol}) + (D_m^{atom}\times D_h^{atom}),
\label{eq:dust_tot}
\end{equation}
along the line of sight from the Sun to each model WR star to provide
a total amount of obscuring dust. Hence we obtain an extinction to each model
star, assuming $A_{K_S}$ is proportional to the amount of dust along the line of
sight, normalised to give $A_{K_S}{=}\,2.42\,$mag towards the GC ($R_G\,{=}\,0$, $z\,{=}\,0$; \citealt{fritz11}). 

\subsection{Quantifying the total population}
By combining our model WR star population (Section \ref{sec:model_stars}) with
the dust function described in the previous section, we are able to generate a
global magnitude distribution for a WR population containing any number of
stars. The total WR star population follows an assessment of which predicted
magnitude distribution most closely reproduces that observed.

Before deducing this number, it is worth reconsidering what the WR stars in our 
model represent. As their absolute magnitudes are based on our
calibrated values for WR stars (Section \ref{sec:absmag_cal}), they represent
what we will refer to as `WR-dominated' systems, i.e., where any companion star(s)
do not affect the systemic magnitude by more that $0.4\,$mag (typical error on
our absolute magnitude calibrations), i.e., $(m_K^{\mbox{\tiny
WR}}-m_K^{sys})\,{>}\,0.4$, corresponding to a WR/system flux ratio of
$\mathrm{F}^{\mbox{\tiny WR}}_K/\mathrm{F}^{sys}_K\,{>}\,0.7$. Also, these model
stars do not represent WO, WN/C or dust-producing WC stars. Therefore, to achieve a
like-for-like comparison to observations, we initially only consider observed 
`WR-dominated' systems, and in Section \ref{sec:local_wr} estimate the 
contribution of neglected WR types.

\subsubsection{Comparison to a magnitude-limited sample}

For comparison to our model K$_S$-band magnitude distributions, we assemble a
magnitude-limited sample of real WR-dominated systems. We note that ${<}\,5\%$ 
of WR stars discovered since the year 2011 are 
brighter than K$_S{=}\,8$\,mag \citep{mauerhan11, shara12, smith12, chene13}. 
Therefore, we adopt this as the current completeness limit in \emph{systemic} magnitude..

\begin{figure}
\includegraphics[width=0.48\textwidth]{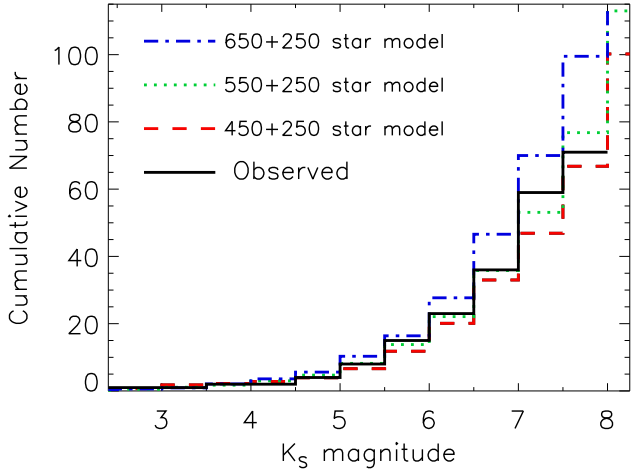}
\caption{The cumulative number of observed WR-dominated systems (black solid line) 
in bins of 0.5 $K_S\,$mag, compared with that of three different model WR populations.}
\label{fig:model_obs}
\end{figure}

Figure \ref{fig:model_obs} presents predicted magnitude distributions for WR
populations containing 450, 550 and 650 stars between
$R_G\,{=}\,3\mbox{--}15\,$kpc plus 250 central stars ($R_G\,{<}\,3\,$kpc). Also
shown is the number of observed WR-dominated systems with systemic
K$_S\,{<}\,8\,$mag. In spite of providing too few systems with
$7.5\,{<}\,{K_S}\,{<}\,8.0$, the best agreement is found with the 
$550\,{+}\,250$ WR star model. Thus we take forward $800{\pm}100$ as the number 
of Galactic WR-dominated systems, with an approximate uncertainty based on the 
comparison shown in Figure \ref{fig:model_obs}.

Assuming a shallower extinction law along lines of sight towards the GC has the
effect of brightening our calibrated $M_{JHK_S}$ values for late-type WN and WC
stars by up to $0.3\,$mag (Section \ref{sec:gc_law}). Upon altering the
magnitudes of the late-type WR stars in our model population by this amount, we
find more consistency with magnitude distributions drawn from models containing
${\sim}100$ fewer WR stars. Hence, the effect of a shallower GC extinction law
is within the uncertainties. 

\subsubsection{Fractions of dusty and companion-dominated WR systems in a volume-limited sample}
\label{sec:local_wr}
Each point in our model WR star population represents a stellar system where a
WR star is the dominant or sole near-IR source. To gain an insight into
how many dusty and companion-dominated WR systems are overlooked by these models
- and hence our initially deduced population of 800 - we construct a volume
limited sample of nearby WR systems. In Appendix \ref{app:vol_lim} we list all
known WR stars within 3kpc of the Sun, where distances are taken from this work
where possible (non-dusty WR stars), or by implementing $\bar{M}_v$-subtype
calibrations of \citet{vdH01}. A $v$-band approach can be used to determine
distances to dusty WC stars as hot dust emission does not contribute to the
continuum flux at these wavelengths. We inspect the near-IR properties of these nearby WR
stars, categorising each as either WR-dominated ($\mathrm{F}^{\mbox{\tiny
WR}}_K/\mathrm{F}^{sys}_K\,{>}\,0.7$), having a significant companion
($\mathrm{F}^{\mbox{\tiny WR}}_K/\mathrm{F}^{sys}_K\,{<}\,0.7$), or dusty WC.  

Of the 72 WR stars in this volume-limited sample, 41 are WC type of which $11$
show evidence of circumstellar dust, indicating that $15\,{\pm}\,5\%$ of WR
stars and $28\,{\pm}\,9\%$ of WC stars display circumstellar dust. A companion
star dominates the near-IR continuum in $11$ of the remaining $61$ non-dusty WR
systems ($18\,{\pm}\,6\%$). Uncertainties on these fractions are calculated
assuming a $\sqrt{N}$ uncertainty on each number count.

The fractions derived from this volume-limited sample imply the previously 
derived population of 800 represents only $82\%$ of the non-dusty population, 
as $\sim$18\% (150) will have an IR bright companion. Furthermore, this non-dusty 
population of 1050 (${=}800{+}150$) represents only
85\% of the total population, as a further $\sim$15\% (150) will be dusty WC 
stars. Two key assumptions underpin this estimate; constant fractions of 
companion-dominated and dust-producing WR stars at all Galactocentric radii. 
We can have confidence in the former, as metallicity is not expected to effect 
the binary fraction. The latter however is almost certainly invalid. The inner 
Galaxy is a preferential environment for WC stars, particularly late-type (WCL) 
which constitute 95\% of WC stars (Table~\ref{tab:ratios}). If 100\% of 
these WCL were dust-producing, this would equate to 31\% of the total WR 
population. Mindful of this potential underestimate in the number of neglected 
systems in the model population, we estimate a total of $1200^{+300}_{-100}$ 
Galactic WR stars. 

\subsection{Expectations from star formation arguments}
By combining the measured Milky Way star formation rate (SFR) with an initial
mass function (IMF), it is possible to derive the average lifetime of the
Wolf-Rayet phase ($\tau_{\mbox{\tiny WR}}$) necessary to sustain a population of
${\sim}$1200 WR stars. Taking the Milky Way SFR to be
$1.9\,$M$_\odot$yr$^{-1}$ \citep{chomiuk11}, adopting a three-part Kroupa IMF
\citep{kroupa03}, and assuming only stars with an initial mass
${>}\,25\,\textrm{M}_\odot$ experience a WR phase, our derived
population can be reproduced with $\tau_{\mbox{\tiny WR}}\,{\simeq}\,0.25\,$Myr. 

This result is broadly consistent with rotating (non-rotating) Geneva models at
solar metallicity \citep{georgy12}, which display $\tau_{\mbox{\tiny
WR}}\,{=}\,0.45\,$Myr ($0.006\,$Myr) at $M_i\,{=}\,32\,\textrm{M}_\odot$
increasing with mass to $0.9\,$Myr ($0.4\,$Myr) at
$M_i\,{=}\,120\,\textrm{M}_\odot$. WR lifetimes as a result of binary evolution
at solar metallicity are predicted to span $\tau_{\mbox{\tiny
WR}}\,{=}\,0.5\,$Myr at $M_i\,{=}\,30\,\textrm{M}_\odot$ to $1.0\,$Myr at
$120\,\textrm{M}_\odot$ \citep{eldridge08}. Although, binary evolution favours lower WR progenitor 
masses, so the assumed progenitor mass limit of ${>}\,25\,M_\odot$ used to 
calculate of $\tau_{\mbox{\tiny WR}}$ based on our population estimate may be 
inappropriate. Previously claimed population sizes exceeding 
$6000$ \citep{vdH01,shara09} are difficult to reconcile with the measured 
Galactic SFR and a progenitor mass limit $M\,{>}\,25\,\textrm{M}_\odot$, as 
WR lifetimes in excess of $1\,$Myr would be required.

The CMZ accounts for an estimated ${\sim}4\mbox{--}5\%$ of
Galactic star formation \citep{longmore13}, yet we estimate it contains
${\sim}250$ (13\%) of the Galactic WR star population. The discrepancy between
these fractions would suggest either we have underestimated WR numbers in the
Galactic disk, or this CMZ star formation rate is insensitive to the
most recent episodes of massive star formation.

\begin{figure}
\includegraphics[width=0.48\textwidth]{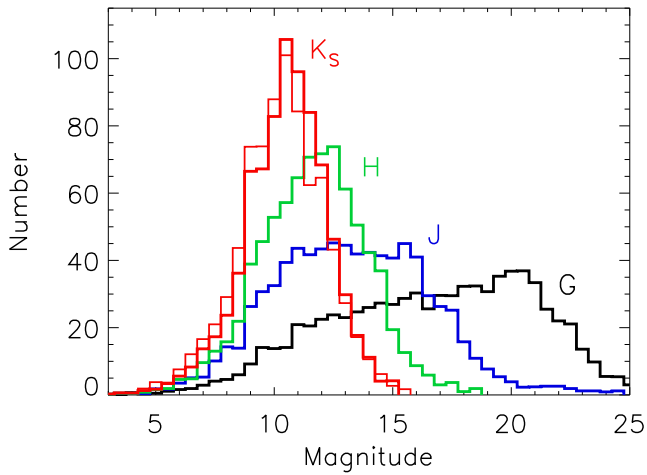}
\caption{Histogram of 2MASS $JHK_S$ and G-band (\textit{Gaia}) magnitudes predicted 
for our preferred model Galactic WR star population. Each distribution shown is 
an average over 10 model repetitions. Two $K_S$-band distributions are plotted, 
the thin (red) line represents a model population where 28\% of WC stars are dust 
forming (WC8d/9d, $M_{K_S}\,{=}\,-6.95$, Table \ref{tab:cal_jhk}). All thick
lines represent populations consisting of WN and non-dusty WC stars. } 
\label{fig:multiband_model}
\end{figure}

\subsection{Implications for future spectroscopic surveys}
Near-IR surveys, both broad and narrow-band, continue to add to our knowledge of
the obscured Galactic Wolf-Rayet population. A crucial question of to 
spectroscopic follow-up campaigns is how deep do spectroscopic surveys
need to go? Figure \ref{fig:multiband_model} shows multi-band magnitude
distributions derived from our favoured model WR star population. We estimate
that to achieve 95\% completeness, spectra of candidate WR stars need to be
taken to a depth of K$_S\,{\simeq}\,13$\,mag - shallower by
${\sim}2.5$ mag than the 95\% limit estimated by \citet{shara09}.

The ESA \textit{Gaia} mission will perform precision astrometry for a billion
stars down to visual magnitudes of $20$, and acquire low-resolution spectra of
objects brighter that magnitude $16$. To investigate the potential of
\textit{Gaia} in the search for and characterisation of WR stars, we derive the
observed G-band distribution of our favoured model population (Figure
\ref{fig:multiband_model}). To do this we utilise magnitude transformations
provided by \citet{jordi10}, $M_v$ for WR stars from \citet{vdH01}, and
$(b-v)_o$ colours from \citet{morris93}. We predict that approximately 600
(${\sim}\,\nicefrac{1}{3}$ of total) WR stars will appear within \textit{Gaia}'s
$6\,{<}\,G\,{<}\,20$ observing range, with ${\sim}\,300$ brighter than the
magnitude limit for spectroscopy. With the known population currently totalling
${\sim}\,635$, \textit{Gaia} is unlikely to discover significant WR populations
via spectroscopy, but the majority of those known will have distances measured
to a significantly higher level of accuracy than is currently possible.


\section{Conclusions}
\label{sec:conc}
We have presented near-IR absolute magnitude-spectral type calibrations for WN, WC and WO type 
Wolf-Rayet stars, based on $126$ examples with known distances (mostly by cluster or OB 
association membership). Applying these calibrations to the rapidly growing 
known Galactic population, we derive distances to a further $246$ WR stars and present 
a 3D map of their locations. We note that approximately half as many WC stars are 
available for calibration as cluster/association members than WN, consistent with 
the findings of \citet{smith14} than WC stars are generally more isolated. This challenges 
the idea that WC stars descend directly from WN, which in turn descend from the most 
massive O-stars. We have shown the heights of 
WR stars from the Galactic midplane to be Cauchy distributed with 
HWHM$\,{=}\,39.2\,$pc, where 12 stars reside at 
$|z|\,{>}\,300\,$pc.
The low binary fraction and a preference for WN8 subtypes in this small sample 
of runaway stars indicates a binary supernova origin for the most extreme examples.

Exploiting the variation of metallicity across the Galactic disk, we have
compared subtype number ratios measured in the inner Galaxy, solar
neighbourhood, and outer Galaxy to the predictions of various
metallicity-dependent evolutionary models. We measure $N_{WC}/N_{WN}$ to be
significantly higher than predicted by evolutionary models including fast
rotation \citep{meynet05}, suggesting that lengthened WNL and shortened WC
phases resulting from stellar rotation are not widely observed at
($\textrm{Z}\,{\gtrsim}\,\textrm{Z}_\odot$). Similarly, a shortened eWNE phase
in such models - particularly at lower metallicity - is not manifest in our
observations, as we observe approximately equal numbers of WNE and WNL stars in
regions of $\textrm{Z}\,{\lesssim}\,\textrm{Z}_\odot$. Single-star models
without rotation \citep{eldridge06} and models that account for the various
effects of binary interaction \citep{eldridge08} reproduce our measurements of
$N_{WC}/N_{WN}$ more appropriately. Hence, to a first approximation a population
consisting of non-rotating single stars and interacting binaries would be
consistent with the $N_{WC}/N_{WN}$ we observe. However, we caution that all
comparisons of this nature are subject to how the physics contained in stellar
models is expected to translate into observable properties, which currently
rests on estimations of surface abundances and temperatures that may not be
appropriate \citep{groh14}, particularly for the transition between eWNE and eWNL
subtypes. 

Consolidating information gained about the spatial distribution, subtype 
variation, and intrinsic IR brightness of WR stars, we have created a scalable 
toy model of the Galactic WR population. By applying a 3D dust distribution to this model
- spatially congruous with the gas content of the Galaxy - we derive observable
properties for populations of various sizes at multiple wavelengths. Comparison
of these model-derived observables to the observed population of non-dusty,
WR-dominated ($m_K^{\mbox{\tiny WR}}\,{-}\,m_K^{sys}\,{>}\,0.4$) systems to a
completeness limit of $m_K^{sys}\,{<}\,8$ indicates a total of ${\simeq}800$
in the Galaxy. Using a volume-limited sample ($d\,{<}\,3$kpc) we
estimate such systems represent ${\sim}\,69$\% of the whole WR population, hence
we claim a Galactic WR population totalling $1200^{+300}_{-100}$. 

We deduce that an average WR phase duration of $0.25\,$Myr is necessary to 
sustain our estimated population, assuming a Kroupa IMF and a constant Milky 
Way star formation rate of $1.9\,\textrm{M}_\odot$yr$^{-1}$. This is compatible 
with the WR phase duration in rotating stellar models 
at solar metallicity \citep{georgy12}.    

Looking to the future, we have used our favoured model WR population 
to estimate a required depth of $K_S\,{<}\,13$ for spectroscopic surveys to achieve
95\% completeness in Galactic WR stars. We have also predicted that the 
ESA \textit{Gaia} mission will not deliver a significant number of 
WR star discoveries via low-resolution spectroscopy, but should provide 
improved distance measurements for the majority of the currently 
recognised population.


\section*{Acknowledgments}
The authors would like to thank he referee (A. F. J. Moffat) for a careful reading and useful comments that improved the paper. We also thank William. D. Vacca for making available high-quality 
spectra of selected WR stars, useful for refining near-IR classification diagnostics. 
We are also grateful to MPhys students Adrian Kus and Katie Baker for constructing 
and originally populating the Wolf-Rayet catalogue database.
This work was supported in part by a UK Science and Technology
Facilities Council studentship (CKR).




\addtocounter{table}{-4}

\begin{table*}
\begin{minipage}{15cm}
\caption{Full version.}
\centering
\begin{tabular}{ l c c c c c c c c c }
\hline\\[-2.0ex]
WR\# & Sp. Type & ref & J & H & K$_S$ & $\bar{\mathrm{A}}_{K_S}$ & $d$(kpc) & R$_G$(kpc) & z(pc) \\[1.0ex]
\hline\\[-2.0ex]
3 & WN3ha & 1 & 10.24 & 10.13 & 10.01  & $0.18\pm0.02$ & $4.53\pm1.15$ & $11.41\pm1.18$ & $-308\pm83$ \\
4 & WC5 & 2 & 8.75 & 8.57 & 7.88 & $0.13\pm0.01$ & $2.69\pm0.49$ & $10.15\pm0.55$ & $-121\pm25$ \\
5 & WC6 & 2 & 8.63 & 8.34 & 7.65 & $0.16\pm0.02$ & $2.69\pm0.84$ & $10.18\pm0.88$ & $-82\pm32$ \\
13 & WC6 & 2 & 10.14 & 9.64 & 8.86 & $0.29\pm0.02$ & $4.42\pm1.38$ & $9.46\pm1.40$ & $-40\pm19$ \\
15 & WC6 & 2 & 7.85 & 7.34 & 6.60 & $0.28\pm0.02$ & $1.57\pm0.49$ & $8.11\pm0.55$ & $-10\pm9$ \\
16 & WN8h & 2 & 6.97 & 6.71 & 6.38 & $0.24\pm0.02$ & $2.77\pm0.46$ & $7.95\pm0.52$ & $-104\pm20$ \\
17 & WC5 & 2 & 9.93 & 9.74 & 9.17 & $0.07\pm0.02$ & $5.02\pm0.91$ & $8.31\pm0.95$ & $-303\pm59$ \\
17-1 & WN5b & 3 & 11.73 & 10.38 & 9.53 & $0.85\pm0.05$ & $5.43\pm0.86$ & $8.58\pm0.89$ & $-41\pm10$ \\
19 & WC5+O9 & 4,5 & 9.75 & 9.13 & 8.53 & $0.20\pm0.02$ & $3.52\pm0.64$ & $7.93\pm0.69$ & $-54\pm13$ \\
19a & WN7(h) & 2 & 9.07 & 8.13 & 7.50 & $0.71\pm0.04$ & $2.41\pm0.47$ & $7.78\pm0.53$ & $-23\pm8$ \\
20 & WN5 & 2 & 11.00 & 10.43 & 9.93 & $0.45\pm0.03$ & $4.65\pm0.72$ & $8.18\pm0.76$ & $-130\pm23$ \\
20b & WN6ha & 2 & 8.65 & 7.80 & 7.18 & $0.75\pm0.04$ & $4.84\pm0.79$ & $8.26\pm0.83$ & $-10\pm5$ \\
27 & WC6 & 2 & 9.88 & 9.17 & 8.30 & $0.44\pm0.03$ & $3.18\pm0.99$ & $7.69\pm1.02$ & $26\pm2$ \\
31c & WC6 & 2 & 11.04 & 10.34 & 9.49 & $0.42\pm0.03$ & $5.56\pm1.74$ & $8.09\pm1.76$ & $-50\pm22$ \\
33 & WC6 & 4 & 10.62 & 10.35 & 9.69 & $0.13\pm0.01$ & $6.98\pm2.18$ & $8.79\pm2.19$ & $251\pm72$ \\
34 & WN5 & 2 & 11.20 & 10.59 & 10.04 & $0.50\pm0.03$ & $4.78\pm0.74$ & $7.79\pm0.78$ & $-96\pm18$ \\
37 & WN4b & 2 & 11.03 & 10.34 & 9.67 & $0.50\pm0.03$ & $6.36\pm1.11$ & $8.29\pm1.14$ & $-97\pm20$ \\
38 & WC4 & 2 & 11.99 & 11.50 & 10.79 & $0.25\pm0.03$ & $6.43\pm0.98$ & $8.32\pm1.01$ & $-84\pm16$ \\
38a & WN5 & 2 & 11.53 & 11.21 & 10.70 & $0.40\pm0.05$ & $6.81\pm1.07$ & $8.49\pm1.10$ & $-90\pm17$ \\
40 & WN8h & 2 & 6.62 & 6.41 & 6.11 & $0.20\pm0.02$ & $2.48\pm0.41$ & $7.42\pm0.48$ & $-190\pm34$ \\
42a & WN5b & 2 & 12.08 & 11.31 & 10.81 & $0.38\pm0.03$ & $12.20\pm1.91$ & $11.93\pm1.93$ & $-84\pm16$ \\
42b & WN4b & 2 & 11.35 & 10.52 & 9.90 & $0.51\pm0.03$ & $7.05\pm1.23$ & $8.48\pm1.25$ & $-168\pm33$ \\
42c & WN5 & 2 & 11.15 & 10.47 & 9.90 & $0.53\pm0.03$ & $4.41\pm0.68$ & $7.59\pm0.73$ & $-10\pm5$ \\
42d & WN5b & 2 & 10.20 & 9.52 & 8.91 & $0.45\pm0.03$ & $4.92\pm0.77$ & $7.70\pm0.81$ & $-21\pm6$ \\
43-1 & WN4b & 6 & 13.16 & 11.57 & 10.47 & $1.13\pm0.06$ & $6.88\pm1.21$ & $8.39\pm1.24$ & $-59\pm14$ \\
44 & WN4 & 2 & 11.16 & 10.89 & 10.48 & $0.45\pm0.03$ & $4.83\pm0.98$ & $7.71\pm1.01$ & $125\pm22$ \\
44a & WN5b & 2 & 12.07 & 11.34 & 10.82 & $0.39\pm0.03$ & $12.18\pm1.91$ & $11.79\pm1.93$ & $-95\pm18$ \\
45 & WC6 & 2 & 10.60 & 9.99 & 9.19 & $0.35\pm0.02$ & $5.03\pm1.57$ & $7.48\pm1.59$ & $-33\pm16$ \\
45a & WN5 & 2 & 11.90 & 11.24 & 10.73 & $0.48\pm0.03$ & $6.66\pm1.03$ & $7.93\pm1.06$ & $75\pm9$ \\
45b & WN4b & 2 & 11.45 & 10.61 & 9.92 & $0.57\pm0.03$ & $6.91\pm1.20$ & $7.98\pm1.23$ & $-28\pm8$ \\
45-3 & WN5b & 3 & 12.57 & 11.57 & 10.94 & $0.57\pm0.04$ & $11.89\pm1.87$ & $11.04\pm1.89$ & $-152\pm27$ \\
45-4 & WN6 & 6 & 11.65 & 10.76 & 10.14 & $0.64\pm0.04$ & $6.73\pm1.33$ & $7.81\pm1.35$ & $-47\pm13$ \\
45c & WN5 & 2 & 11.35 & 10.79 & 10.32 & $0.42\pm0.03$ & $5.66\pm0.88$ & $7.45\pm0.91$ & $-34\pm8$ \\
46a & WN4 & 2 & 12.08 & 11.42 & 10.92 & $0.65\pm0.04$ & $5.38\pm1.09$ & $7.18\pm1.12$ & $-89\pm22$ \\
46-1 & WN6 & 7 & 11.55 & 10.82 & 10.22 & $0.57\pm0.03$ & $7.24\pm1.43$ & $7.89\pm1.45$ & $-7\pm5$ \\
46-8 & WN6 & 8 & 12.02 & 10.70 & 9.85 & $0.96\pm0.05$ & $5.08\pm1.01$ & $7.16\pm1.04$ & $-20\pm8$ \\
46-9* & WN5 & 3 & 12.07 & 10.72 & 9.88 & $0.96\pm0.06$ & $3.60\pm0.56$ & $7.04\pm0.62$ & $4\pm2$ \\
46-2 & WN7h & 7 & 10.56 & 9.64 & 9.03 & $0.68\pm0.04$ & $4.93\pm0.95$ & $7.10\pm0.98$ & $-31\pm10$ \\
46-5 & WN6 & 9 & - & - & 10.92 & $1.18\pm0.13^7$ & $8.63\pm1.97$ & $8.49\pm1.99$ & $-42\pm14$ \\
46-6 & WN7 & 9 & - & - & 10.48 & $1.18\pm0.13^7$ & $9.10\pm2.24$ & $8.76\pm2.25$ & $-45\pm16$ \\
46-15 & WN8 & 10 & 13.13 & 11.03 & 9.84 & $1.51\pm0.08$ & $7.58\pm1.28$ & $7.94\pm1.30$ & $-7\pm4$ \\
46-12 & WN4b & 3 & 13.56 & 12.55 & 11.84 & $0.64\pm0.04$ & $16.19\pm2.82$ & $13.97\pm2.83$ & $-128\pm26$ \\
46-14 & WN5b & 6 & 12.93 & 11.76 & 10.96 & $0.75\pm0.04$ & $11.04\pm1.74$ & $9.68\pm1.76$ & $-91\pm17$ \\
47a & WN8h & 2 & 10.44 & 9.68 & 9.06 & $0.64\pm0.04$ & $7.90\pm1.30$ & $7.67\pm1.33$ & $-159\pm29$ \\
47-1 & WN6 & 7 & 12.18 & 11.22 & 10.56 & $0.70\pm0.04$ & $7.96\pm1.57$ & $7.70\pm1.59$ & $7\pm3$ \\
47b & WN9h & 2 & 10.29 & 9.44 & 8.84 & $0.65\pm0.04$ & $7.96\pm1.23$ & $7.67\pm1.26$ & $-88\pm17$ \\
47c & WC5 & 2 & 11.24 & 10.48 & 9.89 & $0.20\pm0.02$ & $6.60\pm1.20$ & $7.06\pm1.22$ & $-85\pm19$ \\
48-1 & WC7 & 7 & 11.06 & 10.09 & 9.31 & $0.41\pm0.03$ & $5.59\pm0.94$ & $6.62\pm0.97$ & $33\pm2$ \\
48-5 & WN6b & 3 & 13.06 & 11.34 & 10.26 & $1.15\pm0.06$ & $7.14\pm1.24$ & $6.99\pm1.27$ & $-21\pm7$ \\
49 & WN5(h) & 2 & 11.89 & 11.57 & 11.21 & $0.25\pm0.02$ & $9.19\pm1.42$ & $7.96\pm1.44$ & $-386\pm63$ \\
49-1 & WN8 & 10 & 13.82 & 11.88 & 10.67 & $1.48\pm0.08$ & $11.29\pm1.91$ & $9.28\pm1.93$ & $88\pm12$ \\
52 & WC4 & 2 & 8.41 & 8.21 & 7.55 & $0.11\pm0.02$ & $1.54\pm0.23$ & $7.19\pm0.34$ & $142\pm18$ \\
52-1 & WN6 & 6 & 14.77 & 12.69 & 11.55 & $1.42\pm0.08$ & $9.02\pm1.81$ & $7.72\pm1.83$ & $42\pm5$ \\
52-2 & WN6 & 6 & 12.52 & 11.40 & 10.68 & $0.79\pm0.05$ & $8.10\pm1.60$ & $7.15\pm1.62$ & $84\pm13$ \\
54 & WN5 & 2 & 10.85 & 10.48 & 10.09 & $0.29\pm0.02$ & $5.40\pm0.84$ & $6.39\pm0.87$ & $-216\pm37$ \\
55 & WN7 & 2 & 8.77 & 8.49 & 8.01 & $0.36\pm0.03$ & $3.57\pm0.69$ & $6.46\pm0.74$ & $29\pm2$ \\
56 & WC7 & 2 & 11.84 & 11.48 & 10.77 & $0.21\pm0.03$ & $12.02\pm2.03$ & $9.55\pm2.04$ & $-323\pm58$ \\
57 & WC8 & 2 & 9.09 & 8.75 & 8.01 & $0.37\pm0.02$ & $2.98\pm0.52$ & $6.61\pm0.57$ & $-242\pm45$ \\
57-1 & WN7 & 10 & 15.01 & 12.77 & 11.46 & $1.65\pm0.09$ & $9.68\pm1.92$ & $7.77\pm1.93$ & $-64\pm17$ \\
60 & WC8 & 2 & 8.91 & 8.37 & 7.70 & $0.39\pm0.03$ & $2.56\pm0.45$ & $6.63\pm0.51$ & $53\pm6$ \\
60-1 & WC8 & 6 & 15.30 & 12.43 & 10.60 & $2.04\pm0.11$ & $4.56\pm0.82$ & $5.98\pm0.86$ & $34\pm3$ \\
60-5 & WC7 & 3 & 10.84 & 10.12 & 9.39 & $0.32\pm0.03$ & $6.02\pm1.01$ & $5.98\pm1.04$ & $134\pm19$ \\[0.5ex]
\hline
\end{tabular}\par
\label{tab:single-field-wr}
\end{minipage}
\end{table*}

\addtocounter{table}{-1}

\clearpage
\begin{table*}
\begin{minipage}{15cm}
\centering
\caption{(Continued)}
\begin{tabular}{ l c c c c c c c c c }
\hline\\[-2.0ex]
WR\# & Sp. Type  & ref & J & H & K$_S$ & $\bar{\mathrm{A}}_{K_S}$ & $d$(kpc) & R$_G$(kpc) & z(pc) \\[1.0ex]
\hline\\[-2.0ex]
60-4 & WC8 & 10 & 15.58 & 12.71 & 11.00 & $1.93\pm0.11$ & $5.75\pm1.03$ & $6.00\pm1.06$ & $35\pm3$ \\
61 & WN5 & 2 & 10.96 & 10.67 & 10.35 & $0.21\pm0.02$ & $6.32\pm0.98$ & $6.10\pm1.01$ & $-411\pm67$ \\
61-2 & WN5b & 6 & 13.60 & 11.39 & 10.01 & $1.55\pm0.09$ & $4.92\pm0.79$ & $5.88\pm0.83$ & $11\pm1$ \\
61-3 & WC9 & 10 & 13.62 & 11.63 & 10.38 & $1.41\pm0.08$ & $4.79\pm0.98$ & $5.84\pm1.02$ & $37\pm4$ \\
61-1 & WN6 & 8 & 11.68 & 10.43 & 9.61 & $0.91\pm0.05$ & $4.68\pm0.93$ & $5.83\pm0.96$ & $72\pm10$ \\
62 & WN6b & 2 & 9.11 & 8.35 & 7.75 & $0.46\pm0.03$ & $3.08\pm0.53$ & $6.24\pm0.59$ & $-21\pm7$ \\
62b & WN5 & 2 & 12.01 & 11.44 & 10.97 & $0.42\pm0.09$ & $7.64\pm1.41$ & $5.82\pm1.43$ & $-153\pm32$ \\
64 & WC7 & 2 & 12.50 & 12.03 & 11.33 & $0.23\pm0.02$ & $15.35\pm2.59$ & $10.55\pm2.61$ & $775\pm127$ \\
64-1 & WN4b & 6 & 13.84 & 11.83 & 10.64 & $1.32\pm0.07$ & $6.83\pm1.21$ & $5.32\pm1.23$ & $-38\pm10$ \\
67-2 & WC7 & 3 & 10.34 & 9.26 & 8.46 & $0.46\pm0.03$ & $3.68\pm0.62$ & $5.62\pm0.67$ & $-7\pm5$ \\
68a & WN6 & 2 & 9.59 & 9.02 & 8.62 & $0.36\pm0.02$ & $3.80\pm0.75$ & $5.52\pm0.79$ & $-22\pm8$ \\
68-1 & WN4b & 6 & 13.88 & 12.15 & 11.06 & $1.16\pm0.07$ & $8.88\pm1.56$ & $5.56\pm1.58$ & $-11\pm5$ \\
70a & WN6 & 2 & 10.94 & 10.24 & 9.76 & $0.47\pm0.03$ & $6.13\pm1.21$ & $4.26\pm1.24$ & $-72\pm18$ \\
70-1 & WN7 & 8 & 13.84 & 12.39 & 11.46 & $1.10\pm0.06$ & $12.40\pm2.42$ & $7.51\pm2.43$ & $-23\pm8$ \\
70-3 & WC7 & 3 & 10.96 & 9.94 & 9.06 & $0.53\pm0.03$ & $4.68\pm0.79$ & $4.89\pm0.83$ & $9\pm2$ \\
70-4 & WN9h & 11 & 14.58 & 12.22 & 10.87 & $1.72\pm0.10$ & $12.46\pm1.99$ & $7.20\pm2.01$ & $219\pm32$ \\
70-5 & WC9 & 11 & 11.19 & 9.71 & 8.49 & $1.24\pm0.07$ & $2.17\pm0.44$ & $6.31\pm0.51$ & $-2\pm5$ \\
70-6 & WN6b & 3 & 15.16 & 12.51 & 10.97 & $1.80\pm0.10$ & $7.34\pm1.30$ & $4.29\pm1.33$ & $25\pm1$ \\
70-7 & WN6b & 10 & 15.94 & 13.05 & 11.36 & $2.00\pm0.11$ & $7.99\pm1.43$ & $4.27\pm1.45$ & $52\pm6$ \\
70-8 & WN7 & 10 & 15.64 & 12.59 & 10.91 & $2.20\pm0.12$ & $5.84\pm1.16$ & $4.19\pm1.19$ & $105\pm17$ \\
70-2 & WN5b & 3 & 10.88 & 9.52 & 8.66 & $0.86\pm0.05$ & $3.63\pm0.57$ & $5.20\pm0.63$ & $69\pm8$ \\
70-9 & WC8 & 10 & 15.56 & 12.87 & 11.31 & $1.75\pm0.10$ & $7.20\pm1.28$ & $4.10\pm1.31$ & $29\pm2$ \\
70-11 & WN7 & 3 & 12.02 & 10.65 & 9.88 & $0.94\pm0.05$ & $6.47\pm1.25$ & $4.05\pm1.27$ & $96\pm15$ \\
71 & WN6 & 2 & 9.48 & 9.31 & 9.09 & $0.10\pm0.01$ & $5.34\pm1.05$ & $4.93\pm1.09$ & $-689\pm139$ \\
72-3 & WC9 & 12 & 15.62 & 13.05 & 11.53 & $1.82\pm0.10$ & $6.74\pm1.40$ & $3.80\pm1.42$ & $92\pm15$ \\
72-1 & WC9 & 7 & 10.16 & 9.25 & 8.46 & $0.72\pm0.04$ & $2.73\pm0.56$ & $5.76\pm0.61$ & $-4\pm5$ \\
73-1 & WC7 & 3 & 15.05 & 12.79 & 11.54 & $1.18\pm0.07$ & $10.94\pm1.88$ & $5.42\pm1.90$ & $-45\pm11$ \\
74 & WN7 & 2 & 9.73 & 9.22 & 8.80 & $0.40\pm0.03$ & $5.04\pm0.97$ & $4.28\pm1.00$ & $-36\pm11$ \\
74-1 & WN9ha & 10 & 15.50 & 12.28 & 10.59 & $2.37\pm0.13$ & $8.15\pm1.24$ & $3.71\pm1.27$ & $13\pm1$ \\
74-2 & WN7 & 10 & 15.64 & 12.90 & 11.43 & $1.94\pm0.11$ & $8.34\pm1.66$ & $3.64\pm1.68$ & $27\pm2$ \\
75 & WN6b & 2 & 8.60 & 8.24 & 7.84 & $0.19\pm0.02$ & $3.64\pm0.62$ & $5.05\pm0.67$ & $-74\pm16$ \\
75a & WC9 & 2 & 9.96 & 9.14 & 8.50 & $0.56\pm0.03$ & $2.98\pm0.60$ & $5.48\pm0.65$ & $-26\pm9$ \\
75-1 & WC8 & 6 & 13.28 & 11.76 & 10.73 & $0.97\pm0.06$ & $7.89\pm1.40$ & $3.48\pm1.42$ & $57\pm7$ \\
75-14 & WC9 & 3 & 14.50 & 12.37 & 11.22 & $1.37\pm0.08$ & $7.17\pm1.50$ & $3.34\pm1.52$ & $51\pm7$ \\
75b & WC9 & 2 & 9.76 & 9.00 & 8.36 & $0.54\pm0.04$ & $2.82\pm0.57$ & $5.55\pm0.63$ & $36\pm3$ \\
75-15 & WC8 & 12 & - & 13.16 & 11.50 & $1.81\pm0.25$ & $7.66\pm1.61$ & $3.32\pm1.63$ & $57\pm8$ \\
75-6 & WN5b & 10 & 14.65 & 12.80 & 11.76 & $1.15\pm0.07$ & $13.28\pm2.12$ & $6.86\pm2.14$ & $37\pm3$ \\
75-16 & WC8 & 12 & - & 13.49 & 11.83 & $1.81\pm0.25$ & $8.90\pm1.86$ & $3.66\pm1.88$ & $43\pm5$ \\
75-2 & WC8 & 6 & 15.56 & 12.86 & 11.32 & $1.73\pm0.10$ & $7.31\pm1.31$ & $3.23\pm1.33$ & $44\pm4$ \\
75-3 & WC8 & 6 & 15.11 & 12.86 & 11.47 & $1.49\pm0.09$ & $8.75\pm1.57$ & $3.52\pm1.59$ & $43\pm4$ \\
75-4 & WN5b & 6 & 15.00 & 12.70 & 11.34 & $1.55\pm0.09$ & $9.08\pm1.47$ & $3.65\pm1.49$ & $46\pm4$ \\
75-17 & WC8 & 12 & 16.00 & 14.17 & 12.75 & $1.39\pm0.09$ & $16.51\pm3.05$ & $9.73\pm3.06$ & $48\pm5$ \\
75-7 & WC9 & 10 & 14.97 & 12.96 & 11.63 & $1.49\pm0.08$ & $8.21\pm1.69$ & $3.32\pm1.71$ & $46\pm5$ \\
75-8 & WN4b & 3 & 15.00 & 12.44 & 10.90 & $1.78\pm0.10$ & $6.22\pm1.12$ & $3.42\pm1.14$ & $-11\pm6$ \\
75c & WC9 & 13 & 11.62 & 11.12 & 10.52 & $0.44\pm0.03$ & $8.00\pm1.62$ & $2.83\pm1.64$ & $406\pm78$ \\
75d & WC9 & 13 & 10.68 & 9.88 & 9.12 & $0.66\pm0.04$ & $3.78\pm0.77$ & $4.72\pm0.81$ & $90\pm14$ \\
75-10 & WN6b & 3 & 13.75 & 11.70 & 10.46 & $1.38\pm0.08$ & $7.03\pm1.23$ & $3.11\pm1.26$ & $63\pm8$ \\
75-5 & WC8 & 6 & 14.90 & 12.77 & 11.43 & $1.41\pm0.08$ & $8.93\pm1.60$ & $3.57\pm1.62$ & $-48\pm12$ \\
75-12 & WN6 & 10 & 15.02 & 12.85 & 11.49 & $1.63\pm0.09$ & $7.98\pm1.63$ & $3.15\pm1.65$ & $49\pm6$ \\
75-22 & WC9 & 12 & - & 13.72 & 12.03 & $2.01\pm0.30$ & $7.80\pm1.99$ & $3.14\pm2.00$ & $30\pm3$ \\
75-13 & WC8 & 10 & 13.60 & 11.67 & 10.39 & $1.30\pm0.07$ & $5.80\pm1.03$ & $3.48\pm1.06$ & $26\pm1$ \\
75-23 & WC9 & 12 & 10.27 & 9.59 & 8.93 & $0.54\pm0.03$ & $3.67\pm0.75$ & $4.80\pm0.79$ & $49\pm6$ \\
76-1 & WN9 & 10 & 14.84 & 12.00 & 10.42 & $2.05\pm0.11$ & $8.69\pm1.41$ & $3.14\pm1.43$ & $82\pm10$ \\
77-1 & WN7b & 3 & 10.06 & 9.00 & 8.29 & $0.65\pm0.04$ & $4.03\pm0.63$ & $4.45\pm0.67$ & $38\pm3$ \\
77-2 & WN7 & 8 & 10.75 & 9.52 & 8.73 & $0.93\pm0.05$ & $3.83\pm0.74$ & $4.60\pm0.78$ & $5\pm3$ \\
77-5 & WN6 & 12 & 10.56 & 9.71 & 9.14 & $0.58\pm0.03$ & $4.37\pm0.86$ & $4.22\pm0.90$ & $17\pm1$ \\
77-3 & WN6 & 10 & 11.12 & 9.84 & 9.10 & $0.85\pm0.05$ & $3.79\pm0.75$ & $4.64\pm0.79$ & $-19\pm8$ \\
77-4 & WN6 & 3 & 14.14 & 11.85 & 10.59 & $1.58\pm0.09$ & $5.38\pm1.09$ & $3.51\pm1.12$ & $-54\pm15$ \\
77-6 & WN6b & 3 & 13.56 & 11.93 & 10.95 & $1.04\pm0.06$ & $10.27\pm1.78$ & $3.65\pm1.80$ & $120\pm17$ \\
77-7 & WC9 & 12 & - & 13.46 & 12.06 & $1.60\pm0.22$ & $9.51\pm2.17$ & $3.18\pm2.19$ & $61\pm10$ \\
81 & WC9 & 2 & 8.29 & 7.76 & 7.12 & $0.48\pm0.03$ & $1.64\pm0.34$ & $6.47\pm0.42$ & $-55\pm15$ \\
82 & WN7(h) & 2 & 9.48 & 9.04 & 8.69 & $0.32\pm0.02$ & $4.98\pm0.95$ & $3.61\pm0.99$ & $-182\pm39$ \\[1.0ex]
\hline
\end{tabular}\par
\end{minipage}
\end{table*}

\addtocounter{table}{-1}

\clearpage
\begin{table*}
\begin{minipage}{15cm}
\centering
\caption{(Continued)}
\begin{tabular}{ l c c c c c c c c c }
\hline\\[-2.0ex]
WR\# & Sp. Type & ref & J & H & K$_S$ & $\bar{\mathrm{A}}_{K_S}$ & $d$(kpc) & R$_G$(kpc) & z(pc) \\[1.0ex]
\hline\\[-2.0ex]
83 & WN5 & 2 & 10.26 & 9.89 & 9.53 & $0.27\pm0.02$ & $4.20\pm0.65$ & $4.24\pm0.70$ & $-281\pm47$ \\
84 & WN7 & 2 & 9.58 & 8.96 & 8.50 & $0.46\pm0.10$ & $4.28\pm0.93$ & $3.95\pm0.96$ & $4\pm3$ \\
84-2 & WC8 & 12 & 13.98 & 12.20 & 10.98 & $1.21\pm0.07$ & $7.95\pm1.41$ & $1.73\pm1.43$ & $43\pm4$ \\
84-1 & WN9 & 3 & 12.37 & 10.43 & 9.19 & $1.51\pm0.08$ & $6.31\pm1.00$ & $2.28\pm1.03$ & $40\pm3$ \\
84-4 & WN7ha & 3 & 10.88 & 10.00 & 9.49 & $0.69\pm0.04$ & $16.50\pm3.02$ & $8.86\pm3.03$ & $35\pm3$ \\
84-5 & WC9 & 12 & 14.34 & 12.78 & 11.59 & $1.24\pm0.07$ & $9.03\pm1.86$ & $1.97\pm1.87$ & $75\pm11$ \\
88 & WC9 & 2 & 9.03 & 8.56 & 8.05 & $0.36\pm0.03$ & $2.66\pm0.54$ & $5.37\pm0.60$ & $114\pm19$ \\
90 & WC7 & 2 & 6.25 & 6.09 & 5.52 & $0.05\pm0.02$ & $1.14\pm0.19$ & $6.92\pm0.32$ & $-76\pm16$ \\
91 & WN7b & 2 & 9.53 & 8.77 & 8.20 & $0.44\pm0.03$ & $4.25\pm0.66$ & $3.92\pm0.70$ & $-60\pm12$ \\
91-1 & WC7 & 3 & 15.11 & 12.05 & 10.33 & $1.80\pm0.10$ & $4.69\pm0.82$ & $3.41\pm0.85$ & $80\pm11$ \\
92 & WC9 & 2 & 9.50 & 9.22 & 8.82 & $0.21\pm0.02$ & $4.07\pm0.82$ & $4.20\pm0.86$ & $-294\pm63$ \\
93a & WN3 & 2 & 13.88 & 13.22 & 12.72 & $0.65\pm0.02$ & $10.22\pm3.25$ & $2.20\pm3.27$ & $694\pm214$ \\
93b & WO3 & 14 & - & - & 10.17 & $0.73\pm0.04$ & $2.18\pm0.32$ & $5.84\pm0.40$ & $-13\pm5$ \\
94 & WN5 & 2 & 7.09 & 6.19 & 5.91 & $0.32\pm0.02$ & $0.78\pm0.12$ & $7.23\pm0.28$ & $16\pm1$ \\
94-1 & WC9 & 12 & 15.99 & 12.59 & 10.76 & $2.31\pm0.13$ & $3.78\pm0.81$ & $4.24\pm0.85$ & $35\pm3$ \\
98-1 & WC8 & 12 & 13.52 & 11.70 & 10.56 & $1.17\pm0.03$ & $6.67\pm1.18$ & $1.34\pm1.20$ & $45\pm5$ \\
100 & WN7b & 2 & 8.85 & 8.27 & 7.72 & $0.38\pm0.03$ & $3.50\pm0.54$ & $4.51\pm0.60$ & $-58\pm12$ \\
100-3 & WN8 & 12 & 13.90 & 11.52 & 10.17 & $1.71\pm0.05$ & $7.20\pm1.21$ & $0.82\pm1.24$ & $-17\pm6$ \\
100-1 & WN7b & 3 & 15.12 & 12.61 & 11.08 & $1.75\pm0.04$ & $8.75\pm1.37$ & $0.75\pm1.39$ & $12\pm1$ \\
101 & WC8 & 2 & 9.62 & 8.78 & 7.89 & $0.66\pm0.04$ & $2.46\pm0.43$ & $5.54\pm0.50$ & $-42\pm11$ \\
101-3 & WN8 & 3 & 16.67 & 13.08 & 11.09 & $2.61\pm0.05$ & $7.49\pm1.26$ & $0.51\pm1.28$ & $25\pm1$ \\
101-8 & WC9 & 15 & - & 14.67 & 12.70 & $2.31\pm0.11$ & $9.23\pm1.93$ & $1.23\pm1.94$ & $5\pm3$ \\
101-5 & WN6b & 3 & - & 15.21 & 12.22 & $3.68\pm0.17$ & $5.50\pm1.04$ & $2.50\pm1.07$ & $22\pm1$ \\
101-9 & WC9 & 15 & - & 14.62 & 12.02 & $3.17\pm0.13$ & $4.53\pm0.96$ & $3.47\pm0.99$ & $15\pm1$ \\
101-1 & WN9h & 3 & 15.55 & 12.26 & 10.42 & $2.39\pm0.05$ & $7.40\pm1.15$ & $0.60\pm1.18$ & $19\pm0$ \\
101p & WC8 & 3 & 16.32 & 13.05 & 11.20 & $2.16\pm0.06$ & $5.68\pm1.02$ & $2.32\pm1.05$ & $21\pm1$ \\
102 & WO2 & 2 & - & - & 10.62 & $0.50\pm0.10$ & $2.99\pm0.45$ & $5.02\pm0.52$ & $93\pm11$ \\
102-12 & WN9 & 3 & - & 13.61 & 11.14 & $3.20\pm0.14$ & $7.11\pm1.18$ & $0.89\pm1.20$ & $22\pm1$ \\
102b & WN7b & 3 & 15.44 & 12.48 & 10.85 & $1.95\pm0.04$ & $7.18\pm1.11$ & $0.82\pm1.14$ & $14\pm1$ \\
102bd & WC9 & 3 & 16.26 & 13.47 & 11.49 & $2.26\pm0.07$ & $5.40\pm1.14$ & $2.60\pm1.16$ & $15\pm1$ \\
102-17 & WN6b & 3 & 16.33 & 13.27 & 11.43 & $2.16\pm0.04$ & $7.69\pm1.33$ & $0.32\pm1.35$ & $9\pm2$ \\
102-8 & WN9 & 3 & - & 13.74 & 11.24 & $3.23\pm0.15$ & $7.34\pm1.24$ & $0.66\pm1.26$ & $14\pm1$ \\
102-9 & WN9 & 3 & 16.56 & 12.72 & 10.54 & $2.84\pm0.05$ & $6.37\pm0.99$ & $1.63\pm1.02$ & $12\pm1$ \\
102-10 & WN8 & 3 & 17.06 & 13.07 & 10.78 & $2.97\pm0.06$ & $5.48\pm0.92$ & $2.52\pm0.95$ & $12\pm1$ \\
102-19 & WN5 & 12 & 11.84 & 10.95 & 10.32 & $0.65\pm0.04$ & $5.08\pm0.79$ & $2.98\pm0.83$ & $73\pm8$ \\
102-20 & WC9 & 12 & - & 12.38 & 10.89 & $1.74\pm0.26$ & $5.22\pm1.25$ & $2.85\pm1.27$ & $-3\pm5$ \\
102l & WN8 & 2 & 8.83 & 8.10 & 7.57 & $0.56\pm0.04$ & $4.15\pm0.69$ & $3.91\pm0.73$ & $36\pm3$ \\
107 & WN8 & 2 & 9.38 & 8.69 & 8.19 & $0.52\pm0.04$ & $5.62\pm0.93$ & $2.57\pm0.96$ & $4\pm3$ \\
107a & WC6 & 2 & 11.17 & 10.31 & 9.38 & $0.53\pm0.04$ & $5.04\pm1.58$ & $3.09\pm1.60$ & $-17\pm12$ \\
108 & WN9ha & 2 & 7.66 & 7.34 & 7.10 & $0.30\pm0.02$ & $4.25\pm0.60$ & $3.82\pm0.65$ & $-44\pm9$ \\
110 & WN5b & 2 & 7.12 & 6.72 & 6.22 & $0.27\pm0.02$ & $1.55\pm0.24$ & $6.48\pm0.35$ & $30\pm2$ \\
111-1 & WN6 & 7 & 12.58 & 11.06 & 10.31 & $0.91\pm0.05$ & $6.45\pm1.27$ & $2.01\pm1.30$ & $-26\pm9$ \\
111-5* & WN9 & 3 & 13.80 & 11.97 & 10.92 & $1.32\pm0.07$ & $15.29\pm2.41$ & $7.66\pm2.43$ & $-14\pm5$ \\
111-6 & WC9 & 12 & 14.24 & 13.35 & 10.75 & $2.05\pm0.13$ & $4.22\pm0.92$ & $3.99\pm0.95$ & $21\pm1$ \\
111-3 & WC8 & 6 & 11.21 & 9.70 & 8.57 & $1.05\pm0.06$ & $2.81\pm0.49$ & $5.30\pm0.55$ & $26\pm1$ \\
113-1 & WN7 & 7 & 9.09 & 8.30 & 7.76 & $0.58\pm0.04$ & $2.88\pm0.56$ & $5.27\pm0.61$ & $-0\pm4$ \\
114-1 & WN7b & 7 & 12.21 & 11.24 & 10.61 & $0.55\pm0.04$ & $12.25\pm1.91$ & $5.52\pm1.93$ & $230\pm33$ \\
114-2 & WC8 & 12 & 14.45 & 12.90 & 11.69 & $1.13\pm0.07$ & $11.43\pm2.07$ & $4.66\pm2.09$ & $138\pm22$ \\
115-1 & WN6 & 7 & 10.32 & 9.52 & 8.96 & $0.56\pm0.03$ & $4.07\pm0.80$ & $4.32\pm0.84$ & $-21\pm8$ \\
115-2 & WN8 & 10 & 11.53 & 10.13 & 9.28 & $1.02\pm0.06$ & $7.36\pm1.22$ & $2.56\pm1.25$ & $-16\pm6$ \\
115-3 & WN7 & 10 & 9.59 & 8.82 & 8.30 & $0.56\pm0.03$ & $3.72\pm0.71$ & $4.63\pm0.76$ & $-8\pm5$ \\
116 & WN8h & 2 & 8.21 & 7.57 & 6.95 & $0.60\pm0.03$ & $3.07\pm0.50$ & $5.20\pm0.56$ & $3\pm3$ \\
116-2 & WN5 & 6 & 14.07 & 12.75 & 11.93 & $0.94\pm0.05$ & $9.32\pm1.46$ & $3.65\pm1.48$ & $162\pm22$ \\
116-3 & WN6ha & 3 & 10.68 & 9.91 & 9.43 & $0.61\pm0.04$ & $14.53\pm2.36$ & $7.66\pm2.37$ & $-29\pm8$ \\
117-1 & WN7 & 10 & 9.09 & 8.29 & 7.63 & $0.68\pm0.04$ & $2.59\pm0.50$ & $5.67\pm0.56$ & $11\pm2$ \\
118-4 & WC8 & 10 & 12.18 & 10.52 & 9.36 & $1.12\pm0.06$ & $3.92\pm0.68$ & $4.63\pm0.73$ & $-7\pm5$ \\
118-7 & WC9 & 12 & 15.91 & 12.89 & 11.01 & $2.25\pm0.12$ & $4.37\pm0.92$ & $4.42\pm0.95$ & $30\pm2$ \\
118-8 & WC9 & 12 & 15.63 & 13.02 & 11.29 & $2.00\pm0.11$ & $5.55\pm1.16$ & $3.84\pm1.19$ & $23\pm1$ \\
118-6 & WN7 & 10 & 13.64 & 11.73 & 10.59 & $1.42\pm0.08$ & $7.21\pm1.40$ & $3.39\pm1.43$ & $-31\pm10$ \\
118-9 & WC9 & 12 & 14.78 & 13.04 & 11.15 & $1.87\pm0.12$ & $5.54\pm1.20$ & $3.89\pm1.22$ & $29\pm2$ \\
119-2 & WC8 & 10 & 11.83 & 10.49 & 9.56 & $0.83\pm0.05$ & $4.90\pm0.85$ & $4.22\pm0.89$ & $21\pm1$ \\
119-3 & WN7 & 12 & 13.27 & 11.50 & 10.50 & $1.25\pm0.07$ & $7.46\pm1.46$ & $3.47\pm1.48$ & $-40\pm12$ \\[1.0ex]
\hline
\end{tabular}\par
\end{minipage}
\end{table*}

\addtocounter{table}{-1}

\clearpage
\begin{table*}
\begin{minipage}{15cm}
\centering
\caption{(Continued)}
\begin{tabular}{ l c c c c c c c c c }
\hline\\[-2.0ex]
WR\# & Sp. Type & ref & J & H & K$_S$ & $\bar{\mathrm{A}}_{K_S}$ & $d$(kpc) & R$_G$(kpc) & z(pc) \\[1.0ex]
\hline\\[-2.0ex]
120 & WN7 & 2 & 8.90 & 8.41 & 8.01 & $0.38\pm0.03$ & $3.54\pm0.68$ & $5.14\pm0.73$ & $37\pm3$ \\
120-1 & WC9 & 7 & 11.05 & 10.17 & 9.44 & $0.65\pm0.04$ & $4.41\pm0.90$ & $4.56\pm0.93$ & $24\pm1$ \\
120-11 & WC8 & 12 & 12.35 & 11.15 & 10.25 & $0.77\pm0.04$ & $6.95\pm1.21$ & $3.60\pm1.23$ & $-25\pm8$ \\
120-7 & WN7 & 10 & 10.22 & 9.21 & 8.51 & $0.78\pm0.04$ & $3.71\pm0.71$ & $5.00\pm0.76$ & $12\pm2$ \\
120-8 & WN9 & 3 & 12.86 & 10.82 & 9.66 & $1.47\pm0.08$ & $7.97\pm1.26$ & $3.91\pm1.29$ & $60\pm6$ \\
120-3 & WN9 & 16 & 11.95 & 10.22 & 9.16 & $1.29\pm0.07$ & $6.89\pm1.08$ & $3.82\pm1.11$ & $60\pm6$ \\
120-4 & WN9 & 16 & 11.85 & 10.26 & 9.27 & $1.20\pm0.07$ & $7.55\pm1.19$ & $3.85\pm1.21$ & $60\pm6$ \\
120-5 & WC8 & 16 & 12.39 & 10.83 & 9.75 & $1.03\pm0.06$ & $4.89\pm0.85$ & $4.37\pm0.89$ & $47\pm5$ \\
120-6 & WN6 & 11 & 13.51 & 12.83 & 12.32 & $0.48\pm0.03$ & $19.85\pm3.92$ & $13.44\pm3.92$ & $177\pm31$ \\
120-9 & WN7ha & 3 & 15.15 & 12.55 & 11.14 & $1.95\pm0.11$ & $19.74\pm3.81$ & $13.27\pm3.82$ & $-16\pm7$ \\
120-10 & WN7 & 10 & 10.00 & 9.16 & 8.53 & $0.67\pm0.04$ & $3.94\pm0.76$ & $4.97\pm0.80$ & $47\pm5$ \\
121-2 & WN7 & 10 & 14.22 & 12.23 & 11.13 & $1.40\pm0.08$ & $9.29\pm1.84$ & $4.53\pm1.85$ & $8\pm2$ \\
121-3 & WN7 & 10 & 13.81 & 11.87 & 10.73 & $1.43\pm0.08$ & $7.66\pm1.49$ & $3.95\pm1.51$ & $4\pm3$ \\
121-1 & WN7h & 11 & 10.94 & 10.07 & 9.47 & $0.66\pm0.04$ & $6.12\pm1.18$ & $4.27\pm1.21$ & $14\pm1$ \\
121-4 & WC7 & 12 & 14.13 & 12.13 & 10.66 & $1.31\pm0.07$ & $6.89\pm1.18$ & $4.07\pm1.21$ & $-62\pm14$ \\
122-2 & WN9 & 3 & 12.49 & 10.68 & 9.61 & $1.33\pm0.07$ & $8.35\pm1.32$ & $4.68\pm1.34$ & $-13\pm5$ \\
122-3 & WN6 & 3 & 12.81 & 11.06 & 9.99 & $1.27\pm0.07$ & $4.72\pm0.94$ & $4.90\pm0.97$ & $25\pm1$ \\
122-4 & WC8 & 10 & 14.70 & 12.67 & 11.25 & $1.45\pm0.08$ & $8.07\pm1.44$ & $5.17\pm1.46$ & $-45\pm12$ \\
122-5 & WC8 & 10 & 14.25 & 12.27 & 10.99 & $1.31\pm0.08$ & $7.61\pm1.35$ & $5.03\pm1.37$ & $-42\pm11$ \\
123 & WN8 & 2 & 9.52 & 9.28 & 8.92 & $0.26\pm0.02$ & $8.81\pm1.45$ & $4.48\pm1.48$ & $-711\pm120$ \\
123-1 & WN6 & 7 & 10.59 & 9.47 & 8.71 & $0.83\pm0.05$ & $3.20\pm0.63$ & $6.04\pm0.68$ & $28\pm2$ \\
123-4 & WC8 & 12 & 16.31 & 14.38 & 13.68 & $0.80\pm0.09$ & $33.24\pm6.41$ & $27.58\pm6.42$ & $-56\pm15$ \\
123-3 & WN8 & 10 & 12.22 & 10.81 & 9.94 & $1.05\pm0.06$ & $9.83\pm1.63$ & $6.79\pm1.65$ & $110\pm15$ \\
124-8 & WN6 & 12 & 8.61 & 7.99 & 7.80 & $0.18\pm0.02$ & $2.84\pm0.56$ & $6.36\pm0.61$ & $46\pm5$ \\
124-3 & WC7 & 12 & 12.95 & 11.82 & 10.77 & $0.71\pm0.04$ & $9.47\pm1.60$ & $7.11\pm1.61$ & $85\pm11$ \\
124-9 & WC6 & 12 & 13.90 & 12.72 & 11.67 & $0.73\pm0.04$ & $13.17\pm4.12$ & $9.88\pm4.13$ & $195\pm55$ \\
124-10 & WC6 & 12 & 14.35 & 12.59 & 11.33 & $1.06\pm0.06$ & $9.66\pm3.03$ & $7.37\pm3.04$ & $79\pm19$ \\
124-6 & WC6 & 3 & 12.67 & 10.91 & 9.68 & $1.04\pm0.06$ & $4.57\pm1.43$ & $6.08\pm1.45$ & $-11\pm9$ \\
124-7 & WC8 & 12 & 12.83 & 11.02 & 9.58 & $1.40\pm0.08$ & $3.82\pm0.67$ & $6.21\pm0.72$ & $-5\pm4$ \\
124-11 & WN6b & 3 & 12.65 & 11.19 & 10.25 & $0.95\pm0.05$ & $7.77\pm1.35$ & $6.92\pm1.37$ & $112\pm16$ \\
125-1 & WC8 & 7 & 10.20 & 9.61 & 9.07 & $0.28\pm0.02$ & $5.06\pm0.87$ & $6.56\pm0.91$ & $6\pm2$ \\
125-4 & WN7 & 10 & 12.71 & 11.07 & 10.12 & $1.18\pm0.07$ & $6.47\pm1.26$ & $6.61\pm1.28$ & $-1\pm4$ \\
128 & WN4(h) & 2 & 9.97 & 9.84 & 9.62 & $0.25\pm0.02$ & $3.57\pm0.72$ & $6.67\pm0.77$ & $-216\pm48$ \\
129 & WN4 & 2 & 11.08 & 10.72 & 10.40 & $0.41\pm0.03$ & $4.75\pm0.96$ & $7.47\pm0.99$ & $222\pm41$ \\
130 & WN8h & 2 & 8.45 & 7.87 & 7.45 & $0.42\pm0.03$ & $4.16\pm0.69$ & $7.52\pm0.73$ & $89\pm11$ \\
132 & WC6 & 2 & 10.18 & 9.76 & 9.05 & $0.22\pm0.02$ & $4.97\pm1.55$ & $7.80\pm1.57$ & $115\pm30$ \\
142a & WC7 & 17 & 9.27 & 8.09 & 7.12 & $0.64\pm0.04$ & $1.83\pm0.31$ & $7.87\pm0.40$ & $91\pm12$ \\
142-1 & WN6 & 18 & 8.77 & 7.86 & 7.19 & $0.69\pm0.04$ & $1.70\pm0.34$ & $7.93\pm0.42$ & $106\pm17$ \\
147 & WN8(h)+OB & 2 & 6.01 & 4.86 & 4.11 & $0.87\pm0.05$ & $0.73\pm0.12$ & $7.90\pm0.28$ & $15\pm1$ \\
148 & WN8h & 2 & 8.76 & 8.53 & 8.32 & $0.15\pm0.02$ & $7.05\pm1.16$ & $10.64\pm1.20$ & $814\pm131$ \\
149 & WN5 & 2 & 10.62 & 10.06 & 9.61 & $0.40\pm0.03$ & $4.11\pm0.64$ & $8.96\pm0.68$ & $66\pm7$ \\
150 & WC5 & 2 & 10.72 & 10.31 & 9.60 & $0.22\pm0.02$ & $5.71\pm1.03$ & $10.31\pm1.07$ & $-228\pm45$ \\[1.0ex] 
\hline\\[-2.0ex]
\end{tabular}
Spectral types: 
(1)\citet{marchenko04}, 
(2)\citet{vdH01}, 
(3) this work, 
(4)\citet{crowther98},
(5)\citet{williams09b},
(6)\citet{shara09},
(7)\citet{hadfield07},
(8)\citet{mauerhan09},
(9)\citet{kurtev07},
(10)\citet{mauerhan11},
(11)\citet{wachter10},
(12)\citet{shara12},
(13)\citet{hopewell05},
(14)\citet{drew04},
(15)\citet{mauerhan10c},
(16)\citet{mauerhan10b},
(17)\citet{pasquali02},
(18)\citet{littlefield12},
\end{minipage}
\end{table*}


\appendix

\clearpage

\section[]{Recently discovered WR stars}
\label{app:new_wr}

In Table \ref{extraWR1} provide identifcations and coordinates of all 322 Wolf-Rayet stars
discovered between publication of the Annex \citep{vdH06} to the VIIth catalogue
of WR stars \citep{vdH01} and March 2014. Table \ref{extraWR1} includes revised distances
for stars for which high quality near-IR photometry and a well-determined
spectral type (${\pm}\,1$) are available. Following a request to the IAU Working
Group for Massive Stars, a panel comprising P.A. Crowther, W-R Hamann, I.D.
Howarth, K.A. van der Hucht, G. Rauw came up with a set of proposals that was
approved by the Working Group in December 2012. Consequently, a revised
nomenclature scheme has been introduced for Galactic WR stars as follows:

\begin{enumerate}
\item All WR identifications up to the VIIth Catalogue \citep{vdH01} and Annex
\citep{vdH06} remain unchanged since many are in widespread usage in the
literature (e.g. WR\,20, WR\,20a). 
\item All subsequent discoveries are switched from alphabetical (WRXXa, b) to
numerical (WRXX-1, -2) identification, sorted by year/month of discovery, in RA
order if multiple discoveries arise from a single source. By way of example,
three WR stars have been discovered since 2006 between the RA's of WR\,20 and
WR\,21. The first, discovered by \citet{mauerhan11}, is assigned WR\,20-1, while
two further discoveries from \citet*{roman-lopes11a} are assigned WR\,20-2 and
-3 (RA ordered). 
\item Multiple WR stars identified within a single source are indicated with
CAPITAL letters (e.g. WR\,43a, b, c replaced with WR\,43A, B, C). 
\item The current Galactic WR census is maintained at 
{\tt http://pacrowther.staff.shef.ac.uk/WRcat/}
\end{enumerate}

\clearpage

\begin{table*}
\begin{minipage}{13cm}
\centering
\caption{WR stars discovered since the \citet{vdH06} updated catalogue
(up to March 2014). Distances are indicated for the stars that feature in this study.}\label{extraWR1}
\centering
\begin{tabular}{l l c c l l l r}
\hline\\[-2.0ex]
WR & Alias & RA & Dec &  & Sp Type &  & Distance\\ 
 &  & (hh:mm:ss) & ($\pm$dd:mm:ss) & Ref & & Ref & kpc\\[1.0ex] 
\hline\\[-2.0ex]
17-1 & SMG09 668\_4  & 10:16:26.226 & -57:28:05.70  & 1 & WN5b & 1 & 5.4$\pm$0.9\\
20-2 & WR\,20aa, SS215         & 10:23:23.49  & -58:00:20.80  & 19 & O2\,If/WN6 & 19 & ---\\ 
20-1 & MDM11 1         & 10:23:28.80  & -57:46:29.4   & 5 & WN7--8 & 5 & ---\\
20-3 & WR\,20c                 & 10:25:02.60  & -57:21:47.30  & 19 & O2\,If/WN6 & 19 & ---\\
42-1 & WR\,42e, SB04 \#954    & 11:14:45.50 & -61:15:00.1 & 21 & O2\,If/WN6 & 21 & ---\\
43-2 & MTT 58        & 11:15:07.60 & -61:16:54.8 & 22 & O2\,If/WN6 & 22 & ---\\
43-1 & SMG09 740\_21 & 11:16:03.536 & -61:26:58.34  & 1 & WN4b & 1 & 6.9$\pm$1.2\\
44-1 & SMG09 740\_16 & 11:19:42.96 & -61:27:12.40 & 1 & WCE & 1 & -- \\
45-1 & HDM07 1       & 11:42:37.66 & -62:41:19.30 & 2 & WN9--10h & 2 & -- \\
45-2 & SMG09 768\_6  & 11:46:06.66 & -62:47:12.70 & 1 & WN5 & 1 & --\\
45-3 & SMG09 772\_17 & 11:50:04.242 & -62:52:15.44 & 1 & WN5b & 0 & 11.9$\pm$1.9\\
45-4 & SMG09 776\_3 & 11:55:52.116 & -62:45:02.28  & 1 & WN6 & 1 & 6.7$\pm$1.3\\
45-5 & VVV CL009-6  & 11:56:03.78 & -63:18:54.44 & 23 & Of/WN7 & 23 & -- \\
46-1 & HDM07 2       & 12:06:56.480 & -62:38:30.45  & 2 & WN6o & 2 & 7.2$\pm$1.4\\
46-7 & J12100795-6244194 & 12:10:07.95 & -62:44:19.40 & 3 & WC5--7 & 3 & -- \\
46-8 & J12110256-6257476 & 12:11:02.567 & -62:57:47.61 & 3 & WN6 & 3 & 5.1$\pm$1.0\\
46-16 & RMM11 \#5        & 12:11:54.1 & -63:17:04.0 & 24 & WN9 & 24 & -- \\
46-9 & J12121681-6246145 & 12:12:16.814 & -62:46:14.54 & 3 & WN5 & 0 & 3.6$\pm$0.6\\
46-17& VVV CL011-2       & 12:12:41.13 & -62:42:30.71 & 23 & WN9/OIf+ & 23 & --\\
46-10 & SMG09 791\_12c   & 12:13:28.29 & -62:41:42.90 & 1 & WCE & 1 & -- \\
46-2 & HDM07 3 & 12:13:38.790 & -63:08:58.06 & 2 & WN7h & 2 & 4.9$\pm$0.9\\
46-3 & KBG07 4 & 12:14:31.54 & -62:58:54.30 & 4 & WN7--8 & 4 & -- \\
46-4 & KBG07 3 & 12:14:31.76 & 062:58:51.90 & 4 & Ofp/WN & 4 & -- \\
46-5 & KBG07 2 & 12:14:33.090 & -62:58:51.00 & 4 & WN6 & 4 & 8.6$\pm$2.0\\
46-6 & KBG07 1 & 12:14:33.910 & -62:58:48.70 & 4 & WN7 & 4 & 9.1$\pm$2.2\\
46-15 & MDM11 2 & 12:15:12.492 & -62:46:43.89 & 5 & WN8 & 5 & 7.6$\pm$1.3\\
46-11 & SMG09 808\_14  & 12:28:41.91 & -63:25:46.10 & 1 & WCE & 1 & -- \\ 
46-12 & SMG09 808\_23  & 12:28:50.995 & -63:17:00.23 & 1 & WN4b & 0 & 16.2$\pm$2.8\\
46-13 & SMG09 807\_16 & 12:30:03.86 & -62:50:17.10 & 1 & WC7 & 1 & -- \\
46-14 & SMG09 816\_10 & 12:38:18.781 & -63:24:19.74 & 1 & WN5b & 1 & 11.0$\pm$1.7\\
47-1 & HDM07 4 & 12:46:16.140 & -62:57:23.47 & 2 & WN6o & 2 & 8.0$\pm$1.6\\
47-2 & SMG09 832\_25 & 12:55:44.26 & -63:35:50.0 & 1 & WC5--6 & 1 & -- \\  
47-3 & SMG09 856\_13c & 13:03:11.08 & -63:42:16.20 & 1 & WC5--6 & 1 & --\\
47-4 & SMG09 839\_12 & 13:04:50.08 & -63:04:40.20 & 1 & WC5--6 & 1 &  -- \\
48-1 & HDM07 5 & 13:10:12.073 & -62:39:06.57 & 2 & WC7 & 2 & 5.6$\pm$0.9\\
48-6 & MDM11 3 & 13:12:09.059 & -62:43:26.71 & 5 & WN9 & 5, 0 & 4.2$\pm$0.6\\
48-3 & SMG09 845\_34 & 13:12:21.307 & -62:40:12.58 & 1 & WC8 & 1& 4.2$\pm$0.6\\
48-10 & DCT12 D1-2 & 13:12:24.980 & -62:42:00.20 & 6 & WN9h & 0 & 4.2$\pm$0.6\\
48-7 & MDM11 5 & 13:12:25.460 & -62:44:41.70 & 5 & WN8 & 0 & 4.2$\pm$0.6\\
48-4 & SMG09 845\_35, MDM11 6 &  13:12:27.670 & -62:44:22.00 & 1 & WC6 & 0 & 4.2$\pm$0.6\\
48-8 & MDM11 7, DCT12 D1-5 & 13:12:28.50  & -62:41:50.9 & 5, 6 & WNLh & 6 & -- \\
48-9 & MDM11 8, DCT12 D1-1 & 13:12:28.550 & -62:41:43.80 & 5, 6  & WN9h & 5 & 4.2$\pm$0.6 \\
48-5 & SMG09 847\_8 & 13:12:45.354 & -63:05:52.00 & 1 & WN6b  & 0 & 7.1$\pm$1.2 \\
48-2 & DCT12 D2-3 & 13:12:57.700 & -62:40:59.90 & 3 & WC7--8 & 6 & 4.2$\pm$0.6 \\
49-1 & MDM11 9 & 13:14:57.048 & -62:23:53.34 & 5 & WN8 & 5 & 11.3$\pm$1.9 \\
52-1 & SMG09 853\_9 & 13:22:16.082 & -62:30:57.45 & 1 & WN6 & 1 & 9.0$\pm$1.8 \\
52-2 & SMG09 858\_26 & 13:28:15.876 & -62:06:23.57 & 1 & WN6 & 1 & 8.1$\pm$1.6\\
57-1 & MDM11 10 &  13:44:06.952 & -62:45:02.20 & 5 & WN7 & 5 & 9.7$\pm$1.9\\
59-1 & SMG09 883\_18 & 13:52:02.36 & -62:26:46.0 & 1 & WCE & 1 & -- \\
59-2 & SMC09 885\_11 & 13:54:13.45 & -61:50:01.80 & 1 & WC5--6 & 1 & -- \\
60-3 & MDM11 11 & 14:01:15.49 & -62:38:19.9 & 5 & WC7 & 5 & -- \\
60-4 & MDM11 12 & 14:04:36.672 & -61:29:16.52 & 5 & WC8 & 5 & 5.7$\pm$1.0\\
60-5 & WR$\,60$a &  14:06:03.619 & -60:27:29.58 &  7 & WC7    & 0 & 6.0$\pm$1.0\\
60-6 & VVV CL036-9 & 14:09:04.30 & -61:15:53.78 &  23 & WN6 & 23 & --\\
60-1 & SMG09 897\_5 & 14:10:10.015 & -61:15:25.59 & 1 & WC8 & 1 & 4.6$\pm$0.8\\
60-2 & SMG09 903\_15c & 14:12:36.54 & -61:45:32.70 & 1 & WC8 & 1 & --\\
61-2 & SMG09 907\_18 &  14:16:27.372 & -61:17:56.25 & 1 & WN5b & 1  & 4.9$\pm$0.8\\
61-3 & MDM11 13 & 14:20:30.746 & -60:48:22.12 & 5 & WC9 & 5  & 4.8$\pm$1.0\\
61-1 & J14212314-6018041 & 14:21:23.148 & -60:18:04.12 & 3 & WN6 & 3 & 4.7$\pm$0.9\\
62-1 & AX J144701-5919 & 14:46:53.58 & -59:19:38.3 & 25 & WN7-8h & 25 & -- \\[1.0ex]
\hline
\end{tabular}
\end{minipage}
\end{table*}

\addtocounter{table}{-1}

\clearpage
\begin{table*}
\begin{minipage}{13cm}
\centering
\caption{(Continued)}
\begin{tabular}{l l c c l l l r} 
\hline\\[-2.0ex]
WR & Alias &   RA & Dec &  & Sp Type & & Distance\\ 
 &  & (hh:mm:ss) & ($\pm$dd:mm:ss) & Ref & &Ref & kpc\\[1.0ex]
\hline\\[-2.0ex]
64-1 & SMG09 956\_25 &  15:01:30.119 & -59:16:12.06 & 1 & WN4b & 1 & 6.8$\pm$1.2\\
67-3 & G321.0331-0.4274 & 15:15:39.44 & -58:08:16.0 & 26 & WN10 & 26 & -- \\
67-1 & WR67a & 15:16:36.96 & -58:09:58.7 & 8 & WN6h & 8 & -- \\
67-2 & WR67b & 15:17:46.301 & -57:56:59.29 & 8 & WC7 & 0 & 3.7$\pm$0.6\\ 
68-1 & SMG09 979\_11 & 15:20:35.916 & -57:27:11.95 & 1 & WN4b & 1 & 8.9$\pm$1.6\\
70-1 & WMD10 9,     &  15:35:26.528 & -56:04:12.33 & 3 & WN7 & 3 & 12.4$\pm$2.4\\
     & J15352652-5604123 &    &   &   &   &   &  \\
70-3 & SMG09 1011\_24 & 15:43:04.681 & -55:11:12.35 & 1 & WC7 & 0 & 4.7$\pm$0.8\\
70-4 & WMD10 10 & 15:45:59.143 & -53:32:32.50 & 9 & WN9h & 9 & 12.5$\pm$2.0\\
70-5 & WMD10 11b & 15:48:42.105 & -55:07:54.21 & 9 & WC9 & 9 & 2.2$\pm$0.4\\
70-10 & SFZ12 1023-63L & 15:52:09.48 & -54:17:14.5 & 10 & WC7: & 10 & -- \\ 
70-6 & MDM11 14 & 15:53:31.854 & -53:45:44.40 & 5 & WN6b & 0 & 7.3$\pm$1.3\\
70-7 & MDM11 15 & 15:58:49.712 & -52:51:32.46 & 5 & WN6b & 5 & 8.0$\pm$1.4\\
70-8 & MDM11 16 & 15:58:54.933 & -52:02:45.52 & 5 & WN7 & 5 & 5.8$\pm$1.2\\
70-2 & J15595671-5159299 & 15:59:56.715 & -51:59:29.93 & 3 & WN5b    & 0 &  3.6$\pm$0.6\\
70-9 & MDM11 17 & 16:00:23.265 & -52:51:42.32 & 5 & WC8 & 5 & 7.2$\pm$1.3\\
70-11 & SFZ12 1042-25L & 16:00:25.264 & -52:03:29.62 & 10 & WN7 &  0 & 6.5$\pm$1.2\\
70-12 & SFZ12 1038-22L & 16:00:26.41 & -52:11:10.1 & 10 & WC7: & 10 & --\\
72-3 & SFZ12 1054-43L, & 16:10:06.251 & -50:47:58.56 & 10 & WC9 & 10 & 6.7$\pm$1.4\\
     & MDM11 18       &              &             &     &     &     &            \\
72-4 & SFZ12 1051-67L & 16:00:06.67 & -51:47:24.5 & 10 & WC7:&10 & --\\
72-1  & HDM07 6 & 16:11:39.271 & -52:05:45.81 & 2 & WC9 & 2 & 2.7$\pm$0.6\\
72-2 & SMG09 1053\_27 & 16:11:43.70 & -51:10:16.6 & 1 & WC8 & 1 & -- \\
73-1 & SMG09 1059\_34 & 16:14:37.237 & -51:26:26.33 & 1 & WC7 & 0 & 10.9$\pm$1.9\\
74-1 & MDM11 19 & 16:20:51.436 & -50:04:03.33 & 5 & WN9ha & 5 & 8.1$\pm$1.2\\
74-3 & SFZ12 1077-55L & 16:24:22.70 & -49:00:42.3 & 10 & WC6: & 10 & --\\
74-2 & MDM11 20 & 16:24:23.294 & -49:21:29.59 & 5 & WN7 & 5 & 8.3$\pm$1.7\\
75-1 & SMG09 1081\_21 & 16:24:58.868 & -48:56:52.45 & 1 & WC8 & 1 & 7.9$\pm$1.4 \\
75-14 & SFZ12 1085-72L & 16:27:42.390 & -48:30:34.20 & 10 & WC9 &  0 & 7.2$\pm$1.5\\
75-15 & SFZ12 1085-69L & 16:28:40.260 & -48:18:12.95 & 10 & WC8 & 10 & 7.7$\pm$1.6\\
75-6 & MDM11 21 & 16:28:53.428 & -48:33:39.41 & 5 & WN5b & 5 & 13.3$\pm$2.1\\
75-16 & SFZ12 1085-83L & 16:29:35.838 & -48:19:34.20 & 10 &  WC8 & 10 & 8.9$\pm$1.9\\
75-25 & VVV CL073-2 & 16:30:23.73 & -48:13:05.48 & 23 & WN9/O4--6If+ & 23 & -- \\
75-26 & VVV CL073-4 & 16:30:23.98 & -48:13:05.48 & 23 & WN7 & 23 & --\\
75-2 & SMG09 1093\_34 & 16:31:29.234 & -47:56:16.40 & 1 & WC8 & 1 & 7.3$\pm$1.3\\
75-30 & KSF14 1089-1117 & 16:31:37.79 & -48:14:55.3 & 27 & WN9 & 27 & -- \\
75-3 & SMG09 1093\_33 & 16:31:49.062 & -47:56:04.47 & 1 &  WC8 & 1 & 8.8$\pm$1.6\\
75-27 & VVV CL074-2 & 16:32;05.27 & -47:49:14.25 & 23 & WC8 & 23 & --\\
75-28 & VVV CL074-3 & 16:32:05.46 & -47:49:28.10 & 23 & WN8 & 23 & --\\
75-29 & VVV CL1074-9 & 16:32:05.93 & -47:49:30.92 & 23 & WN7/O4--6If+ & 23 & --\\
75-4  & SMG09 1093\_53, & 16:32:12.986 & -47:50:35.88 & 1 &  WN5b & 1 & 9.1$\pm$1.5\\
      & WMD10 16       &               &              &   &       &   &           \\
75-17 & SFZ12 1093-138L &  16:32:15.223 & -47:56:12.71 & 10 & WC8: & 10 & 16.5$\pm$3.0\\
75-7 & MDM11 22 & 16:32:22.051 & -47:47:42.60 & 5 & WC9 & 5 & 8.2$\pm$1.7\\
75-31 & KSF14 1093-1765 & 16:32:25.70 & -47:50:46.1 & 27 & WN6 & 27 & --\\
75-18 & SFZ12 1093-140LB & 16:32:48.55 & -47:45:06.2 & 25 & WN9 & 10 & --\\
75-19 & SFZ12 1093-140L & 16:32:49.78 & -47:44:31.4 & 10 & WC7: & 10 & --\\
75-8 & MDM11 23 & 16:33:11.207 & -48:19:41.26 & 5 & WN4b   & 0 & 6.2$\pm$1.1\\
75-20 & SFZ12 1091-46L & 16:33:14.06 & -48:17:37.2 & 10 & WC8 & 10 & --\\ 
75-9 & MDM11 24, & 16:33:45.45 & -47:51:29.1 & 5 & WC8-9d? & 5 & --\\
     & SFZ12 1093-59L &        &             &   &         &   &    \\
75-21 & SFZ12 1095-189L & 16:33:48.13 & -47:52:52.8 & 10 & WC7: & 10 & --\\
75-10 & MDM11 25, & 16:34:57.467 & -47:04:12.95 & 5 & WN5-6b & 5 & 7.0$\pm$1.2 \\
      & SFZ12 1097-156L &        &              &    &       &   &             \\
75-11 & MDM11 26 & 16:35:05.55 & -47:17:13.5 & 5 & WC9d? & 5 & --\\
75-5 & SMG09 1096\_22,  & 16:35:23.317 & -48:09:18.09 & 1 & WC8 & 1 & 8.9$\pm$1.6\\
     & SFZ12 1095-98L   &              &               &   &    &   &            \\
75-12 & MDM11 27 & 16:35:38.882 & -47:09:13.09 & 5 & WN6 & 5 & 8.0$\pm$1.6\\
75-22 & SFZ12 1097-71L & 16:35:44.347 & -47:19:42.28 & 10 & WC9 & 10 & 7.8$\pm$2.0\\
75-13 & MDM11 28, & 16:35:51.169 & -47:19:51.54 & 5 & WC8 & 5 & 5.8$\pm$1.0\\
      & SFZ12 1097-34L &         &              &   &     &   &         \\[1.0ex]
\hline
\end{tabular}
\end{minipage}
\end{table*}

\addtocounter{table}{-1}

\clearpage
\begin{table*}
\begin{minipage}{13cm}
\centering
\caption{(Continued)}
\begin{tabular}{l l c c l l l r} 
\hline\\[-2.0ex]
WR & Alias & RA & Dec &  & Sp Type &  & Distance\\ 
   &  & (hh:mm:ss) & ($\pm$dd:mm:ss) & Ref & & Ref & kpc\\[1.0ex] 
\hline\\[-2.0ex]
75-23 & SFZ12 1106-31L & 16:37:23.991 & -46:26:28.73 & 10 & WC9 & 10 & 3.7$\pm$0.7 \\
75-24 & SFZ12 1105-76L & 16:38:20.18 & -46:23:43.8 & 10 & WC8 & 10 & --\\
76-10 & SFZ12 1109-74L & 16:40:17.12 & -46:20:09.7 & 10 & WC7: & 10 &--\\
76-2 & Mercer 81-5 & 16:40:28.35 & -46:23:25.6 & 28 & WN7--8 & 29 & --\\
76-3 & Mercer 81-8 & 16:40:28.94 & -46:23:27.1 & 28 & WN8: & 29, 0 & --\\
76-4 & Mercer 81-6 & 16:40:29.32 & -46:23:11.6 & 28 & WN9ha & 29, 0 & --\\
76-5 & Mercer 81-9 & 16:40:29.32 & -46:23:38.2 & 28 & WN9ha & 29, 0 & --\\
76-6 & Mercer 81-7 & 16:40:29.60 & -46:23:25.6 & 28 & WN7--8 & 29 & --\\
76-7 & Mercer 81-2 & 16:40:29.65 & -46:23:29.1 & 28 & WN7--8 & 29 & --\\
76-8 & Mercer 81-4 & 16:40:29.65 & -46:23:28.7 & 28 & (WN7--8) & 0 & --\\
76-9 & Mercer 81-3 & 16:40:30.08 & -46:23:11.4 & 28 & WN7--8 & 28 & --\\ 
76-1 & MDM11 29 & 16:40:50.787 & -45:51:23.11 & 5 & WN9 & 5 & 8.7$\pm$1.4\\
77-5 & SFZ12 1115-197L & 16:43:40.369 & -45:57:57.60 & 10 & WN6 & 10 & 4.4$\pm$0.9\\
77-1 & J16441069-4524246 & 16:44:10.695 & -45:24:24.70 & 3 & WN7b& 0 & 4.0$\pm$0.6\\
77-2 & J16465342-4535590 & 16:46:53.428 & -45:35:59.02 & 3 & WN7 & 3 & 3.8$\pm$0.7\\
77-3 & MDM11 30 & 16:47:46.036 & -45:59:04.93 & 5 & WN6 & 5 & 3.8$\pm$0.8\\
77-4 & MDM11 31 & 16:48:27.584 & -46:09:27.20 & 5 & WN6 & 0 & 5.4$\pm$1.1\\
77-6 & SFZ12 1138-133L & 16:51:19.330 & -43:26:55.27 & 10 & WN6b&  0 & 10.3$\pm$1.8\\
77-7 & SFZ12 1133-59L &  16:51:29.702 & -43:53:35.52 & 10 & WC9 & 10 & 9.5$\pm$2.2\\
78-1 & KSF14 1139-49EA & 16:54:08.46 & -43:49:25.3 & 27 & WC6: & 27 & --\\
82-2 & KSF14 1178-66B & 17:07:23.95 & -39:19:54.4 & 27 & WC9 & 27 & --\\
82-1 & SFZ12 1168-91L & 17:09:32.64 & -41:29:47.3 & 10 & WC7: & 10 & --\\
83-1 & SFZ12 1179-129L & 17:11:00.84 & -39:49:31.2 & 10 & WC6: & 10 & -- \\
84-2 & SFZ12 1181-82L & 17:11:28.502 & -39:13:16.8 & 10 & WC8 & 10 & 7.9 $\pm$1.4\\
84-6 & DBS03 179 \#15 & 17:11:31.80 & -39:10:46.8 & 30 & WN8--9 & 30 & --\\
84-7 & DBS03 179 \#4  & 17:11:31.88 & -39:10:46.9 & 30 & Ofpe/WN9 & 30 & --\\
84-1 & MDM11 32, & 17:11:33.037 & -39:10:40.05 & 5 & WN9    &    0 & 6.3$\pm$1.0\\
     & DBS03 179 \#20 &        &               &   &        &      &            \\
84-3 & SFZ12 1181-81L & 17:11:36.12 & -39:11:07.9 & 10 & WC8 & 10 & --\\
84-4 & SFZ12 1181-211L & 17:11:46.133 & -39:20:27.78 & 10 & WN7ha&  0 & 16.5$\pm$3.0\\
84-11 & KSF14 1176-B49 & 17:12:34.87 & -40:37:13.8 & 27 & WN9h & 27 & --\\
84-5 & SFZ12 1189-110L & 17:14:09.551 & -38:11:20.90 & 10 & WC9 & 10 & 9.0$\pm$1.9\\
84-8 & VVV CL099-4 & 17:14:24.71 & -38:09:49.35 & 23 & WN6+O & 23 & --\\
84-9 & VVV CL099-5 & 17:14:25.42 & -38:09:50.40 & 23 & WN6 & 23 & --\\
84-10 & VVV CL099-7 & 17:14:25.66 & -38:09:53.72 & 23 & WC8 & 23 & --\\
85-1 & KSF14 1198-6EC8 & 17:15:55.9 & -37:12:12.0 & 27 & WC6: & 27 & --\\
91-1 & SMG09 1222\_15 & 17:22:40.741 & -35:04:52.95 & 1 & WC7 & 0 & 4.7$\pm$0.8\\
94-1 & SFZ12 1245-23L & 17:33:33.220 & -32:36:16.40 & 10 & WC9 & 10 & 3.8$\pm$0.8\\
98-2 & KSF14 1256-1483A & 17:40:59.35 & -32:11:22.3 & 27 & WN9 & 27 & --\\ 
98-1 & SFZ12 1269-166L & 17:41:13.512 & -30:03:40.98 & 10 & WC8 & 10 & 6.7$\pm$1.2\\
100-3 & SFZ12 1275-184L & 17:44:06.89 & -30:01:13.2 & 10 & WN8 & 10 & --\\
100-2 & MCD10 2 & 17:45:06.91 & -29:12:02.1 & 12 & WC9?d & 12 & --\\ 
100-1 & J174508.9-291218 & 17:45:08.900 & -29:12:18.00 & 11 & WN7b &  0 & 8.7$\pm$1.4\\
101-7 & MCD10 4 & 17:45:09.74 & -29:14:14.6 & 12 & WC9?d & 12 & --\\
101-2 & J174516.1-284909 & 17:45:16.10 & -28:49:09.0 & 11 & WN9 & 11 & --\\
101-3 & J174516.7-285824 & 17:45:16.726 & -28:58:25.10 & 11 & WN8o    &  0 & 7.5$\pm$1.3\\
101-4 & J174519.1-290321 & 17:45:19.16 & -29:03:21.78 & 11 & WC9d & 11 & --\\
101-8 & MCD10 3 & 17:45:21.870 & -29:11:59.43 & 12 & WC9 & 12 & 9.2$\pm$1.9\\
101-5 & J174522.6-285844 & 17:45:22.600 & -28:58:44.00 & 11 & WN6b    &  0 & 5.5$\pm$1.0\\
101-6 & J174528.6-285605 & 17:45:28.60 & -28:56:05.0 & 11 & WN8-9h & 11 & --\\
101-9 & MCD10 8 & 17:45:32.518 & -29:04:57.93 & 12 & WC9 & 12 & 4.5$\pm$1.0\\
101-1 & Edd-1, J174536.1-285638 & 17:45:36.120 & -28:56:38.70 & 31 & WN9h    &  0 & 7.4$\pm$1.1 \\
102-12 & MCD10 11 & 17:45:48.617 & -28:49:42.64 & 12 & WN8--9h & 12 & 7.1$\pm$1.2\\
102-4 & J174550.2-284911 & 17:45:50.20 & -28:49:11.0 & 11 & WN8-9h & 11 & --\\
102-13 & MCD10 19 & 17:45:50.29 & -28:57:26.2 & 12 & WC9 & 12 & -- \\
102-5 & J174550.6-285617 & 17:45:50.60 & -28:56:17.00 & 11 & WN7 & 11 & --\\
102-14 & MCD10 12 & 17:45:53.18 & -28:49:36.0 & 12 & WN8-9h & 12 & --\\
102-6 & J174555.3-285126 & 17:45:55.300 & -28:51:26.00 & 11 & WN5--6b & 11 &  \\
102-15 & MCD10 14 & 17:46:02.48 & -28:54:12.8 & 12 & WC9 & 12 & --\\
102-2 & LHO09 76 & 17:46:14.15 & -28:49:35.4 & 14 & WC9d & 14 & \\[1.0ex]
\hline
\end{tabular}
\end{minipage}
\end{table*}

\addtocounter{table}{-1}

\clearpage
\begin{table*}
\begin{minipage}{13cm}
\centering
\caption{(Continued)}
\begin{tabular}{l l c c l l l r} 
\hline\\[-2.0ex]
WR & Alias & RA & Dec &  & Sp Type &  & Distance\\ 
   &  & (hh:mm:ss) & ($\pm$dd:mm:ss) & Ref & & Ref & kpc\\[1.0ex] 
\hline\\[-2.0ex]
102-3 & LHO09 79, qF250 & 17:46:15.39 & -28:49:34.6 & 14& WC9d & 14 & \\
102-16 & MCD10 16 & 17:46:17.57 & -28:53:03.7 & 12 & WN8-9h & 12 & --\\
102-7 & J174617.7-285007 & 17:46:17.70 & -28:50:07.0 & 11 & WC9d & 11 & --\\
102-17 & MCD10 17 & 17:46:23.838 & -28:48:11.26 & 12 & WN6b &  0 & 7.7$\pm$1.3\\
102-1 & J174645.3-281546 & 17:46:45.34 & -28:15:46.10 & 32 & (WC) & 32 & --\\
102-8 & J174656.3-283232 & 17:46:56.295 & -28:32:32.50 & 11 & WN9h    &  0 & 7.3$\pm$1.2\\
102-18 & MCD10 15 & 17:46:09.73 & -28:55:31.9 & 12 & WN8-9h & 12 & --\\
102-9 & J174711.4-283006 & 17:47:11.471 & -28:30:06.99 & 11 & WN9h    &  0 & 6.4$\pm$1.0\\
102-10 & J174712.2-283121 & 17:47:12.250 & -28:31:21.56 & 11 & WN8o    &  0 & 5.5$\pm$0.9\\
102-11 & J174713.0-282709 & 17:47:13.00 & -28:27:09.00 & 11 & WN7--8h & 11 & --\\
102-19 & SFZ12 1322-220L & 17:55:20.211 & -24:07:38.41 & 10 & WN5 & 10 & 5.1$\pm$0.8\\
102-23 & KSF14 1319-3BC0 & 17:57:16.87 & -25:23:13.8 & 27 & WC7: 27 & --\\
102-22 & FSZ14 WR1327-14AF & 17:59:02.27 & -24:17:00.1 & 33 & WC7 & 33 & --\\
102-20 & SFZ12 1327-25L & 17:59:02.897 & -24:20:50.61 & 10 & WC9 & 10 & 5.2$\pm$1.3\\
102-24 & KSF14 1338-2B3 & 17:59:07.99 & -22:36:43.0 & 27 & WN9 & 27 & --\\
102-21 & SFZ12 1342-208L & 17:59:48.22 & -22:14:52.4 & 10 & WN6 & 10 & --\\
104-1 & KSF14 1343-69E & 18:02:22.35 & -22:38:00.3 & 27 & WN8--9 & 27 & --\\
105-1 & FSZ14 WR1343-193E & 18:02:46.23 & -22:36:39.7 & 33 & WN6 & 33 & --\\ 
105-2 & KSF14 1343-284 & 18:03:28.37 & -22:22:58.9 & 27 & WN8--9 & 27 & --\\
108-2 & KSF14 1353-160A & 18:05:35.60 & -21:04:23.3 & 27 & WC8--9 & 27 & --\\
108-3 & KSF14 1366-438 & 18:05:55.27 & -19:29:44.1 & 27 & WN7--8 & 27 & --\\
108-1 & FSZ14 WR1361-1583 & 18:07:05.16 & -20:15:16.1 & 33 & WN9 & 33 & --\\ 
111-8 & KSF14 1367-638 & 18:09:06.22 & -19:54:27.2 & 27 & WN9 & 27 & --\\
111-1 & HDM07 7 &  18:09:45.057 & -20:17:10.35 & 2 & WN6o & 2 & 6.4$\pm$1.3\\
111-9 & KSF14 1381-19L & 18:12:02.0 & -18:06:55.0 & 27 & WC9 & 27 & --\\
111-5 & MDM11 33 & 18:12:41.102 & -18:26:30.46 & 5 & WN9    & 0 & 15.3$\pm$2.4\\
111-6 & SFZ12 1381-20L & 18:12:57.580 & -18:01:24.36 & 10 & WC9 & 10 & 4.2$\pm$0.9\\
111-2 & HDM07 8, MDF11 \#4 & 18:13:14.200 & -17:53:43.50 & 2 & WN7b & 2 & 3.6$\pm$0.7\\
111-4 & MDM11 34, MDF11 \#7 & 18:13:22.480 & -17:53:50.30 & 5 & WN7   & 0 & 3.6$\pm$0.7\\  
111-3 & SMG09 1385\_24 & 18:13:42.476 & -17:28:12.21 & 1 & WC8 & 1 & 2.8$\pm$0.5\\
111-10 & KSF14 1389-4AB6 & 18:14:14.09 & -17:21:02.6 & 27 & WC7 & 27 & --\\
111-11 & KSF14 1389-1F5D & 18:14:17.37 & -17:21:54.2 & 27 & WN8 & 27 & --\\
111-7 & SFZ12 1395-86L & 18:16:02.36 & -16:53:59.4 & 10 & WC8 & 10 & --\\
113-1 & HDM07 9 & 18:19:22.194 & -16:03:12.38 & 2 & WN7o & 2 & 2.9$\pm$0.6\\
113-3 & KSF14 1430-AB0 & 18:21:02.92 & -12:27:45.8 & 27 & WN4-7 & 27 & --\\
113-2 & SMG09 1425\_47 & 18:23:03.42 & -13:00:00.4 & 1 & WC5-6 & 1 & --\\
114-2 & SFZ12 1434-43L & 18:23:32.333 & -12:03:58.57 & 10 & WC8 & 10 & 11.4$\pm$2.1 \\
114-1 & HDM07 10, KSF14 1446-B1D & 18:25:00.241 & -10:33:23.63 & 2 & WN7b   & 2 & 12.2$\pm$1.9\\
115-1 & HDM07 11 & 18:25:53.094 & -13:28:32.41 & 2 & WN6o & 2 & 4.1$\pm$0.8\\
115-2 & MDM11 35, SFZ12 1431-34L & 18:25:53.617 & -12:50:03.19 & 5 & WN8 & 5 & 7.4$\pm$1.2\\
115-3 & MDM11 36 &  18:26:06.116 & -13:04:10.47 & 5 & WN7 & 5 & 3.7$\pm$0.7\\
116-1 & J18281180-1025424 & 18:28:11.80 & -10:25:42.4 & 3 & WC9+OBI & 3 & --\\
116-4 & KSF14 1443-760 & 18:28:33.39 & -11:46:44.2 & 27 & WN9h & 27 & --\\
116-2 & SMG09 1462\_54 & 18:29:33.847 & -08:39:02.10 & 1 & WN5o& 1 & 9.3$\pm$1.5 \\
116-3 & MDM11 37 & 18:30:53.206 & -10:19:37.09 & 5 & WN6ha&0 & 14.5$\pm$2.4 \\
117-2 & KSF14 1457-673 & 18:31:06.65 & -09:48:01.4 & 27 & WC9 & 27 & --\\
117-1 & MWC10-XGPS14, &  18:31:16.531 & -10:09:25.01 & 20 & WN7    &  5 & 2.6$\pm$0.5\\
      & MDM11 38      &               &              &    &        &    &           \\
118-4 & MDM11 39, SFZ12 1463-7L & 18:33:47.637 & -09:23:07.71 & 5 & WC8 & 5,10 & 3.9$\pm$0.7\\
118-7 & SFZ12 1477-55L & 18:35:47.658 & -07:17:50.07 & 10 & WC9 & 10 & 4.4$\pm$0.9 \\
118-1 & MDI09 Quartet 5 & 18:36:16.33 & -07:05:17.0 & 15 & WC9d& 15,0&\\
118-2 & MDI09 Quartet 2 & 18:36:16.69 & -07:04:59.50 & 15 & WN9 & 15 & --\\
118-3 & MDI09 Quartet 1 & 18:36:17.29 & -07:05:07.30 & 15 & WN9 & 15 & --\\
118-10 & KSF14 1485-6C4 & 18:36:55.53 & -06:31:02.1 & 27 & WN6 & 27 & --\\
118-5 & MDM11 40 & 18:37:51.49 & -06:08:41.7 & 5 & WC9d & 5 & --\\
118-11 & KSF14 1485-844 & 18:37:51.82 & -06:31:19.1 & 27 & WN8 & 27 & --\\
118-8 & SFZ12 1487-80L & 18:38:00.479 & -06:26:46.23 & 10 & WC9 & 10 & 5.5$\pm$1.2\\
118-6 & MDM11 41, SFZ12 1483-212L & 18:38:27.169 & -07:10:44.79 & 5  & WN7 & 5, 10 & 7.2$\pm$1.4\\
118-9 & SFZ12 1489-36L & 18:38:38.931 & -06:00:16.01 & 10 & WC9 & 10 & 5.5$\pm$1.2 \\
119-2 & MDM11 42, SFZ12 1493-9L & 18:39:34.581 & -05:44:23.08 & 5 & WC8 & 5, 10 & 4.9$\pm$0.9\\[1.0ex]
\hline
\end{tabular}
\end{minipage}
\end{table*}

\addtocounter{table}{-1}

\clearpage
\begin{table*}
\begin{minipage}{13cm}
\centering
\caption{(Continued)}
\begin{tabular}{l l c c l l l r} 
\hline\\[-2.0ex]
WR & Alias & RA & Dec &  & Sp Type &  & Distance\\ 
   &  & (hh:mm:ss) & ($\pm$dd:mm:ss) & Ref & & Ref & kpc\\[1.0ex]
\hline\\[-2.0ex]
119-4 & KSF14 1495-1D8A & 18:39:40.60 & -05:35:17.6 & 27 & WC8--9 & 27 & --\\
119-5 & KSF14 1495-705 & 18:39:41.19 & -05:57:36.3 & 27 & WN8 & 27 & --\\
119-3 & SFZ12 1487-212L & 18:39:42.533 & -06:41:46.37 & 10 & WN7 & 10 & 7.5$\pm$1.5 \\
119-1 & HDM07 12 & 18:40:08.66 & -03:29:31.10 & 2 & WN7o & 2 & --\\ 
120-16 & KSF14 1514-AA0 & 18:41:06.79 & -02:56:01.0 & 27 & WC8 & 27 & --\\ 
120-1 & HDM07 13 & 18:41:10.70 & -04:51:27.0 & 2 & WC9 & 2 & --\\ 
120-11 & SFZ12 1495-32L & 18:41:23.348 & -05:40:58.39 & 10 & WC8 & 10 & 7.0$\pm$1.2\\ 
120-7 & MDM11 43, SFZ12 1503-160L & 18:41:34.071 & -05:04:01.26 & 5 & WN7 & 5, 10 & 3.7$\pm$0.7\\ 
120-2 & SMG09 1509\_29, SCB12 2w01 & 18:41:48.49 & -04:00:12.4 &  1 & WC7 & 1, 34 & --\\
120-8 & MDM11 44 & 18:42:02.551 & -03:56:26.37 & 5 & WN9  & 0 & 8.0$\pm$1.3\\
120-3 & WMD47, WR120bb & 18:42:06.308 & -03:48:22.48 & 13 & WN9h & 13, 37 & 6.9$\pm$1.1\\
120-4 & WMD48, WR120bc & 18:42:08.271 & -03:51:02.91 & 13 & WN9h & 13, 37 & 7.6$\pm$1.2\\
120-5 & J18420846-0349352, & 18:42:08.467 & -03:49:35.22 & 13 & WC8 & 13 & 4.9$\pm$0.9 \\
      & SCB12 2w02     &                 &               &    &     &    &           \\
120-14 & SCB12 2w03 & 18:42:22.17 & -03:05:39.6 & 34 & WC8 & 34 & --\\
120-17 & KSF14 1509-2E64 & 18:42:26.61 & -03:56:36.0 & 27 & WC9 & 27 & --\\ 
120-6 & WMD10 50, PN G029.0+0.04 & 18:42:46.921 & -03:13:17.25 & 9 & WN6o& 9 & 19.9$\pm$3.9\\
120-15 & SCB12 2w04 & 18:43:17.23 & -03:08:56.6 & 34 & WC8 & 34 & --\\
120-12 & SFZ12 1513-111L, SCB12 2w05 & 18:43:17.28 & -03:20:23.7 & 10 & WC8 & 34 & --\\
120-9 & MDM11 45 & 18:43:32.575 & -04:04:19.02 & 5 & WN7ha& 0& 19.7$\pm$3.8\\
120-13 & SFZ12 1522-55L & 18:43:39.65 & -02:29:35.9 & 10 & WC9 & 10 & --\\
120-10 & MDM11 46, SFZ12 1517-138L & 18:43:58.034 & -02:45:17.22 & 5& WN7 & 5 & 3.9$\pm$0.8 \\
121-2 & MDM11 47 &  18:44:51.576 & -03:21:49.61 & 5 & WN7 & 5 & 9.3$\pm$1.8\\
121-3 & MDM11 48 & 18:44:51.610 & -03:27:43.75 & 5 & WN7 & 5 & 7.7$\pm$1.5\\
121-11 & KSF14 1525-2352 & 18:45:14.63 & -02:05:05.7 & 27 & WC8: & 27 & --\\ 
121-7 & SCB12 2w06, KSF14 1519-E43 & 18:45:49.90 & -02:59:56.3 & 34 & WC7-8 & 34, 27 & -- \\
121-12 & KSF14 1530-8FA & 18:46:00.97 & -01:14:35.0 & 27 & WN5 & 27 & --\\
121-5 & SFZ12 1527-13L, SCB12 2w07 & 18:47:38.33 & -02:06:38.9 & 10 & WC8 & 10, 34 & --\\
121-8 & SCB12 2w08 & 18:47:57.34 & -01:27:36.9 & 34 & WC8 & 34 & --\\
121-9 & SCB12 2w09 & 18:48:24.50 & -02:06:16.2 & 34 & WC8 & 34 & --\\
121-1 & WR121b, WMD10 52 & 18:49:27.336 & -01:04:20.79 & 16 & WN7h & 16 & 6.1$\pm$1.2\\
121-4 & MDM11 49, SFZ12 1528-15L & 18:49:32.305 & -02:24:27.08 & 5 & WC7    & 0 & 6.9$\pm$1.2\\
121-10 &SCB12 2w10, KSF14 1541-3C8 & 18:50:02.77 & -00:32:08.1 & 34 & WC8 & 34, 27 & --\\
121-13 & KSF14 1541-197C & 18:50:37.54 & -00:01:21.1 & 27 & WC8 & 27 & --\\
121-6  & SFZ12 1536-180L & 18:51:10.77 & -01:30:03.4 & 10 & WN5 & 10 & --\\
121-14 & KSF14 1547-E0B & 18:51:33.07 & -00:13:41.3 & 27 & WN4 & 27 & --\\
121-15 & KSF14 1544-FA4 & 18:51:33.08 &+00:13:41.2 & 27 & WN5 & 27 & --\\ 
121-16 & KSF14 1547-1DF2 & 18:51:38.98 & -00:10:08.1 & 27 & WN8: & 27 & --\\
122-6 & SFZ12 1551-19L & 18:52:32.97 & +00:14:26.8 & 10 & WC8: & 10 & --\\
122-12 & KSF14 1553-9E8 & 18:52:33.12 & +00:47:41.8 & 27 & WN9h & 27 & --\\
122-2 & MDM11 51 & 18:52:43.699 & +00:08:41.58 & 5 & WN9   & 0 & 8.3$\pm$1.3\\
122-13 & KSF14 1547-1488 & 18:52:57.20 & +00:02:54.1 & 27 & WN5 & 27 & --\\
122-14 & KSF14 1553-15DF & 18:53:02.56 & +01:10:22.7 & 27 & WC8 & 27 & --\\
122-3 & MDM11 52 & 18:54:03.125 & +01:24:50.84 & 5 & WN6    & 0 & 4.7$\pm$0.9\\
122-7 & SFZ12 1563-66L & 18:55:44.44 & +01:36:43.9 & 10 & WC8: 10 & --\\
122-8 & SFZ12 1563-89L & 18:56:02.04 & +01:36:32.9 & 10 & WC7: & 10 & --\\ 
122-9 & SFZ12 1567-51L & 18:56:07.90 & +02:20:49.0 & 10 & WC7: & 10 & --\\
122-11 & FSZ14 WR1583-B73 & 19:00:05.09 & +03:47:27.1 & 33 & WN6 & 33 & --\\
122-1 & J190015.86+000517.3 & 19:00:15.86 & +00:05:17.3 & 35 & WC8 & 35 & --\\
122-10 & SFZ12 1583-64L & 19:00:59.99 & +03:55:35.6 & 10 & WC7: & 10 & --\\
122-4 & MDM11 53, SFZ12 1583-48L & 19:01:26.614 & +03:51:55.34 & 5 & WC8 & 5, 10 & 8.1$\pm$1.4\\
122-5 & MDM11 54, SFZ12 1583-47L & 19:01:27.119 & +03:51:54.22 & 5 & WC8 & 5, 10 & 7.6$\pm$1.4\\
122-15 & KSF14 1602-9AF & 19:02:42.32 & +06:54:44.4 & 27 & WN6 & 27 & --\\
123-6 & KSF14 1603-11AD & 19:04:20.14 & +06:07:52.2 & 27 & WN5 & 27 & --\\
123-4 & SFZ12 1603-75L & 19:04:33.490 & +06:05:18.50 & 10 & WC8 & 10 & 33.2$\pm$6.4\\
123-7 & KSF14 1609-1C95 & 19:06:10.68 & +07:19:13.3 & 27 & WC9 & 27 & --\\
123-8 & KSF14 1626-4FC8 & 19:06:33.66 & +09:07:20.8 & 27 & WC6 & 27 & --\\
123-2 & SMG09 1613\_21 & 19:06:36.53 & +07:29:52.40 & 1 & WCE &1  & --\\
123-1 & HDM07 14 & 19:08:17.975 & +08:29:10.56 & 2 & WN6 & 2 &  3.2$\pm$0.6\\
123-3 & MDM11 55 &  19:08:38.093 & +09:28:21.06 & 5 & WN8 & 5 & 9.8$\pm$1.6\\[1.0ex]
\hline
\end{tabular}
\end{minipage}
\end{table*}

\addtocounter{table}{-1}

\clearpage
\begin{table*}
\begin{minipage}{13cm}
\centering
\caption{(Continued)}
\begin{tabular}{l l c c l l l r} 
\hline\\[-2.0ex]
WR & Alias & RA & Dec &  & Sp Type &  & Distance\\ 
   &  & (hh:mm:ss) & ($\pm$dd:mm:ss) & Ref & & Ref & kpc\\[1.0ex] 
\hline\\[-2.0ex]
123-9 & KSF14 1629-14D6 & 19:10:06.40 & +09:45:25.7 & 27 & WN9h & 27 & --\\
123-5 & SCB12 2w11, & 19:10:11.53 & +08:58:39.6 & 34 & WC7 & 34, 27 & --\\
      &  KSF14 1627-A6D &         &             &    &     &       &   \\
124-1 & MDI09 Glimpse 20-6 & 19:12:24.140 & +09:57:29.10 &  15 & WC8&  0 & 4.4$\pm$0.7\\
124-13 & KSF14 1635-AD8 & 19:13:19.19 & +09:55:29.0 & 27 & WN6 & 27 & --\\
124-8 & SFZ12 1650-96L & 19:13:23.731 & +11:43:26.81 & 10 & WN6 & 10 & 2.8$\pm$0.6\\
124-14 & KSF14 1653-FFE & 19:14:40.73 & +11:54:15.4 & 27 & WN5--6 & 27 & --\\
124-15 & KSF14 1651-BB4 & 19:15:37.26 & +11:25:26.3 & 27 & WN5 & 27 & --\\
124-16 & KSF14 1647-1E70 & 19:15:52.52 & +11:12:59.7 & 27 & WC8: & 27 & --\\
124-3 & MDM11 56, SFZ12 1657-51L &  19:16:18.383 & +12:46:49.36 & 5 & WC7    & 0 & 9.5$\pm$1.6\\
124-17 & KSF14 1659-212 & 19:17:22.20 & +12:13:09.2 & 27 & WN9 & 27 & --\\
124-9 & SFZ12 1670-57L & 19:17:32.805 & +14:08:27.98 & 10 & WC6: & 10 & 13.2$\pm$4.1\\
124-4 & MDM11 57, SFZ12 1652-24L & 19:17:41.21 & +11:29:18.9 & 5 & WC7 & 5 & --\\
124-18 & KSF14 1669-3DF & 19:18:31.35 & +13:43:39.4 & 27 & WN9h & 27 & --\\
124-10 & SFZ12 1669-24L & 19:18:31.705 & +13:43:17.84 & 10 & WC6: & 10 & 9.7$\pm$3.0\\
124-19 & KSF14 1660-1169 & 19:20:02.46 & +12:08:20.3 & 27 & WC6: & 27 & --\\
124-5 & MDM11 58 & 19:20:29.32 & +14:12:06.1 & 5 & WC8-9d? & 5 & --\\
124-2 & SMG09 1671\_5 & 19:20:40.38 & +13:50:35.2 & 1 & WC8 & 1 & --\\
124-12 & FSZ14 WR1667-D00 & 19:20:50.59 & +13:18:41.1 & 33 & WN7 & 33 & --\\
124-6 & MDM11 59, SFZ12 1675-17L & 19:22:53.616 & +14:08:49.82 & 5 & WC6 & 0 & 4.6$\pm$1.4 \\
124-7 & MDM11 60, SFZ12 1675-10L & 19:22:54.462 & +14:11:28.01 & 5 & WC8    & 0 & 3.8$\pm$0.7\\
124-11 & SFZ12 1698-70L & 19:24:46.914 & +17:14:25.18 & 10 & WN6b&  0 & 7.8$\pm$1.3\\
124-20 & KSF14 1697-38F & 19:25:18.12 & +17:02:15.9 & 27 & WC9 & 27 & --\\
124-21 & KSF14 1702-23L & 19:26:08.00 & +17:46:23.0 & 27 & WC8 & 27 & --\\
124-22 & KSF14 1695-2B7 & 19:27:17.98 & +16:05:24.6 & 27 & WC9 & 27 & --\\
125-4 & MDM11 61 & 19:30:05.304 & +17:46:01.10 & 5 & WC8d & 5 & 6.5$\pm$1.3 \\
125-3 & HKB10 2, Mercer 23 \#2 &  19:30:13.820 & +18:32:00.33 & 17 & WN7ha   &  0 & 6.5$\pm$0.3\\
125-2 & J193038.84+183909.8 & 19:30:38.84 & +18:39:09.8 & 35 & WN8-9 & 35 & --\\
125-1 & HMD07 15 & 19:33:44.016 & +19:22:47.54 & 2 & WC8 & 2 & 5.1$\pm$0.9\\
138-1 & WR138a, HBHA 4202-22 & 20:17:08.12 & +41:07:27.0 & 36 & WN8-9h & 36 & --\\
142-1 & HBH$\alpha$ 4203-27 & 20:28:14.552 & +43:39:25.74 & 18 & WN6o & 18 & 1.7$\pm$0.3\\[1.0ex]
\hline\\[-1.0ex]
\end{tabular}
(0) This work,
(1)\citet{shara09}, 
(2)\citet{hadfield07}, 
(3)\citet*{mauerhan09}, 
(4)\citet{kurtev07}, 
(5)\citet{mauerhan11}, 
(6)\citet{davies12a}, 
(7)\citet{roman-lopes11b}, 
(8)\citet{roman-lopes11c}, 
(9)\citet{wachter10}, 
(10)\citet{shara12}, 
(11)\citet{mauerhan10a}, 
(12)\citet{mauerhan10c}, 
(13)\citet{mauerhan10b}, 
(14) \citet{liermann09}, 
(15)\citet{messineo09}, 
(16)\citet{gvaramadze10}, 
(17)\citet{hanson10}, 
(18)\citet{littlefield12}. 
(19) \citet{roman-lopes11a}, 
(20) \citet{Motch10}
(21)\citet{roman-lopes12}, 
(22)\citet{roman-lopes13}, 
(23)\citet{chene13},
(24) \citet{rahman11},
(25) \citet{anderson11},
(26) \citet{marston13},
(27) \citet{kanarek14},
(28) \citet{davies12b},
(29) \citet{delaFuente13},
(30) \citet{borissova12},
(31) \citet{mikles06},
(32) \citet{hyodo08},
(33) \citet{faherty14}
(34) \citet{smith12},
(35) \citet{corradi10},
(36) \citet{gvaramadze09},
(37) \citet{burgemeister13}
\end{minipage}
\end{table*}

\clearpage

\section[]{WR stars and clusters with distance ambiguity}
\label{app:distance}

\subsubsection{Pismis 24 and WR$\,93$}
\begin{figure}
\includegraphics[width=0.48\textwidth]{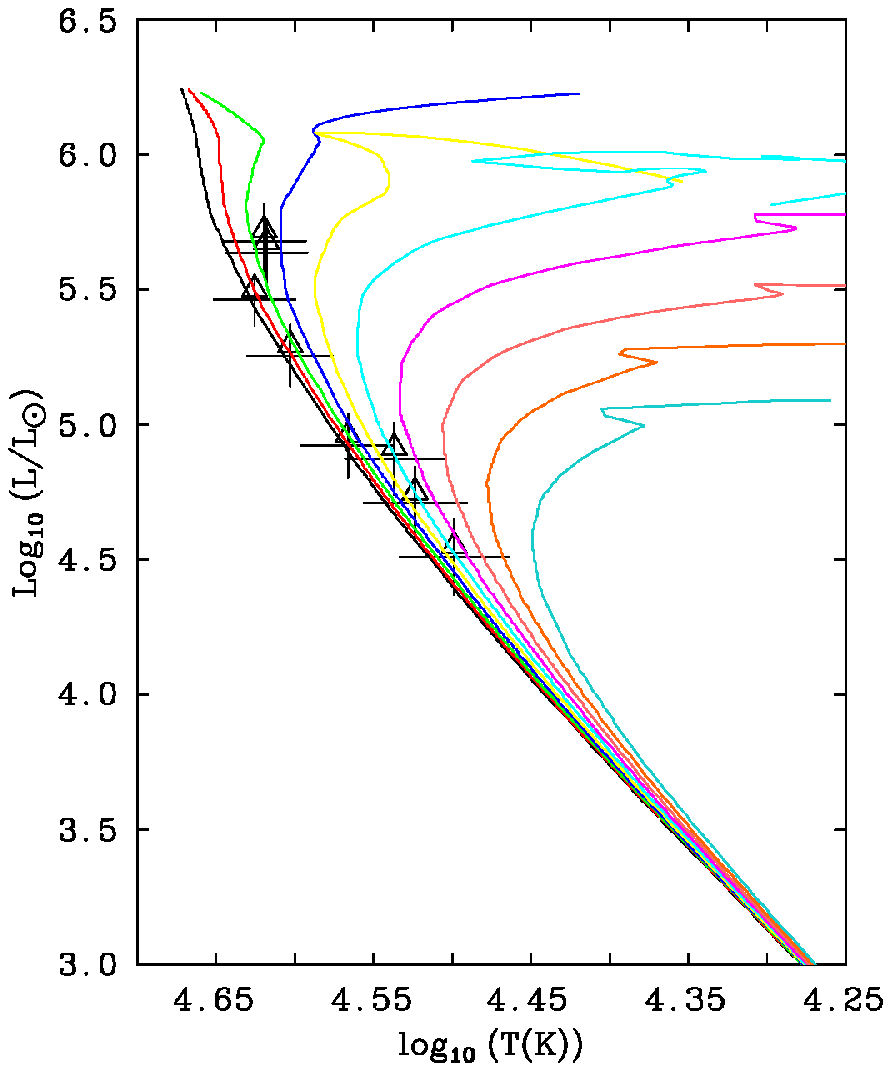}
\caption{Hertzsprung-Russell diagram showing positions of the 8 brightest (V-band, \citealt{massey01a}) O-stars in the Pismis 24 open cluster (triangles). Stars are individually dereddened and shown at a distance modulus of 11.5, with Z=0.02 iochrones \citep{lejeune01} for ages (from left to right) 6.00, 6.09, 6.19, 6.30, 6.40, 6.50, 6.59, 6.69, 6.80, and $6.90\,$Myr (left to right, solid lines).}
\label{afig:pi24}
\end{figure}

The open cluster Pismis 24 (Pi$\,24$) contains the WC7+O binary system WR$\,93$, and has
conflicting distance meaurements in the literature, i.e., $2.56\,{\pm}\,0.10\,$kpc
\citep*{massey01a} and $1.7\,{\pm}\,0.2\,$kpc \citep{fang12}. In Figure \ref{afig:pi24} we
plot \citet{lejeune01} isochrones along with the positions of the $8$
brightest O-stars in Pi$\,24$. We take photometry and spectral types for these
O-stars from \citet{massey01a} except for Pi$\,24\mbox{--}1$, which has since been resolved
into two components, Pi$\,24\mbox{--}1\,$NE and Pi$\,24\mbox{--}1\,$SW \citep{maiz07}, with an optical/near
ultra-violet $\Delta m\,{\sim}\,0.1$. We adjust the \citeauthor{massey01a} photometry for
Pi$\,24\mbox{--}1$ to account for its binary nature, and adopt spectral types of O3.5If* and
O4III(f+) for Pi$\,24\mbox{--}1\,$NE and Pi$\,24\mbox{--}1\,$SW respectively \citep{maiz07}. To calculate an
extinction to each of the 8 O-stars, we evaluate the $E_{B-V}$ colour excess
assuming \citet{martins06} intrinsic colours and an $R_V{=}3.1$ extinction
law. Finally, by taking O-star temperatures and bolometric corrections from
\citet*{martins05}, we see that best agreement with the isochrones is found at a
distance modulus of $11.5\,{\pm}\,0.2$, corresponding to $d\,{=}\,2.00^{+0.19}_{-0.17}\,$kpc. 

\subsubsection{Westerlund 2}
The distance to the young massive open cluster Westerlund 2 - probable host of the O$3$If*/WN$6$+O3If*/WN6 binary WR$\,20$a \citep*{rauw11} and WR$\,20$b (WN6ha) - 
remains controvertial. Literature values range from $2.5\,$kpc to $8\,$kpc, 
bringing the membership of the very luminous WR$\,20$a into doubt. \citet{rauw07} 
used the light curve of WR$\,20$a and the knowledge that both stars are of identical 
spectral type to derive a distance of $8.0\,{\pm}\,1.0\,$kpc for this binary system. 
These authors also derive a distance to the Westerlund 2 cluster of $8.0\,{\pm}\,1.4\,$kpc 
from spectro-photometry of cluster O-stars and use this agreement as evidence for 
membership of WR$\,20$a. However, these distances are derived on the asumption of an 
$R_V\,{=}\,3.1$ extinction law, yielding an average $A_V\,{=}\,4.68$ for the cluster. 
\citet{carraro12} have claimed an anomalous extinction law along this line of sight with an average $R_V\,{=}\,3.8{\pm}\,0.2$. These authors use the spectra of \citeauthor{rauw07} to obtain $A_V\,{=}\,7.1{\pm}\,1.2$, 
corresponding to a much smaller distance of $3.02\,{\pm}\,0.52\,$kpc. The membership of 
WR$\,20$a as two ${\sim}\,80\,M_{\odot}$ stars is unfeasible at such a small distance.

WR$\,20$b showns no evidence of binarity in the spectroscopic and photometric 
monitoring of \citet{rauw07}. The asssumed intrinsic near-IR colour of WN6ha stars (Table 
\ref{tab:int_col}) provides $A_{K_S}\,{=}\,0.75\,{\pm}\,0.05$ ($A_V\,{\sim}\,6.8$), favouring 
the higher extinction and lower distance estimate for Westerlund 2. Until the 
issue of the form of the extinction law to Westerlund 2 is settled, we consider 
neither WR$\,20$a or WR$\,20$b as members of the cluster. We include the stars of WR$\,20$a 
in our calibration at the binary orbit-derived distance of $8.0\,$kpc.

\subsubsection{The Galactic Centre}
The Arches and Quintuplet clusters are found $11.6^\prime$ and $13.1^\prime$ from 
Sgr A* respectively, which is itself surrounded by a cluster of massive stars. The distance 
to the GC ($R_o$), and these three clusters by association, has been the 
subject of considerable study. The first direct parallax 
measurement of a GC object - the star forming region Sgr B2 - was presented 
by \citet{reid09} giving a distance of $7.9\,{\pm}\,0.8\,kpc$; from kinematic arguments  
Sgr B2 is estimated to be $0.13\,$kpc nearer than the GC. The most recent 
determinations of $R_o$ are summerised by \citet{gillessen13}, and are converging on $8\,$kpc. Acknowledging the spread in measurements that 
still remains, we assume a distance of $8.0\,{\pm}\,0.25$kpc for these three clusters. 

\subsubsection{The G305 complex}
There are 9 currently identified WR stars located within the boundary of the
giant HII region G305.4+0.1, of which only 4 reside in the two central clusters
Danks $1\,\&\,2$ (see Figure 16 of \citealt{mauerhan11}, and \citealt{davies12a}). This
is rather surprising, as one may expect to find these stars - as 
descendants of massive progenitors - at the centre of these star clusters due to
relaxation of their orbits. However, in the dense environment of a cluster a
massive star is more likely to encounter other massive stars or binaries,
resulting in possible ejection. The apparent concentration of these 5
non-cluster WR stars in G305 around the younger Danks 1 cluster \citep{davies12a}
supports a dynamical ejection scenario. 

Located only ${\sim}2.3^{\prime}$ from the Danks 1 cluster, WR$\,48\mbox{--}4$ is the
faintest WR star located within G305. \citet{mauerhan11} note that by assuming a
WC7 spectral type and applying the K$_S$-band absolute magnitude-subtype calibations
of \citet{crowther06b}, this star appears to be twice as distant as the two
central clusters yet is reddened by a similar amount. We consider the high 
IR-derived distance to WR$48\mbox{--}4$ as insignificant evidence for a chance 
superposition with G305, and included these 9 WRs in the calibration 
sample at the distance derived for the two central clusters.



\section[]{Single WR star emission line strengths}
\label{app:em_lines}

In Table~C1 we present average strengths for the most prominent lines in the spectra of each WR spectral type, gathered from published spectra of single WR stars. We use these values to calculate 
J and K$_S$-band continuum flux ratios in cases where WR emission lines are diluted by an 
unknown companion. An uncertainty of $0.1\,$dex is assumed on each averaged equivalent 
width, in accordance with the majority of studies from which individual measurements are taken. 
\\
\hfill

\centering

\tablefirsthead{%
\hline\\[-2.0ex]
Spectral & Star & Equivalent & ref & Average \\ type & (WR\#) & \mbox{width (\AA)} & & (\AA)
\\[1.0ex]
\hline\\[-2.0ex]}

\tablehead{%
\hline\\[-2.0ex]
\multicolumn{5}{l}{\small\sl Table C1 continued}\\[0.5ex]
\hline\\[-2.0ex]
Spectral & Star & Equivalent & ref & Average \\ type & (WR\#) & \mbox{width (\AA)} & & (\AA)
\\[1.0ex]
\hline\\[-2.0ex]}

\tabletail{%
\hline\\[-2.0ex]
\multicolumn{5}{r}{\small\sl Continued...}\\[0.5ex]
\hline\\[-2.0ex]}
\tablelasttail{\hline}
\tablecaption{Emission line strengths (Equivalent widths, \AA) in (apparently) single WR stars.
}%

\begin{supertabular}{l l l l l}

\multicolumn{5}{c}{\textbf{He\,{\sc ii} $\mathbf{\,1.012\mu}$m}}      \\[1ex]
WN7    & 55    & 109.5 & a,b &  \textbf{89}   \\
        & 82    & 75.9  &  b   &            \\
        & 84    & 134.6  &  b   &            \\
        & 120    & 70.8 & c,d &                 \\
WN6    & 115    & 126  & c,d & \textbf{119}   \\
        & 85     & 112   & a  &             \\
WN4-5    & 83    & 192.4 & b  & \textbf{221}    \\
        & 54    & 236.2   & b &  \\
        & 61    & 243.8  & b &  \\
        & 149   & 234 & a &  \\
        & 129   & 200 & a &  \\
WN4b    & 1     & 377 & e & \textbf{412} \\
       & 18    & 447 & a & \\
WN5-6b  & 75    & 327 & b & \textbf{375} \\
        & 110   & 423 & a,d & \\
WN7b    & 77sc  & 280 & f & \\
WC8     & 135   & 93     & e &  \textbf{93} \\
        & 57    & 121.7  & b &  \\
        & 60    & 111.1  & b &  \\
        & 118-4  & 80    & g &  \\
        & 119-2 & 93     & g &  \\
        & 53    & 61.7   & a &  \\
WC7     & 56    & 143.8  & b  & \textbf{147}  \\
       & 64    & 145.8  &  b  &   \\
        & 90    &    162   & b  &  \\
     & 124-3     & 138    & g &  \\
WC6  & 154       & 205   & e & \textbf{179}  \\
     & 107a      & 182     & b &  \\
     & 23        & 151   & a &   \\
WC4-5  & 111      & 218 & e & \textbf{225} \\
      & 17        & 263   & a & \\
      & 52        & 234   & b & \\
     & 150       & 186   & a & \\[3ex]
\multicolumn{5}{c}{\textbf{He\,{\sc i} $\mathbf{\,1.083\mu}$m}}      \\[1ex]
WN8  & 116 & 480 & e & \textbf{308} \\
      & 130 & 200 & e & \\
      & 16  & 245 & d & \\ 
WN7 & 55 & 296 & b  & \textbf{291} \\
     & 82 & 309 & b  & \\
    & 84 & 267 & b &  \\
WN6 & 115 & 170 & d & \textbf{170} \\
WN4-5 & 83 & 95 & b & \textbf{114} \\
      & 54 & 106 & b &  \\
      & 61 & 142 & b & \\
WN7b & 77sc & 930 & f & \textbf{930} \\
WN5-6b & 75 & 661 & b & \textbf{712} \\
        & 110 & 763 & d  & \\
WN4b  & 1 & 452 & e & \\
WC8  & 57    & 221 & b  & \textbf{412} \\
      & 60    & 353 & b & \\
      & 118-4 & 537 & g & \\
      & 119-2 & 600 & g & \\
      & 135   & 344 & e & \\
WC7  & 56   & 221 & b & \textbf{227} \\
      & 64   & 171 & b & \\
     & 90   & 301 & b & \\
      & 124-3 & 215 & g & \\
WC6  & 5    & 276 & e & \textbf{232} \\
     & 107a & 200 & b & \\
      & 154  & 220 & e & \\
WC4-5 & 52  & 149 & b & \textbf{169} \\
      & 111  & 189 & e & \\[3ex]
\multicolumn{5}{c}{\textbf{He\,{\sc i-ii} $\mathbf{\,2.164\mu}$m}}      \\[1ex]             
WN8-9 & 16  & 75 & d & \textbf{73} \\      
    & 105 & 63 & d & \\
   & 116 & 106 & e & \\                    
    & 130 & 46 & e & \\
WN7 & 55 & 33 & b & \textbf{43} \\
    & 120 & 52 & d & \\
WN6 & 24 & 34 & d & \textbf{35} \\
     & 115 & 36 & d & \\
WN4-5 & 129 & 50 & h & \textbf{53} \\
      & 149 & 55 & h & \\
WN7b & 77sc & 85 & f & \\
WN6b & 75 & 74 & b & \\
WN4b & 1  & 54 & e & \\[3ex]
\multicolumn{5}{c}{\textbf{He\,{\sc ii} $\mathbf{\,2.189\mu}$m}}   \\[1ex] 
WN7 & 120 & 39 & d & \textbf{35} \\
      & 55 & 31  & b & \\
WN6 & 115 & 58 & d & \textbf{42} \\
      & 24  & 25 & d & \\
WN4-5 & 129 & 110 & h & \textbf{105} \\
       & 149 & 99 & h & \\[3ex]
\multicolumn{5}{c}{\textbf{C\,{\sc iv} $\mathbf{\,2.071\mbox{--}2.084\mu}$m}}   \\[1ex]              
WC8 & 48-2 & 581 & b & \textbf{466} \\
     & 77g  & 430  & f & \\
     & 118-4 & 303 & g & \\
     & 119-2 & 433 & g & \\
     & 135 & 583 & e & \\
WC7 & 67-2 & 783 & i & \textbf{937} \\
     & 90 & 825 & b & \\
    & 124-3 & 1202 & g & \\
WC6 & 5 & 1339 & e & \textbf{1530} \\
     & 48-4 & 1862 & b & \\
     & 154 & 1388 & e  & \\
WC4-5 & 52  & 1169 & b  & \textbf{1391} \\
     & 111 & 1613 & e & \\[3ex]
\multicolumn{5}{c}{\textbf{C\,{\sc iii} $\mathbf{\,2.104{-}2.115\mu}$m}}   \\[1ex]             
WC8 & 48-2 & 220 & b & \textbf{209} \\
     & 77g  & 198 & f & \\            
    & 118-4 & 173 & g & \\
     & 119-2 & 260 & g & \\
     & 135 & 192 & e & \\
WC7 & 67-2 & 234 & i & \textbf{293} \\
     & 90  & 224 & b & \\
    & 124-3 & 422 & g & \\
WC6 & 5 & 228 & e & \textbf{301} \\
    & 48-4 & 385 & b & \\
     & 154 & 289 & e & \\
WC4-5 & 52 & 277 & b & \textbf{290} \\
      & 111 & 302 & e & \\[1.0ex]
\end{supertabular}
\vspace{1.0ex}

(a)\citet*{conti90}, (b)P.A. Crowther, \citetext{priv.\ comm.}, (c)\citet{howarth92}, 
(d)\citet{crowther96}, (e)W.D.~Vacca, \citetext{priv.\ comm.}, (f)\citet{crowther06b}, 
(g)\citet{mauerhan11}, (h)\citet*{figer97}, (i)\citet{roman-lopes11c}


\clearpage

\section[]{A volume-limited sample of WR stars}
\label{app:vol_lim}

We present a volume-limited ($d\,{<}\,3\,$kpc) sample of Galactic Wolf-Rayet stars in Table~D1, in which
distances are either from this work, or are derived by adopting $M_v$ (by subtype) and $A_v$ from \citet{vdH01}.
\\[0.2pt]

\centering
\tablefirsthead{%
\hline\\[-2.0ex]
WR\# & Spectral type & Distance (kpc) & Class
\\[1.0ex]
\hline\\[-2.0ex]}

\tablehead{%
\hline\\[-2.0ex]
\multicolumn{3}{l}{\small\sl Table D1 continued}\\[0.5ex]
\hline\\[-2.0ex]
WR\# & Spectral type & Distance (kpc) & Class
\\[1.0ex]
\hline\\[-2.0ex]}

\tabletail{
\hline\\[-2.0ex]
\multicolumn{3}{r}{\small\sl Continued...}\\[0.5ex]
\hline\\[-2.0ex]
}
\tablelasttail{\hline}
\tablecaption{Closest ($d\,{<}\,3$kpc) WR stars listed in ascending heliocentric distance. 
In the fourth column we class each system as WR-dominated ($m_K^{\mbox{\tiny WR}}\,{-}\,m_K^{sys}\,{>}\,0.4$; WR), 
companion-dominated (C), or dust-producing (D). References are given for spectral types of stars not appearing in Tables 1--3 or 8 
(online).
}%
\begin{supertabular}{l l l l}
11 & WC8+O7.5III & $0.34\pm0.08$ & C \\
147 & WN8(h)+OB & $0.73\pm0.12$ & WR \\
94 & WN5 & $0.78\pm0.12$ & WR \\ 
90 & WC7 & $1.15\pm0.19$  & WR \\
136 & WN6b(h) & $1.3\pm0.2$ & WR \\ 
137 & WC7+O9 & $1.3\pm0.2$ & D \\ 
139 & WN5+O6III-V & $1.3\pm0.2$ & C \\ 
141 & WN5+O5III-V & $1.3\pm0.2$ & C \\ 
143 & WC4+Be & $1.33\pm0.33$ & C \\ 
138 & WN5+OB & $1.38\pm0.26$ & C \\ 
144 & WC4 & $1.40\pm0.08$ & WR \\ 
52 & WC4 & $1.54\pm0.23$ & WR \\ 
110 & WN5b & $1.55\pm0.24$ & WR \\
9 & WC5+O7$^\star$ & $1.57\pm0.58^\star$ & C \\ 
15 & WC6 & $1.57\pm0.49$ & WR \\
81 & WC9 & $1.64\pm0.34$ & WR \\
78 & WN7 & $1.64\pm0.03$ & WR \\
79 & WC7+O5-8V & $1.64\pm0.03$ & C \\
79a & WN9ha & $1.64\pm0.03$ & WR \\
140 & WC7pd+O5fcIII-I$^1$ & $1.67\pm0.03^2$ & D \\ 
142-1 & WN6 & $1.70\pm0.34$ & WR \\
6 & WN4b & $1.80\pm0.27$ & WR \\
121 & WC9d$^\star$ & $1.8\pm0.4^\star$ & D \\ 
142a & WC7 & $1.83\pm0.31$ & WR \\
105 & WN9 & $1.9\pm0.2$ & WR \\
111 & WC5 & $1.9\pm0.2$ & WR \\
134 & WN6b & $1.9\pm0.2$ & WR \\
135 & WC8 & $1.9\pm0.2$ & WR \\
86 & WC7+B0III & $1.97\pm0.47$ & WR \\
113 & WC8d+O8-9$^\star$ & $2.0\pm0.2$ & D \\
14 & WC7 & $2.0\pm0.1$ & WR \\
93 & WC7+O7-9 & $2.0\pm0.2$ & WR \\
114 & WC5 & $2.05\pm0.09$ & WR \\
115 & WN6 & $2.05\pm0.09$ & WR \\
70 & WC9vd+B0I$^\star$  & $2.1\pm0.4^\star$ & D \\
47 & WN6+O5.5 & $2.13\pm0.44$ & WR \\
133 & WN5+O9I & $2.14\pm0.07$ & C \\
70-5 & WC9 & $2.17\pm0.45$ & WR \\
48 & WC6+O6-7V$^3$... & ${<}\,2.3^\star$ & C \\ 
   & ...+O9.5/B0Iab$^\star$ &     &   \\
1 & WN4b & $2.3\pm0.5$ & WR \\
103 & WC9d$^\star$ & $2.3\pm0.5^\star$ & D \\
59 & WC9d$^\star$ & $2.3\pm0.5^\star$ & D \\
2 & WN2b+B$^4$ & $2.4\pm0.8$ & WR \\
106 & WC9d$^\star$ & $2.4\pm0.6^\star$ & D \\
19a & WN7 & $2.41\pm0.47$ & WR \\
101 & WC8 & $2.46\pm0.43$ & WR \\
40 & WN8h & $2.48\pm0.41$ & WR \\
95 & WC9d$^\star$ & $2.5\pm0.5$ & D \\
60 & WC8 & $2.55\pm0.45$ & WR \\
155 & WN6+O9II-Ib & $2.56\pm0.56$ & WR \\
117-1 & WN7 & $2.59\pm0.50$ & WR \\
18 & WN4b & $2.6\pm0.2$ & WR \\
22 & WN7ha+O9III-V & $2.6\pm0.2$ & WR \\
23 & WC6 & $2.6\pm0.2$ & WR \\
24 & WN6ha & $2.6\pm0.2$ & WR \\
25 & O2.5If*/WN6$^5$+OB & $2.6\pm0.2$ & C \\
104 & WC9d+B0.5V$^\star$ & $2.6\pm0.7$ & D \\
88 & WC9 & $2.67\pm0.54$ & WR \\
4 & WC5 & $2.69\pm0.49$ & WR \\
5 & WC6 & $2.69\pm0.84$ & WR \\
72-1 & WC9 & $2.73\pm0.56$ & WR \\    
16 & WN8h & $2.77\pm0.46$ & WR \\
85 & WN6 & $2.8\pm1.1$ & WR \\
111-3 & WC8 & $2.80\pm0.49$ & WR \\
69 & WC9d+OB$^\star$ & $2.8\pm0.6^\star$ & D \\
75b & WC9 & $2.82\pm0.58$ & WR \\
124-8 & WN6 & $2.84\pm0.56$ & WR \\
113-1 & WN7 & $2.88\pm0.56$ & WR \\
151 & WN4+O5V & $2.93\pm0.65$ & WR \\
42 & WC7+O7V & $2.96\pm0.53$ & C \\
57 & WC8 & $2.97\pm0.52$ & WR \\
75a & WC9 & $2.98\pm0.60$ & WR \\[1.0ex]
\end{supertabular}
\vspace{1.0ex}

$^\star$\citet{vdH01},\\
(1) \citet{fahed11},\\
(2) \citet{monnier11},\\
(3) \citet{hill02},\\
(4) \citet{chene14}, \\
(5) \citet{crowther11}.


\bsp

\label{lastpage}


\begin{thebibliography}{}
\bibliographystyle{mn2e}

\bibitem[\protect\citeauthoryear{{Anderson}, {Gaensler}, {Kaplan} et al.}{{Anderson} et al.}{2011}]{anderson11}
{Anderson} G.~E. et al., 2011, ApJ 727, 105


\bibitem[\protect\citeauthoryear{{Arnal}, {Cappa}, {Rizzo} \&
  {Cichowolski}}{{Arnal} et~al.}{1999}]{arnal99}
{Arnal} E.~M.,  {Cappa} C.~E.,  {Rizzo} J.~R.,    {Cichowolski} S.,  1999, AJ,
  118, 1798

\bibitem[\protect\citeauthoryear{{Balser}, {Rood}, {Bania} \&
  {Anderson}}{{Balser} et~al.}{2011}]{balser11}
{Balser} D.~S.,  {Rood} R.~T.,  {Bania} T.~M.,    {Anderson} L.~D.,  2011, ApJ,
  738, 27

\bibitem[\protect\citeauthoryear{{Baume}, {V{\'a}zquez} \& {Carraro}}{{Baume}
  et~al.}{2004}]{baume04}
{Baume} G.,  {V{\'a}zquez} R.~A.,    {Carraro} G.,  2004, MNRAS, 355, 475

\bibitem[\protect\citeauthoryear{{Benjamin}, {Churchwell}, {Babler}, {Bania},
  {Clemens}, {Cohen}, {Dickey}, {Indebetouw} \& {Jackson}}{{Benjamin}
  et~al.}{2003}]{benjamin03}
{Benjamin} R.~A. et al.,  2003, PASP, 115, 953

\bibitem[\protect\citeauthoryear{{Bibby}, {Crowther}, {Furness} \&
  {Clark}}{{Bibby} et~al.}{2008}]{bibby08}
{Bibby} J.~L.,  {Crowther} P.~A.,  {Furness} J.~P.,    {Clark} J.~S.,  2008,
  MNRAS, 386, L23

\bibitem[\protect\citeauthoryear{{Blaauw}}{{Blaauw}}{1961}]{blaauw61}
{Blaauw} A.,  1961, Bull. Astron. Inst. Netherlands, 15, 265


\bibitem[\protect\citeauthoryear{{Bohannan} \& {Crowther}}{{Bohannan} \& {Crowther}}{1999}]{bohannan99}
{Bohannan} B.,  {Crowther} P.~A.,  1999, ApJ, 511, 374

\bibitem[\protect\citeauthoryear{{Bohlin}, {Savage} \& {Drake}}{{Bohlin} et~al.}{1978}]{bohlin78} Bohlin, R.~C., Savage, 
B.~D., \& Drake, J.~F., 1978, ApJ, 224, 132

\bibitem[\protect\citeauthoryear{{Borissova} et al.}{{Borissova} et al.}{2012}]{borissova12}
{Borissova}, J.  {Georgiev}, L.  {Hanson}, M.~M. et al. 2012, A\&A 546, A110


\bibitem[\protect\citeauthoryear{{Bressert} et al.}{{Bressert} et al.}{2010}]{bressert10}
{Bressert}, E., {Bastian}, N., {Gutermuth}, R. et al. 2010, MNRAS 409, L54


\bibitem[\protect\citeauthoryear{{Breysacher}, {Azzopardi} \&
  {Testor}}{{Breysacher} et~al.}{1999}]{breysacher99}
{Breysacher} J.,  {Azzopardi} M.,    {Testor} G.,  1999, A\&As, 137, 117

\bibitem[\protect\citeauthoryear{{Bronfman}, {Casassus}, {May} \&
  {Nyman}}{{Bronfman} et~al.}{2000}]{bronfman00}
{Bronfman} L.,  {Casassus} S.,  {May} J.,    {Nyman} L.-{\AA}.,  2000, A\&A,
  358, 521


\bibitem[\protect\citeauthoryear{{Burgemeister}, {Gvaramadze}, {Stringfellow}, 
{Kniazev}, {Todt}, \& {Hamann}}{{Burgemeister} et al.}{2013}]{burgemeister13}
{Burgemeister}, S. {Gvaramadze}, V.~V., {Stringfellow}, G.~S., {Kniazev}, A.~Y.
{Todt}, H., {Hamann}, W.-R., 2013, MNRAS, 429, 3305


\bibitem[\protect\citeauthoryear{{Cappa}, {Goss}, {Niemela} \&
  {Ostrov}}{{Cappa} et~al.}{1999}]{cappa99}
{Cappa} C.~E.,  {Goss} W.~M.,  {Niemela} V.~S.,    {Ostrov} P.~G.,  1999, AJ,
  118, 948

\bibitem[\protect\citeauthoryear{{Cappa}, {Vasquez}, {Pineault} \&
  {Cichowolski}}{{Cappa} et~al.}{2010}]{cappa10}
{Cappa} C.~E.,  {Vasquez} J.,  {Pineault} S.,    {Cichowolski} S.,  2010,
  MNRAS, 403, 387

\bibitem[\protect\citeauthoryear{{Carraro}, {Turner}, {Majaess} \&
  {Baume}}{{Carraro} et~al.}{2012}]{carraro12}
{Carraro} G.,  {Turner} D.,  {Majaess} D.,    {Baume} G.,  2012, in Proc. IAU
  Symp. 289, Beijing, arXiv:1209.2080

\bibitem[\protect\citeauthoryear{{Chen{\'e}}, {Borissova}, {Bonatto},
  {Majaess}, {Baume}, {Clarke}, {Kurtev} \& {Schnurr} O.}{{Chen{\'e}}
  et~al.}{2013}]{chene13}
{Chen{\'e}} A.-N.,  {Borissova} J.,  {Bonatto} C.,  {Majaess} D.~J.,  {Baume}
  G.,  {Clarke} J.~R.~A.,  {Kurtev} R.,    {Schnurr} O.,  2013, A\&A,
  549, A98

\bibitem[\protect\citeauthoryear{{Chen{\'e}}, {Moffat}, {St-Louis}, {Schnurr}, {Crowther}, {Artiguau}, {Alecian}, {Wade}}{{Chen{\'e}} et~al.}{2014}]{chene14}
{Chen{\'e}} A-N., {Moffat} A.~F.~J., {St-Louis} N., {Schnurr} O., {Crowther} P.~A., {Artiguau} E., {Alecian} E., {Wade} G.~A., 2014, MNRAS submitted

\bibitem[\protect\citeauthoryear{{Chomiuk} \& {Povich}}{{Chomiuk} \&
  {Povich}}{2011}]{chomiuk11}
{Chomiuk} L.,  {Povich} M.~S.,  2011, AJ, 142, 197

\bibitem[\protect\citeauthoryear{{Clark}, {Negueruela}, {Crowther} \&
  {Goodwin}}{{Clark} et~al.}{2005}]{clark05}
{Clark} J.~S.,  {Negueruela} I.,  {Crowther} P.~A.,    {Goodwin} S.~P.,  2005,
  A\&A, 434, 949

\bibitem[\protect\citeauthoryear{{Cohen}, {Parker} \& {Green}}{{Cohen}
  et~al.}{2005}]{cohen05}
{Cohen} M.,  {Parker} Q.~A.,    {Green} A.~J.,  2005, MNRAS, 360, 1439

\bibitem[\protect\citeauthoryear{{Conti}, {Massey} \& {Vreux}}{{Conti}
  et~al.}{1990}]{conti90}
{Conti} P.~S.,  {Massey} P.,    {Vreux} J.-M.,  1990, ApJ, 354, 359


\bibitem[\protect\citeauthoryear{{Corradi}, {Valentini}, {Munari}, et al.}{{Corradi} et al.}{2010}]{corradi10}
{Corradi}, R.~L.~M., {Valentini}, M., {Munari}, U. et al., 2010, A\&A 509, A41


\bibitem[\protect\citeauthoryear{{Corti}, {Bosch} \& {Niemela}}{{Corti}
  et~al.}{2007}]{corti07}
{Corti} M.,  {Bosch} G.,    {Niemela} V.,  2007, A\&A, 467, 137

\bibitem[\protect\citeauthoryear{{Cotera}, {Simpson}, {Erickson}, {Colgan},
  {Burton} \& {Allen}}{{Cotera} et~al.}{1999}]{cotera99}
{Cotera} A.~S.,  {Simpson} J.~P.,  {Erickson} E.~F.,  {Colgan} S.~W.~J.,
  {Burton} M.~G.,    {Allen} D.~A.,  1999, ApJ, 510, 747

\bibitem[\protect\citeauthoryear{{Crowther}}{{Crowther}}{2003}]{crowther03}
{Crowther} P.~A.,  2003, Ap\&SS, 285, 677

\bibitem[\protect\citeauthoryear{{Crowther}}{{Crowther}}{2007}]{crowther07}
{Crowther} P.~A.,  2007, ARA\&A, 45, 177

\bibitem[\protect\citeauthoryear{{Crowther} \& {Smith}}{{Crowther} \&
  {Smith}}{1996}]{crowther96}
{Crowther} P.~A.,  {Smith} L.~J.,  1996, A\&A, 305, 541

\bibitem[\protect\citeauthoryear{{Crowther} \& {Walborn}}{{Crowther} \&
  {Walborn}}{2011}]{crowther11}
{Crowther} P.~A.,  {Walborn} N.~R.,  2011, MNRAS, 416, 1311

\bibitem[\protect\citeauthoryear{{Crowther}, {Smith} \& {Hillier}}{{Crowther}
  et~al.}{1995{\natexlab{b}}}]{crowther95b}
{Crowther} P.~A.,  {Smith} L.~J.,    {Hillier} D.~J.,  1995\natexlab{b}, A\&A, 302, 457

\bibitem[\protect\citeauthoryear{{Crowther}, {De Marco} \& {Barlow}}{{Crowther}
  et~al.}{1998}]{crowther98}
{Crowther} P.~A.,  {De Marco} O.,    {Barlow} M.~J.,  1998, MNRAS, 296, 367

\bibitem[\protect\citeauthoryear{{Crowther}, {Morris} \& {Smith}}{{Crowther}
  et~al.}{2006{\natexlab{a}}}]{crowther06a}
{Crowther} P.~A.,  {Morris} P.~W.,    {Smith} J.~D.,  2006{\natexlab{a}}, ApJ, 636, 1033

\bibitem[\protect\citeauthoryear{{Crowther} et 
al.(1995)}{{Crowther} et~al.}{1995{\natexlab{a}}}]{crowther95a} Crowther, P.~A., 
Smith, L.~J., Hillier, D.~J., \& Schmutz, W., 1995\natexlab{a}, A\&A, 293, 427 

\bibitem[\protect\citeauthoryear{{Crowther}, {Dessart}, {Hillier}, {Abbott} \&
  {Fullerton}}{{Crowther} et~al.}{2002}]{crowther02}
{Crowther} P.~A.,  {Dessart} L.,  {Hillier} D.~J.,  {Abbott} J.~B.,
  {Fullerton} A.~W.,  2002, A\&A, 392, 653

\bibitem[\protect\citeauthoryear{{Crowther}, {Hadfield}, {Clark}, {Negueruela}
  \& {Vacca}}{{Crowther} et~al.}{2006{\natexlab{b}}}]{crowther06b}
{Crowther} P.~A.,  {Hadfield} L.~J.,  {Clark} J.~S.,  {Negueruela} I.,
  {Vacca} W.~D.,  2006{\natexlab{b}}, MNRAS, 372, 1407


\bibitem[\protect\citeauthoryear{{Davies}, {Clark}, {Trombley}, {Figer},
  {Najarro}, {Crowther}, {Kudritzki}, {Thompson}, {Urquhart} \&
  {Hindson}}{{Davies} et~al.}{2012\natexlab{a}}]{davies12a}
{Davies} B., et~al., 2012\natexlab{a}, MNRAS, 419, 1871


\bibitem[\protect\citeauthoryear{{Davies}, {de La Fuente}, {Najarro}}{{Davies} et al.}{2012}]{davies12b}
{Davies}, B., {de La Fuente}, D., {Najarro}, F., {Hinton}, J.~A., 
	{Trombley}, C., {Figer}, D.~F., {Puga}, E. 2012, MNRAS 419, 1860

\bibitem[\protect\citeauthoryear{{Davis}, {Moffat} \& {Niemela}}{{Davis} et~al.}{1981}]{davis81} Davis, A.~B., Moffat, 
A.~F.~J., \& Niemela, V.~S., 1981, ApJ, 244, 528 


\bibitem[\protect\citeauthoryear{{de La Chevroti{\`e}re}, {Moffat} \&
  {Chen{\'e}}}{{de La Chevroti{\`e}re} et~al.}{2011}]{chevrotiere11}
{de La Chevroti{\`e}re} A.,  {Moffat} A.~F.~J.,    {Chen{\'e}} A.-N.,  2011,
  MNRAS, 411, 635


\bibitem[\protect\citeauthoryear{{de La Fuente}, {Najarro}, {Davies}, {Figer}}{{de La Fuente}, {Najarro}, {Davies}, \& {Figer}}{2013}]{delaFuente13}
{de la Fuente}, D., {Najarro}, F., {Davies}, B., {Figer}, D.~F. 2013, in: Highlights of Spanish Astrophysics VII, p.534, {Sociedad Espa{\~n}ola de Astronom{\'\i}a}, arXiv:1210.1781


\bibitem[\protect\citeauthoryear{{De Marco} \& {Schmutz}}{{De Marco} \&
  {Schmutz}}{1999}]{demarco99}
{De Marco} O.,  {Schmutz} W.,  1999, A\&A, 345, 163

\bibitem[\protect\citeauthoryear{{Depew}, {Parker}, {Miszalski}, {De Marco},
  {Frew}, {Acker}, {Kovacevic} \& {Sharp}}{{Depew} et~al.}{2011}]{depew11}
{Depew} K.,  {Parker} Q.~A.,  {Miszalski} B.,  {De Marco} O.,  {Frew} D.~J.,
  {Acker} A.,  {Kovacevic} A.~V.,    {Sharp} R.~G.,  2011, MNRAS, 414, 2812

\bibitem[\protect\citeauthoryear{{Djura{\v s}evi{\'c}}, {Zakirov},
  {Eshankulova} \& {Erkapi{\'c}}}{{Djura{\v s}evi{\'c}}
  et~al.}{2001}]{djurasevic01}
{Djura{\v s}evi{\'c}} G.,  {Zakirov} M.,  {Eshankulova} M.,    {Erkapi{\'c}}
  S.,  2001, A\&A, 374, 638

\bibitem[\protect\citeauthoryear{{Dong}, {Wang} \& {Morris}}{{Dong}
  et~al.}{2012}]{dong12}
{Dong} H.,  {Wang} Q.~D.,    {Morris} M.~R.,  2012, MNRAS, 425, 884

\bibitem[\protect\citeauthoryear{{Dray}, {Dale}, {Beer}, {Napiwotzki} \&
  {King}}{{Dray} et~al.}{2005}]{dray05}
{Dray} L.~M.,  {Dale} J.~E.,  {Beer} M.~E.,  {Napiwotzki} R.,    {King} A.~R.,
  2005, MNRAS, 364, 59

\bibitem[\protect\citeauthoryear{{Dray}, {Tout}, {Karakas} \&
  {Lattanzio}}{{Dray} et~al.}{2003}]{dray03}
{Dray} L.~M.,  {Tout} C.~A.,  {Karakas} A.~I.,    {Lattanzio} J.~C.,  2003,
  MNRAS, 338, 973

\bibitem[\protect\citeauthoryear{{Drew} et al.(2004)}{{Drew} et~al.}{2004}]{drew04} Drew, J.~E., Barlow, 
M.~J., Unruh, Y.~C., Parker, Q.~A., Wesson R., Pierce, M.~J., Masheder, M.~R.~W., Phillipps, S., 2004, MNRAS, 351, 206 

\bibitem[\protect\citeauthoryear{{Drissen}, {Lamontagne}, {Moffat}, {Bastien}
  \& {Seguin}}{{Drissen} et~al.}{1986}]{drissen86}
{Drissen} L.,  {Lamontagne} R.,  {Moffat} A.~F.~J.,  {Bastien} P.,    {Seguin}
  M.,  1986, ApJ, 304, 188

\bibitem[\protect\citeauthoryear{{Eikenberry}, {Matthews}, {LaVine}, {Garske},
  {Hu}, {Jackson}, {Patel}, {Barry}, {Colonno}, {Houck}, {Wilson}, {Corbel} \&
  {Smith}}{{Eikenberry} et~al.}{2004}]{eikenberry04}
{Eikenberry} S.~S. et al., 2004, ApJ, 616, 506

\bibitem[\protect\citeauthoryear{{Eldridge}, {Izzard} \& {Tout}}{{Eldridge}
  et~al.}{2008}]{eldridge08}
{Eldridge} J.~J.,  {Izzard} R.~G.,    {Tout} C.~A.,  2008, MNRAS, 384, 1109

\bibitem[\protect\citeauthoryear{{Eldridge} \& {Vink}}{{Eldridge} \&
  {Vink}}{2006}]{eldridge06}
{Eldridge} J.~J.,  {Vink} J.~S.,  2006, A\&A, 452, 295

\bibitem[\protect\citeauthoryear{{Eldridge} et al.(2013)}{{Eldridge} et~al.}{2013}]{eldridge13} Eldridge, J.~J., 
Fraser, M., Smartt, S.~J., Maund, J.~R., \& Crockett, R.~M., 2013, MNRAS, 436, 774 

\bibitem[\protect\citeauthoryear{{Espinoza}, {Selman} \& {Melnick}}{{Espinoza}
  et~al.}{2009}]{espinoza09}
{Espinoza} P.,  {Selman} F.~J.,    {Melnick} J.,  2009, A\&A, 501, 563

\bibitem[\protect\citeauthoryear{{Esteban} \& {Peimbert}}{{Esteban} \&
  {Peimbert}}{1995}]{esteban95}
{Esteban} C.,  {Peimbert} M.,  1995, A\&A, 300, 78

\bibitem[\protect\citeauthoryear{{Esteban} \& {Rosado}}{{Esteban} \&
  {Rosado}}{1995}]{esteban95b}
{Esteban} C.,  {Rosado} M.,  1995, A\&A, 304, 491

\bibitem[\protect\citeauthoryear{{Fahed} \&  {Moffat}}{{Fahed} \& {Moffat}}{2012}]{fahed12}
{Fahed}, R., {Moffat}, A.~F.~J., 2012, MNRAS 424, 1601

\bibitem[\protect\citeauthoryear{{Fahed} et al.(2011)}{{Fahed} et~al.}{2011}]{fahed11} Fahed, R., Moffat, 
A.~F.~J., Zorec, J., et al., 2011, MNRAS, 418, 2 

\bibitem[\protect\citeauthoryear{{Faherty}, {Shara}, {Zurek}, {Kanarek}, \& {Moffat}}{{Faherty} et al.}{2014}]{faherty14}
{Faherty}, J.~K., {Shara}, M.~M., {Zurek}, D., {Kanarek}, G., {Moffat}, A.~F.~J. 2014, AJ 147, 115

\bibitem[\protect\citeauthoryear{{Fang}, {van Boekel}, {King}, {Henning},
  {Bouwman}, {Doi}, {Okamoto}, {Roccatagliata} \& {Sicilia-Aguilar}}{{Fang}
  et~al.}{2012}]{fang12}
{Fang} M., et~al., 2012, A\&A, 539, A119

\bibitem[\protect\citeauthoryear{{Ferri{\`e}re}}{{Ferri{\`e}re}}{2008}]{ferriere08}
{Ferri{\`e}re} K.,  2008, Astronomische Nachrichten, 329, 992

\bibitem[\protect\citeauthoryear{{Figer}, {McLean} \& {Morris}}{{Figer}
  et~al.}{1999}]{figer99}
{Figer} D.~F.,  {McLean} I.~S.,    {Morris} M.,  1999, ApJ, 514, 202

\bibitem[\protect\citeauthoryear{{Figer}, {McLean} \& {Najarro}}{{Figer}
  et~al.}{1997}]{figer97}
{Figer} D.~F.,  {McLean} I.~S.,    {Najarro} F.,  1997, ApJ, 486, 420

\bibitem[\protect\citeauthoryear{{Fritz}, {Gillessen}, {Dodds-Eden}, {Lutz},
  {Genzel}, {Raab}, {Ott}, {Pfuhl}, {Eisenhauer} \& {Yusef-Zadeh}}{{Fritz}
  et~al.}{2011}]{fritz11}
{Fritz} T.~K. et al., 2011, ApJ, 737, 73

\bibitem[\protect\citeauthoryear{{Fujii} \& {Portegies Zwart}}{{Fujii} \&
  {Portegies Zwart}}{2011}]{fujii11}
{Fujii} M.~S.,  {Portegies Zwart} S.,  2011, Science, 334, 1380

\bibitem[\protect\citeauthoryear{{Gamen}, {Gosset}, {Morrell}, {Niemela}, {Sana}, {Naz{\'e}}, {Rauw}, {Barb{\'a}} \& {Solivella}}{Gamen et~al.}{2006}]{gamen06} {Gamen} R., et~al. 2006, A\&A, 460, 777 

\bibitem[\protect\citeauthoryear{{Gamen} et 
al.(2009)}{{Gamen} et al.}{2009}]{gamen09} Gamen, R.~C., Fern{\'a}ndez-Laj{\'u}s, E., Niemela, V.~S., \& Barb{\'a}, R.~H., 2009, A\&A, 506, 1269 

\bibitem[\protect\citeauthoryear{{Gamen} et~
al.(2014)}{{Gamen} et~al.}{2014}]{gamen14} Gamen, R.~C, Collado, A., Barb{\'a}, R., Chen{\'e}, A.-N., \& St-Louis, N., 2014, A\&A, 562, AA13 

\bibitem[\protect\citeauthoryear{{Garmany} \& {Stencel}}{{Garmany} \&
  {Stencel}}{1992}]{garmany92}
{Garmany} C.~D.,  {Stencel} R.~E.,  1992, A\&AS, 94, 211

\bibitem[\protect\citeauthoryear{{Georgy}, {Ekstr{\"o}m}, {Meynet}, {Massey},
  {Levesque}, {Hirschi}, {Eggenberger} \& {Maeder}}{{Georgy}
  et~al.}{2012}]{georgy12}
{Georgy} C.,  {Ekstr{\"o}m} S.,  {Meynet} G.,  {Massey} P.,  {Levesque} E.~M.,
  {Hirschi} R.,  {Eggenberger} P.,    {Maeder} A.,  2012, A\&A, 542, A29

\bibitem[\protect\citeauthoryear{{Gillessen}, {Eisenhauer}, {Fritz}, {Pfuhl},
  {Ott} \& {Genzel}}{{Gillessen} et~al.}{2013}]{gillessen13}
{Gillessen} S.,  {Eisenhauer} F.,  {Fritz} T.~K.,  {Pfuhl} O.,  {Ott} T.,
  {Genzel} R.,  2013, in {de Grijs} R.,  ed., IAU Symposium Vol.~289,
pp 29--35

\bibitem[\protect\citeauthoryear{{Gosset} et al.(2001)}{{Gosset} et~al.}{2001}]{gosset01} Gosset, E., Royer, P., 
Rauw, G., Manfroid, J., \& Vreux, J.-M., 2001, MNRAS, 327, 435 

\bibitem[\protect\citeauthoryear{{Groh}, {Meynet}, {Ekstr{\"o}m} \& {Georgy}}
  {{Groh} et~al.}{2014}]{groh14} 
  {Groh} J.~H., {Meynet} G., {Ekstr{\"o}m} S, {Georgy} C., 2014, A\&A, 564, A30

\bibitem[\protect\citeauthoryear{{Gvaramadze} et al.(2009)}{{Gvaramadze} et al.}{2009}]{gvaramadze09}
{Gvaramadze}, V.~V., et~al., 2009, MNRAS 400, 524

\bibitem[\protect\citeauthoryear{{Gvaramadze}, {Kniazev}, {Hamann},
  {Berdnikov}, {Fabrika} \& {Valeev}}{{Gvaramadze} et~al.}{2010}]{gvaramadze10}
{Gvaramadze} V.~V.,  {Kniazev} A.~Y.,  {Hamann} W.-R.,  {Berdnikov} L.~N.,
  {Fabrika} S.,    {Valeev} A.~F.,  2010, MNRAS, 403, 760

\bibitem[\protect\citeauthoryear{{Hadfield}, {van Dyk}, {Morris}, {Smith},
  {Marston} \& {Peterson}}{{Hadfield} et~al.}{2007}]{hadfield07}
{Hadfield} L.~J.,  {van Dyk} S.~D.,  {Morris} P.~W.,  {Smith} J.~D.,  {Marston}
  A.~P.,    {Peterson} D.~E.,  2007, MNRAS, 376, 248

\bibitem[\protect\citeauthoryear{{Hamann}, {Gr{\"a}fener} \&
  {Liermann}}{{Hamann} et~al.}{2006}]{hamann06}
{Hamann} W.-R.,  {Gr{\"a}fener} G.,    {Liermann} A.,  2006, A\&A, 457, 1015

\bibitem[\protect\citeauthoryear{{Hanson}, {Kurtev}, {Borissova}, {Georgiev},
  {Ivanov}, {Hillier} \& {Minniti}}{{Hanson} et~al.}{2010}]{hanson10}
{Hanson} M.~M.,  {Kurtev} R.,  {Borissova} J.,  {Georgiev} L.,  {Ivanov} V.~D.,
   {Hillier} D.~J.,    {Minniti} D.,  2010, A\&A, 516, A35

\bibitem[\protect\citeauthoryear{{Harayama}, {Eisenhauer} \&
  {Martins}}{{Harayama} et~al.}{2008}]{harayama08}
{Harayama} Y.,  {Eisenhauer} F.,    {Martins} F.,  2008, ApJ, 675, 1319

\bibitem[\protect\citeauthoryear{{Hill}, {Moffat} \& {St-Louis}}{{Hill} et~al.}{2002}]{hill02} Hill, G.~M., Moffat, 
A.~F.~J., \& St-Louis, N., 2002, MNRAS, 335, 1069 

\bibitem[\protect\citeauthoryear{{Hillenbrand}, {Massey}, {Strom} \&
  {Merrill}}{{Hillenbrand} et~al.}{1993}]{hillenbrand93}
{Hillenbrand} L.~A.,  {Massey} P.,  {Strom} S.~E.,    {Merrill} K.~M.,  1993,
  AJ, 106, 1906

\bibitem[\protect\citeauthoryear{{Homeier}, {Blum}, {Pasquali}, {Conti} \&
  {Damineli}}{{Homeier} et~al.}{2003}]{homeier03}
{Homeier} N.~L.,  {Blum} R.~D.,  {Pasquali} A.,  {Conti} P.~S.,    {Damineli}
  A.,  2003, A\&A, 408, 153

\bibitem[\protect\citeauthoryear{{Hopewell} et al.(2005)}{{Hopewell} et~al.}{2005}]{hopewell05} Hopewell, E.~C., 
Barlow, M.~J., Drew, J.~E., et al., 2005, MNRAS, 363, 857 

\bibitem[\protect\citeauthoryear{{Hopkins}, {Quataert} \& {Murray}}{{Hopkins}
  et~al.}{2011}]{hopkins11}
{Hopkins} P.~F.,  {Quataert} E.,    {Murray} N.,  2011, MNRAS, 417, 950

\bibitem[\protect\citeauthoryear{{Howarth} \& {Schmutz}}{{Howarth} \&
  {Schmutz}}{1992}]{howarth92}
{Howarth} I.~D.,  {Schmutz} W.,  1992, A\&A, 261, 503

\bibitem[\protect\citeauthoryear{{Howarth} \& {Schmutz}}{{Howarth} \&
  {Schmutz}}{1995}]{howarth95}
{Howarth} I.~D.,  {Schmutz} W.,  1995, A\&A, 294, 529

\bibitem[\protect\citeauthoryear{{Humphreys} \& {Larsen}}{{Humphreys} \&
  {Larsen}}{1995}]{humphreys95}
{Humphreys} R.~M.,  {Larsen} J.~A.,  1995, AJ, 110, 2183

\bibitem[\protect\citeauthoryear{{Hur}, {Sung} \& {Bessell}}{{Hur}
  et~al.}{2012}]{hur12}
{Hur} H.,  {Sung} H.,    {Bessell} M.~S.,  2012, AJ, 143, 41

\bibitem[\protect\citeauthoryear{{Hyodo}, {Tsujimoto}, {Koyama} et al.}{{Hyodo} et al.}{2008}]{hyodo08}
{Hyodo}, Y.  {Tsujimoto}, M.  {Koyama}, K.  {Nishiyama}, S.  
	{Nagata}, T.  {Sakon}, I.  {Murakami}, H.  {Matsumoto}, H. 2008, PASJ 60, 173

\bibitem[\protect\citeauthoryear{{Indebetouw}, {Mathis}, {Babler}, {Meade},
  {Watson}, {Whitney}, {Wolff}, {Wolfire} \& {Cohen}}{{Indebetouw}
  et~al.}{2005}]{indebetouw05}
{Indebetouw} R.,  {Mathis} J.~S.,  {Babler} B.~L.,  {Meade} M.~R.,  {Watson}
  C.,  {Whitney} B.~A.,  {Wolff} M.~J.,  {Wolfire} M.~G.,    {Cohen} M.,
  2005, ApJ, 619, 931

\bibitem[\protect\citeauthoryear{{Isserstedt}, {Moffat} \&
  {Niemela}}{{Isserstedt} et~al.}{1983}]{isserstedt83}
{Isserstedt} J.,  {Moffat} A.~F.~J.,    {Niemela} V.~S.,  1983, A\&A, 126, 183

\bibitem[\protect\citeauthoryear{{Jordi}, {Gebran}, {Carrasco}, {de Bruijne},
  {Voss}, {Fabricius}, {Knude}, {Vallenari}, {Kohley} \& {Mora}}{{Jordi}
  et~al.}{2010}]{jordi10}
{Jordi} C., et~al., 2010, A\&A, 523, A48

\bibitem[\protect\citeauthoryear{{Kanarek}, {Shara}, {Faherty}, {Zurek}, \& {Moffat}}
{{Kanarek} et al.}{2014}]{kanarek14}
{Kanarek}, G.~C., {Shara}, M.~M., {Faherty}, J.~K., {Zurek}, D., {Moffat}, A.~F.~J. 2014, AJ, submitted, arXiv:1403.0975


\bibitem[\protect\citeauthoryear{{Kendrew}, {Simpson}, {Bressert}, {Povich},
  {Sherman}, {Lintott}, {Robitaille}, {Schawinski} \& {Wolf-Chase}}{{Kendrew}
  et~al.}{2012}]{kendrew12}
{Kendrew} S., et~al., 2012, ApJ, 755, 71

\bibitem[\protect\citeauthoryear{{Kothes} \& {Dougherty}}{{Kothes} \&
  {Dougherty}}{2007}]{kothes07}
{Kothes} R.,  {Dougherty} S.~M.,  2007, A\&A, 468, 993

\bibitem[\protect\citeauthoryear{{Koumpia} \& {Bonanos}}{{Koumpia} \&
  {Bonanos}}{2012}]{koumpia12}
{Koumpia} E.,  {Bonanos} A.~Z.,  2012, A\&A, 547, A30

\bibitem[\protect\citeauthoryear{{Kroupa} \& {Weidner}}{{Kroupa} \&
  {Weidner}}{2003}]{kroupa03}
{Kroupa} P.,  {Weidner} C.,  2003, ApJ, 598, 1076

\bibitem[\protect\citeauthoryear{{Kurtev}, {Borissova}, {Georgiev}, {Ortolani}
  \& {Ivanov}}{{Kurtev} et~al.}{2007}]{kurtev07}
{Kurtev} R.,  {Borissova} J.,  {Georgiev} L.,  {Ortolani} S.,    {Ivanov}
  V.~D.,  2007, A\&A, 475, 209

\bibitem[\protect\citeauthoryear{{Lada} \& {Lada}}{{Lada} \& {Lada}}{2003}]{lada03}
{Lada} C.~J., {Lada}, E.~A. 2003, ARA\&A 41, 57

\bibitem[\protect\citeauthoryear{{Lejeune} \& {Schaerer}}{{Lejeune} \&
  {Schaerer}}{2001}]{lejeune01}
{Lejeune} T.,  {Schaerer} D.,  2001, A\&A, 366, 538


\bibitem[\protect\citeauthoryear{{L{\'e}pine} et al.(2001)}{{L{\'e}pine} et~al.}{2001}]{lepine01} L{\'e}pine, S., 
Wallace, D., Shara, M.~M., Moffat, A.~F.~J., 
\& Niemela, V.~S., 2001, AJ, 122, 3407 


\bibitem[\protect\citeauthoryear{{Liermann}, {Hamann} \& {Oskinova}}{{Liermann}
  et~al.}{2009}]{liermann09}
{Liermann} A.,  {Hamann} W.-R.,    {Oskinova} L.~M.,  2009, A\&A, 494, 1137

\bibitem[\protect\citeauthoryear{{Littlefield}, {Garnavich}, {''Howie''
  Marion}, {Vink{\'o}}, {McClelland}, {Rettig} \& {Wheeler}}{{Littlefield}
  et~al.}{2012}]{littlefield12}
{Littlefield} C.,  {Garnavich} P.,  {''Howie'' Marion} G.~H.,  {Vink{\'o}} J.,
  {McClelland} C.,  {Rettig} T.,    {Wheeler} J.~C.,  2012, AJ, 143, 136

\bibitem[\protect\citeauthoryear{{Longmore}, {Bally}, {Testi}, {Purcell},
  {Walsh}, {Bressert}, {Pestalozzi}, {Molinari}, {Ott}, {Cortese}, {Battersby},
  {Murray}, {Lee}, {Kruijssen}, {Schisano} \& {Elia}}{{Longmore}
  et~al.}{2013}]{longmore13}
{Longmore} S.~N., et~al., 2013, MNRAS, 429, 987

\bibitem[\protect\citeauthoryear{{Lortet}, {Testor} \& {Niemela}}{{Lortet}
  et~al.}{1984}]{lortet84}
{Lortet} M.~C.,  {Testor} G.,    {Niemela} V.,  1984, A\&A, 140, 24

\bibitem[\protect\citeauthoryear{{Lundstr{\"o}m} \& {Stenholm}}{{Lundstr{\"o}m}
  \& {Stenholm}}{1984}]{lundstrom84}
{Lundstr{\"o}m} I.,  {Stenholm} B.,  1984, A\&As, 58, 163

\bibitem[\protect\citeauthoryear{{Neugent}, {Massey} \& {Morell}}{{Neugent} et~al.}{2012}]
{neugent12} Neugent, K.~F., Massey, 
P., \& Morrell, N., 2012, AJ, 144, 162 

\bibitem[\protect\citeauthoryear{{Maeder} \& {Lequeux}}{{Maeder} \&
  {Lequeux}}{1982}]{maeder82}
{Maeder} A.,  {Lequeux} J.,  1982, A\&A, 114, 409

\bibitem[\protect\citeauthoryear{{Maeder} \& {Meynet}}{{Maeder} \&
  {Meynet}}{2000}]{maeder00}
{Maeder} A.,  {Meynet} G.,  2000, ARA\&A, 38, 143

\bibitem[\protect\citeauthoryear{{Ma{\'{\i}}z Apell{\'a}niz}, {Walborn},
  {Morrell}, {Niemela} \& {Nelan}}{{Ma{\'{\i}}z Apell{\'a}niz}
  et~al.}{2007}]{maiz07}
{Ma{\'{\i}}z Apell{\'a}niz} J.,  {Walborn} N.~R.,  {Morrell} N.~I.,  {Niemela}
  V.~S.,    {Nelan} E.~P.,  2007, ApJ, 660, 1480

\bibitem[\protect\citeauthoryear{{Malchenko} \& {Tarasov}}{{Malchenko} \&
  {Tarasov}}{2009}]{malchenko09}
{Malchenko} S.~L.,  {Tarasov} A.~E.,  2009, Astrophysics, 52, 235

\bibitem[\protect\citeauthoryear{{Marchenko}, {Moffat} \&
  {Crowther}}{{Marchenko} et~al.}{2010}]{marchenko10}
{Marchenko} S.~V.,  {Moffat} A.~F.~J.,    {Crowther} P.~A.,  2010, ApJL, 724,
  L90

\bibitem[\protect\citeauthoryear{{Marchenko}, {Moffat}, {Crowther},
  {Chen{\'e}}, {De Serres}, {Eenens}, {Hill}, {Moran} \& {Morel}}{{Marchenko}
  et~al.}{2004}]{marchenko04}
{Marchenko} S.~V. et al., 2004, MNRAS, 353, 153


\bibitem[\protect\citeauthoryear{{Marchenko}, {Moffat} \& {Eenens}}{{Marchenko}
  et~al.}{1998}]{marchenko98}
{Marchenko} S.~V.,  {Moffat} A.~F.~J.,    {Eenens} P.~R.~J.,  1998, PASP, 110,
  1416

\bibitem[\protect\citeauthoryear{{Marchenko}, {Moffat} \&
  {Koenigsberger}}{{Marchenko} et~al.}{1994}]{marchenko94}
{Marchenko} S.~V.,  {Moffat} A.~F.~J.,    {Koenigsberger} G.,  1994, ApJ, 422,
  810

\bibitem[\protect\citeauthoryear{{Marston}, {Mauerhan}, {Van Dyk}, {Cohen} \& {Morris}}{{Marston} et~al.}{2013}]{marston13}
{Marston}, A., {Mauerhan}, J.~C.m {Van Dyk}, S., {Cohen}, M., {Morris}, P. 2013, in: {\it Massive Stars: From Alpha to Omega}, Rhodes, Greece, arXiv:1309.1584

\bibitem[\protect\citeauthoryear{{Mart{\'{\i}}n}, {Cappa} \&
  {Testori}}{{Mart{\'{\i}}n} et~al.}{2007}]{martin07}
{Mart{\'{\i}}n} M.~C.,  {Cappa} C.~E.,    {Testori} J.~C.,  2007, Rev. Mex.
  Astron. Astrofis., 43, 243

\bibitem[\protect\citeauthoryear{{Martins}, {Hillier}, {Paumard}, {Eisenhauer},
  {Ott} \& {Genzel}}{{Martins} et~al.}{2008}]{martins08}
{Martins} F.,  {Hillier} D.~J.,  {Paumard} T.,  {Eisenhauer} F.,  {Ott} T.,
  {Genzel} R.,  2008, A\&A, 478, 219

\bibitem[\protect\citeauthoryear{{Martins} \& {Plez}}{{Martins} \&
  {Plez}}{2006}]{martins06}
{Martins} F.,  {Plez} B.,  2006, A\&A, 457, 637

\bibitem[\protect\citeauthoryear{{Martins}, {Schaerer} \& {Hillier}}{{Martins}
  et~al.}{2005}]{martins05}
{Martins} F.,  {Schaerer} D.,    {Hillier} D.~J.,  2005, A\&A, 436, 1049

\bibitem[\protect\citeauthoryear{{Massey} \& {Conti}}{{Massey} \&
  {Conti}}{1983}]{massey83}
{Massey} P.,  {Conti} P.~S.,  1983, ApJ, 273, 576

\bibitem[\protect\citeauthoryear{{Massey} \& {Duffy}}{{Massey} \&
  {Duffy}}{2001{\natexlab{b}}}]{massey01b}
{Massey} P.,  {Duffy} A.~S.,  2001{\natexlab{b}}, ApJ, 550, 713

\bibitem[\protect\citeauthoryear{{Massey}, {DeGioia-Eastwood} \&
  {Waterhouse}}{{Massey} et~al.}{2001\natexlab{a}}]{massey01a}
{Massey} P.,  {DeGioia-Eastwood} K.,    {Waterhouse} E.,  2001\natexlab{a}, AJ, 121, 1050

\bibitem[\protect\citeauthoryear{{Massey}, {Olsen} \& {Parker}}{{Massey}
  et~al.}{2003}]{massey03}
{Massey} P.,  {Olsen} K.~A.~G.,    {Parker} J.~W.,  2003, PASP, 115, 1265

\bibitem[\protect\citeauthoryear{{Massey}, {Neugent}, {Morrell} \& {Hillier}}{{Massey} et~al.}{2014}]
{massey14} Massey, P., Neugent, 
K.~F., Morrell, N., \& Hillier, D.~J., 2014, ApJ, 788, 83 

\bibitem[\protect\citeauthoryear{{Mauerhan}, {van Dyk} \& {Morris}}{{Mauerhan}
  et~al.}{2009}]{mauerhan09}
{Mauerhan} J.~C.,  {van Dyk} S.~D.,    {Morris} P.~W.,  2009, PASP, 121, 591

\bibitem[\protect\citeauthoryear{{Mauerhan}, {Muno}, {Morris}, {Stolovy} \&
  {Cotera}}{{Mauerhan} et~al.}{2010a}]{mauerhan10a}
{Mauerhan} J.~C.,  {Muno} M.~P.,  {Morris} M.~R.,  {Stolovy} S.~R.,    {Cotera}
  A.,  2010, ApJ, 710, 706

\bibitem[\protect\citeauthoryear{{Mauerhan}, {Wachter}, {Morris}, {Van Dyk} \&
  {Hoard}}{{Mauerhan} et~al.}{2010\natexlab{b}}]{mauerhan10b}
{Mauerhan} J.~C.,  {Wachter} S.,  {Morris} P.~W.,  {Van Dyk} S.~D.,    {Hoard}
  D.~W.,  2010\natexlab{b}, ApJL, 724, L78
  
\bibitem[\protect\citeauthoryear{{Mauerhan}, {Cotera}, {Dong}, {Morris},
  {Wang}, {Stolovy} \& {Lang}}{{Mauerhan} et~al.}{2010\natexlab{c}}]{mauerhan10c}
{Mauerhan} J.~C.,  {Cotera} A.,  {Dong} H.,  {Morris} M.~R.,  {Wang} Q.~D.,
  {Stolovy} S.~R., {Lang} C.,  2010\natexlab{c}, ApJ, 725, 188

\bibitem[\protect\citeauthoryear{{Mauerhan}, {Van Dyk} \& {Morris}}{{Mauerhan}
  et~al.}{2011}]{mauerhan11}
{Mauerhan} J.~C.,  {Van Dyk} S.~D.,    {Morris} P.~W.,  2011, AJ, 142, 40


\bibitem[\protect\citeauthoryear{{Melena}, {Massey}, {Morrell} \&
  {Zangari}}{{Melena} et~al.}{2008}]{melena09}
{Melena} N.~W.,  {Massey} P.,  {Morrell} N.~I.,    {Zangari} A.~M.,  2008, AJ,
  135, 878

\bibitem[\protect\citeauthoryear{{Mel'Nik} \& {Dambis}}{{Mel'Nik} \&
  {Dambis}}{2009}]{melnik09}
{Mel'Nik} A.~M.,  {Dambis} A.~K.,  2009, MNRAS, 400, 518

\bibitem[\protect\citeauthoryear{{Messineo}, {Davies}, {Figer}, {Kudritzki},
  {Valenti}, {Trombley}, {Najarro} \& {Rich}}{{Messineo}
  et~al.}{2011}]{messineo11}
{Messineo} M.,  {Davies} B.,  {Figer} D.~F.,  {Kudritzki} R.~P.,  {Valenti} E.,
   {Trombley} C.,  {Najarro} F.,    {Rich} R.~M.,  2011, ApJ, 733, 41

\bibitem[\protect\citeauthoryear{{Messineo}, {Davies}, {Ivanov}, {Figer},
  {Schuller}, {Habing}, {Menten} \& {Petr-Gotzens}}{{Messineo}
  et~al.}{2009}]{messineo09}
{Messineo} M.,  {Davies} B.,  {Ivanov} V.~D.,  {Figer} D.~F.,  {Schuller} F.,
  {Habing} H.~J.,  {Menten} K.~M.,    {Petr-Gotzens} M.~G.,  2009, ApJ, 697,
  701

\bibitem[\protect\citeauthoryear{{Meynet} \& {Maeder}}{{Meynet} \&
  {Maeder}}{2005}]{meynet05}
{Meynet} G.,  {Maeder} A.,  2005, A\&A, 429, 581

\bibitem[\protect\citeauthoryear{{Mikles} et al.}{{Mikles} et al.}{2006}]{mikles06}
{Mikles}, V.~J., {Eikenberry}, S.~S., {Muno}, M.~P., 
	{Bandyopadhyay}, R.~M., {Patel}, S. 2006, ApJ 651, 408

\bibitem[\protect\citeauthoryear{{Miszalski} et al.}{{Miszalski}
  et~al.}{2012}]{miszalski12}
{Miszalski} B.,  {Crowther} P.~A.,  {De Marco} O.,  {K{\"o}ppen} J.,  {Moffat}
  A.~F.~J.,  {Acker} A.,    {Hillwig} T.~C.,  2012, MNRAS, 423, 934

\bibitem[\protect\citeauthoryear{{Moffat}}{{Moffat}}{1989}]{moffat89} Moffat, A.~F.~J., 1989, ApJ, 
347, 373 


\bibitem[\protect\citeauthoryear{{Moffat} \& {Shara}}{{Moffat} \& {Shara}}{1983}]{moffat83} {Moffat}, A.~F.~J. {Shara}, M.~M., 1983, ApJ 273, 544

\bibitem[\protect\citeauthoryear{{Moffat}, {Lamontagne} \& {Seggewiss}}{{Moffat} et~al.}{1982}]
{moffat82} Moffat, A.~F.~J., Lamontagne, R., \& Seggewiss, W., 1982, A\&A, 114, 135 

\bibitem[\protect\citeauthoryear{{Moffat}, {Shara} \& {Potter}}{{Moffat}
  et~al.}{1991}]{moffat91}
{Moffat} A.~F.~J.,  {Shara} M.~M.,    {Potter} M.,  1991, AJ, 102, 642

\bibitem[\protect\citeauthoryear{{Monnier} et al.}{{Monnier} et al.}{2011}]{monnier11}
{Monnier} J.~D., {Zhao} M., {Pedretti} E. et al. 2011, ApJ, 742, 1


\bibitem[\protect\citeauthoryear{{Morris}, {Brownsberger}, {Conti}, {Massey} \&
  {Vacca}}{{Morris} et~al.}{1993}]{morris93}
{Morris} P.~W.,  {Brownsberger} K.~R.,  {Conti} P.~S.,  {Massey} P.,    {Vacca}
  W.~D.,  1993, ApJ, 412, 324
  
\bibitem[\protect\citeauthoryear{{Motch} et al.}{{Motch} et al.}{2010}]{Motch10}
{Motch}, C., {Warwick}, R., {Cropper}, M.~S. et al. 2010, A\&A 523, A92

\bibitem[\protect\citeauthoryear{{Nakanishi} \& {Sofue}}{{Nakanishi} \&
  {Sofue}}{2003}]{nakanishi03}
{Nakanishi} H.,  {Sofue} Y.,  2003, PASJ, 55, 191

\bibitem[\protect\citeauthoryear{{Nakanishi} \& {Sofue}}{{Nakanishi} \&
  {Sofue}}{2006}]{nakanishi06}
{Nakanishi} H.,  {Sofue} Y.,  2006, PASJ, 58, 847

\bibitem[\protect\citeauthoryear{{Nishiyama}, {Tamura}, {Hatano}, {Kato},
  {Tanab{\'e}}, {Sugitani} \& {Nagata}}{{Nishiyama} et~al.}{2009}]{nishiyama09}
{Nishiyama} S.,  {Tamura} M.,  {Hatano} H.,  {Kato} D.,  {Tanab{\'e}} T.,
  {Sugitani} K.,    {Nagata} T.,  2009, ApJ, 696, 1407

\bibitem[\protect\citeauthoryear{{Oskinova}}{{Oskinova}}{2005}]{oskinova05}
{Oskinova} L.~M.,  2005, MNRAS, 361, 679

\bibitem[\protect\citeauthoryear{{Paladini}, {Davies} \& {De Zotti}}{{Paladini}
  et~al.}{2004}]{paladini04}
{Paladini} R.,  {Davies} R.~D.,    {De Zotti} G.,  2004, MNRAS, 347, 237

\bibitem[\protect\citeauthoryear{{Parker}, {Phillipps}, {Pierce}, {Hartley},
  {Hambly}, {Read}, {MacGillivray}, {Tritton} \& {Cass}}{{Parker}
  et~al.}{2005}]{parker05}
{Parker} Q.~A., et~al., 2005, MNRAS, 362, 689

\bibitem[\protect\citeauthoryear{{Pasquali} et 
al.(2002)}{{Pasquali} et~al}{2002}]{pasquali02} Pasquali, A., Comer{\'o}n, F., Gredel, R., Torra, J., \& Figueras, F., 2002, A\&A, 396, 533 

\bibitem[\protect\citeauthoryear{{Penny} \& {Gies}}{{Penny} \&
  {Gies}}{2009}]{penny09}
{Penny} L.~R.,  {Gies} D.~R.,  2009, ApJ, 700, 844

\bibitem[\protect\citeauthoryear{{Pollock}, {Haberl} \& {Corcoran}}{{Pollock}
  et~al.}{1995}]{pollock95}
{Pollock} A.~M.~T.,  {Haberl} F.,    {Corcoran} M.~F.,  1995, in {van der
  Hucht} K.~A.,  {Williams} P.~M.,  eds, Wolf-Rayet Stars: Binaries; Colliding
  Winds; Evolution Vol.~163 of IAU Symposium, {The ROSAT PSPC survey of the
  Wolf-Rayet stars}.
p.~512

\bibitem[\protect\citeauthoryear{{Poveda}, {Ruiz} \& {Allen}}{{Poveda}
  et~al.}{1967}]{poveda67}
{Poveda} A., {Ruiz} J., {Allen} C., 1967, Bol. Obs. Tonantzintla Tacubaya, 4, 86

\bibitem[\protect\citeauthoryear{{Przybilla}, {Butler}, {Becker} \&
  {Kudritzki}}{{Przybilla} et~al.}{2006}]{przybilla06}
{Przybilla} N.,  {Butler} K.,  {Becker} S.~R.,    {Kudritzki} R.~P.,  2006,
  A\&A, 445, 1099

\bibitem[\protect\citeauthoryear{{Puls}, {Vink} \& {Najarro}}{{Puls}
  et~al.}{2008}]{puls08}
{Puls} J.,  {Vink} J.~S.,    {Najarro} F.,  2008, A\&ARv, 16, 209

\bibitem[\protect\citeauthoryear{{Rahman}, {Moon}, {Matzner}}
{{Rahman}, {Moon}, \& {Matzner}}{2011}]{rahman11}
{Rahman}, M., {Moon}, D.-S., {Matzner}, C.~D. 2011, ApJ 743, L28

\bibitem[\protect\citeauthoryear{{Rauw}, {Crowther}, {De Becker}, {Gosset},
  {Naz{\'e}}, {Sana}, {van der Hucht}, {Vreux} \& {Williams}}{{Rauw}
  et~al.}{2005}]{rauw05}
{Rauw} G., et~al., 2005, A\&A, 432, 985

\bibitem[\protect\citeauthoryear{{Rauw}, {Manfroid}, {Gosset}, {Naz{\'e}},
  {Sana}, {De Becker}, {Foellmi} \& {Moffat}}{{Rauw} et~al.}{2007}]{rauw07}
{Rauw} G.,  {Manfroid} J.,  {Gosset} E.,  {Naz{\'e}} Y.,  {Sana} H.,  {De
  Becker} M.,  {Foellmi} C.,    {Moffat} A.~F.~J.,  2007, A\&A, 463, 981

\bibitem[\protect\citeauthoryear{{Rauw}, {Sana} \& {Naz{\'e}}}{{Rauw}
  et~al.}{2011}]{rauw11}
{Rauw} G.,  {Sana} H.,    {Naz{\'e}} Y.,  2011, A\&A, 535, A40

\bibitem[\protect\citeauthoryear{{Reid}, {Menten}, {Zheng}, {Brunthaler} \&
  {Xu}}{{Reid} et~al.}{2009}]{reid09}
{Reid} M.~J.,  {Menten} K.~M.,  {Zheng} X.~W.,  {Brunthaler} A.,    {Xu} Y.,
  2009, ApJ, 705, 1548

\bibitem[\protect\citeauthoryear{{Reid}, {McClintock}, {Narayan}, {Gou},
  {Remillard} \& {Orosz}}{{Reid} et~al.}{2011}]{reid11}
{Reid} M.~J.,  {McClintock} J.~E.,  {Narayan} R.,  {Gou} L.,  {Remillard}
  R.~A.,    {Orosz} J.~A.,  2011, ApJ, 742, 83

\bibitem[\protect\citeauthoryear{{Rieke} \& {Lebofsky}}{{Rieke} \&
  {Lebofsky}}{1985}]{rieke85}
{Rieke} G.~H.,  {Lebofsky} M.~J.,  1985, ApJ, 288, 618

\bibitem[\protect\citeauthoryear{{Rolleston}, {Trundle} \&
  {Dufton}}{{Rolleston} et~al.}{2002}]{rolleston02}
{Rolleston} W.~R.~J.,  {Trundle} C.,    {Dufton} P.~L.,  2002, A\&A, 396, 53

\bibitem[\protect\citeauthoryear{{Rolleston}, {Venn}, {Tolstoy} \&
  {Dufton}}{{Rolleston} et~al.}{2003}]{rolleston03}
{Rolleston} W.~R.~J.,  {Venn} K.,  {Tolstoy} E.,    {Dufton} P.~L.,  2003,
  A\&A, 400, 21

\bibitem[\protect\citeauthoryear{{Roman-Lopes}}{{Roman-Lopes}}{2011\natexlab{b}}]{roman-lopes11b}
{Roman-Lopes} A.,  2011b, ISRNA\&A, 2011\natexlab{E}, 8

\bibitem[\protect\citeauthoryear{{Roman-Lopes}}{{Roman-Lopes}}{2011\natexlab{c}}]{roman-lopes11c}
{Roman-Lopes} A.,  2011\natexlab{c}, MNRAS, 410, 161

\bibitem[\protect\citeauthoryear{{Roman-Lopes}}{{Roman-Lopes}}{2012}]{roman-lopes12}
{Roman-Lopes} A.,  2012, MNRAS, 427, L65

\bibitem[\protect\citeauthoryear{{Roman-Lopes}}{{Roman-Lopes}}{2013}]{roman-lopes13}
{Roman-Lopes} A.,  2013, MNRAS, 433, 712

\bibitem[\protect\citeauthoryear{{Roman-Lopes}, {Barba}, \& {Morrell}}{{Roman-Lopes} et~al.}{2011\natexlab{a}}]{roman-lopes11a}
{Roman-Lopes} A., {Barba}, R.~H., {Morrell}, N.~I., 2011\natexlab{a}, MNRAS 416, 501

\bibitem[\protect\citeauthoryear{{Rygl}, {Brunthaler}, {Sanna}, {Menten},
  {Reid}, {van Langevelde}, {Honma}, {Torstensson} \& {Fujisawa}}{{Rygl}
  et~al.}{2012}]{rygl12}
{Rygl} K.~L.~J., et~al., 2012, A\&A, 539, A79

\bibitem[\protect\citeauthoryear{{Sana}, {Gosset}, {Rauw}, {Sung} \&
  {Vreux}}{{Sana} et~al.}{2006}]{sana06}
{Sana} H.,  {Gosset} E.,  {Rauw} G.,  {Sung} H.,    {Vreux} J.-M.,  2006, A\&A,
  454, 1047

\bibitem[\protect\citeauthoryear{{Sander}, {Hamann} \& {Todt}}{{Sander}
  et~al.}{2012}]{sander12}
{Sander} A.,  {Hamann} W.-R.,    {Todt} H.,  2012, A\&A, 540, A144

\bibitem[\protect\citeauthoryear{{Schaerer} \& {Vacca}}{{Schaerer} \&
  {Vacca}}{1998}]{schaerer98}
{Schaerer} D.,  {Vacca} W.~D.,  1998, ApJ, 497, 618

\bibitem[\protect\citeauthoryear{{Schnurr}, {Casoli}, {Chen{\'e}}, {Moffat} \&
  {St-Louis}}{{Schnurr} et~al.}{2008}]{schnurr08}
{Schnurr} O.,  {Casoli} J.,  {Chen{\'e}} A.-N.,  {Moffat} A.~F.~J.,
  {St-Louis} N.,  2008, MNRAS, 389, L38
  
\bibitem[\protect\citeauthoryear{{Schweickhardt}, {Schmutz}, {Stahl}, {Szeifert} \& {Wolf}}{{Schweickhardt} et~al.}{1999}]{schweickhardt99}
{Schweickhardt} J.,  {Schmutz} W., {Stahl} O., {Szeifert} Th., {Wolf} B.,  1999, A\&A, 347, 127

\bibitem[\protect\citeauthoryear{{Shara}, {Smith}, {Potter} \&
  {Moffat}}{{Shara} et~al.}{1991}]{shara91}
{Shara} M.~M.,  {Smith} L.~F.,  {Potter} M.,    {Moffat} A.~F.~J.,  1991, AJ,
  102, 716

\bibitem[\protect\citeauthoryear{{Shara}, {Moffat}, {Smith}, {Niemela},
  {Potter} \& {Lamontagne}}{{Shara} et~al.}{1999}]{shara99}
{Shara} M.~M.,  {Moffat} A.~F.~J.,  {Smith} L.~F.,  {Niemela} V.~S.,  {Potter}
  M.,    {Lamontagne} R.,  1999, AJ, 118, 390

\bibitem[\protect\citeauthoryear{{Shara}, {Moffat}, {Gerke}, {Zurek},
  {Stanonik}, {Doyon}, {Artigau}, {Drissen} \& {Villar-Sbaffi}}{{Shara}
  et~al.}{2009}]{shara09}
{Shara} M.~M., et~al., 2009, AJ, 138, 402

\bibitem[\protect\citeauthoryear{{Shara}, {Faherty}, {Zurek}, {Moffat},
  {Gerke}, {Doyon}, {Artigau} \& {Drissen}}{{Shara} et~al.}{2012}]{shara12}
{Shara} M.~M.,  {Faherty} J.~K.,  {Zurek} D.,  {Moffat} A.~F.~J.,  {Gerke} J.,
  {Doyon} R.,  {Artigau} E.,    {Drissen} L.,  2012, AJ, 143, 149

\bibitem[\protect\citeauthoryear{{Shetty} \& {Ostriker}}{{Shetty} \&
  {Ostriker}}{2008}]{shetty08}
{Shetty} R.,  {Ostriker} E.~C.,  2008, ApJ, 684, 978

\bibitem[\protect\citeauthoryear{{Skrutskie}, {Cutri}, {Stiening}, {Weinberg},
  {Schneider}, {Carpenter}, {Beichman} \& {Capps}}{{Skrutskie}
  et~al.}{2006}]{skrutskie06}
{Skrutskie} M.~F.,  {Cutri} R.~M.,  {Stiening} R.,  {Weinberg} M.~D.,
  {Schneider} S.,  {Carpenter} J.~M.,  {Beichman} C.,    {Capps} R., et~al.,
  2006, AJ, 131, 1163

\bibitem[\protect\citeauthoryear{{Smith}, {Cushing}, {Barletta}, {McCarthy},
  {Kulesa} \& {Van Dyk}}{{Smith} et~al.}{2012}]{smith12}
{Smith} J.~D.~T.,  {Cushing} M.,  {Barletta} A.,  {McCarthy} D.,  {Kulesa} C.,
    {Van Dyk} S.~D.,  2012, AJ, 144, 166

\bibitem[\protect\citeauthoryear{{Smith}}{{Smith}}{1968}]{smith68}
{Smith} L.~F.,  1968, MNRAS, 141, 317

\bibitem[\protect\citeauthoryear{{Smith}, {Shara} \& {Moffat}}{{Smith}
  et~al.}{1990}]{smith90}
{Smith} L.~F.,  {Shara} M.~M.,    {Moffat} A.~F.~J.,  1990, ApJ, 358, 229

\bibitem[\protect\citeauthoryear{{Smith}, {Shara} \& {Moffat}}{{Smith}
  et~al.}{1996}]{smith96}
{Smith} L.~F.,  {Shara} M.~M.,    {Moffat} A.~F.~J.,  1996, MNRAS, 281, 163


\bibitem[\protect\citeauthoryear{{Smith}}{{Smith}}{2006}]{smith06}
{Smith} N.,  2006, ApJ, 644, 1151

\bibitem[\protect\citeauthoryear{{Smith} \& {Tombleson}}{{Smith} \& {Tombleson}}{2014}]{smith14}
{Smith} N., {Tombleson} R., 2014, MNRAS, accepted, arXiv:1406.7431

\bibitem[\protect\citeauthoryear{{Stead} \& {Hoare}}{{Stead} \&
  {Hoare}}{2009}]{stead09}
{Stead} J.~J.,  {Hoare} M.~G.,  2009, MNRAS, 400, 731

\bibitem[\protect\citeauthoryear{{Stock} \& {Barlow}}{{Stock} \&
  {Barlow}}{2010}]{stock10}
{Stock} D.~J.,  {Barlow} M.~J.,  2010, MNRAS, 409, 1429


\bibitem[\protect\citeauthoryear{{Todt}, {Kniazev}, {Gvaramadze}, {Hamann},
  {Buckley}, {Crause}, {Crawford}, {Gulbis} \& et al.}{{Todt}
  et~al.}{2013}]{todt13}
{Todt} H., et al., 2013, MNRAS, 430, 2302

\bibitem[\protect\citeauthoryear{{Tovmassian}, {Navarro} \&
  {Cardona}}{{Tovmassian} et~al.}{1996}]{tovmassian96}
{Tovmassian} H.~M.,  {Navarro} S.~G.,    {Cardona} O.,  1996, AJ, 111, 306

\bibitem[\protect\citeauthoryear{{Turner}}{{Turner}}{1980}]{turner80}
{Turner} D.~G.,  1980, ApJ, 235, 146

\bibitem[\protect\citeauthoryear{{Turner}, {Rohanizadegan}, {Berdnikov} \&
  {Pastukhova}}{{Turner} et~al.}{2006}]{turner06}
{Turner} D.~G.,  {Rohanizadegan} M.,  {Berdnikov} L.~N.,    {Pastukhova} E.~N.,
   2006, PASP, 118, 1533

\bibitem[\protect\citeauthoryear{{Tuthill}, {Monnier}, {Lawrance}, {Danchi},
  {Owocki} \& {Gayley}}{{Tuthill} et~al.}{2008}]{tuthill08}
{Tuthill} P.~G.,  {Monnier} J.~D.,  {Lawrance} N.,  {Danchi} W.~C.,  {Owocki}
  S.~P.,    {Gayley} K.~G.,  2008, ApJ, 675, 698

\bibitem[\protect\citeauthoryear{{Underhill} \& {Hill}}{{Underhill} \&
  {Hill}}{1994}]{underhill94}
{Underhill} A.~B.,  {Hill} G.~M.,  1994, ApJ, 432, 770

\bibitem[\protect\citeauthoryear{{van der Hucht}}{{van der
  Hucht}}{2001}]{vdH01}
{van der Hucht} K.~A.,  2001, New.Astron.Rev, 45, 135

\bibitem[\protect\citeauthoryear{{van der Hucht}}{{van der
  Hucht}}{2006}]{vdH06}
{van der Hucht} K.~A.,  2006, A\&A, 458, 453

\bibitem[\protect\citeauthoryear{{van Leeuwen}}{{van
  Leeuwen}}{2007}]{leeuwen07}
{van Leeuwen} F.,  2007, A\&A, 474, 653

\bibitem[\protect\citeauthoryear{{Varricatt} 
\& {Ashok}}{{Varricatt}\&{Ashok}}{2006}]{varricatt06} Varricatt, W.~P., \& Ashok, N.~M., 2006, MNRAS, 365, 127


\bibitem[\protect\citeauthoryear{{V{\'a}zquez} \& {Baume}}{{V{\'a}zquez} \&
  {Baume}}{2001}]{vazquez01}
{V{\'a}zquez} R.~A.,  {Baume} G.,  2001, A\&A, 371, 908

\bibitem[\protect\citeauthoryear{{V{\'a}zquez}, {Baume}, {Feinstein},
  {Nu{\~n}ez} \& {Vergne}}{{V{\'a}zquez} et~al.}{2005}]{vazquez05}
{V{\'a}zquez} R.~A.,  {Baume} G.~L.,  {Feinstein} C.,  {Nu{\~n}ez} J.~A.,
  {Vergne} M.~M.,  2005, A\&A, 430, 471

\bibitem[\protect\citeauthoryear{{Vazquez}, {Will}, {Prado} \&
  {Feinstein}}{{Vazquez} et~al.}{1995}]{vazquez95}
{Vazquez} R.~A.,  {Will} J.-M.,  {Prado} P.,    {Feinstein} A.,  1995, A\&AS,
  111, 85

\bibitem[\protect\citeauthoryear{{Wachter}, {Mauerhan}, {Van Dyk}, {Hoard},
  {Kafka} \& {Morris}}{{Wachter} et~al.}{2010}]{wachter10}
{Wachter} S.,  {Mauerhan} J.~C.,  {Van Dyk} S.~D.,  {Hoard} D.~W.,  {Kafka} S.,
     {Morris} P.~W.,  2010, AJ, 139, 2330

\bibitem[\protect\citeauthoryear{{Wegner}}{{Wegner}}{2006}]{wegner06}
{Wegner} W.,  2006, MNRAS, 371, 185

\bibitem[\protect\citeauthoryear{{Williams} et al.(1990\natexlab{a})}{{Williams} et al.}{1990\natexlab{a}}]{williams90a} Williams, P.~M., van 
der Hucht, K.~A., Pollock, A.~M.~T., Florkowski, D.~R., van der Woerd, H., Wamsteker, W.~M., 1990\natexlab{a}, MNRAS, 243, 662 

\bibitem[\protect\citeauthoryear{{Williams}, {van der Hucht}, {Sandell} \&
  {The}}{{Williams} et~al.}{1990\natexlab{b}}]{williams90b}
{Williams} P.~M.,  {van der Hucht} K.~A.,  {Sandell} G.,    {The} P.~S.,  1990\natexlab{b},
  MNRAS, 244, 101

\bibitem[\protect\citeauthoryear{{Williams} et al.(1992)}{{Williams} et~al.}{1992}]{williams92} Williams, P.~M., van 
der Hucht, K.~A., Bouchet, P., et al., 1992, MNRAS, 258, 461 

\bibitem[\protect\citeauthoryear{{Williams}, {Kidger}, {van der Hucht},
  {Morris}, {Tapia}, {Perinotto}, {Morbidelli}, {Fitzsimmons}, {Anthony},
  {Caldwell}, {Alonso} \& {Wild}}{{Williams} et~al.}{2001}]{williams01}
{Williams} P.~M., et~al., 2001, MNRAS, 324, 156

\bibitem[\protect\citeauthoryear{{Williams}, {Rauw} \& {van der Hucht}}{{Williams}, {Rauw} \& {van der Hucht}}{2009\natexlab{b}}]{williams09b} Williams, P.~M., Rauw, 
G., \& van der Hucht, K.~A., 2009{\natexlab{b}}, MNRAS, 395, 2221


\bibitem[\protect\citeauthoryear{{Wright}, {Eisenhardt}, {Mainzer}, {Ressler},
  {Cutri}, {Jarrett}, {Kirkpatrick}, {Padgett} \& {McMillan}}{{Wright}
  et~al.}{2010}]{wright10}
{Wright} E.~L., et~al., 2010, AJ, 140, 1868

\bibitem[\protect\citeauthoryear{{Yusof}, {Hirschi}, {Meynet}, {Crowther},
  {Ekstr{\"o}m}, {Frischknecht}, {Georgy}, {Abu Kassim} \& {Schnurr}}{{Yusof}
  et~al.}{2013}]{yusof13}
{Yusof} N., et~al., 2013, MNRAS, 433, 1114

\end{thebibliography}
\end{document}